\documentclass[reprint,eqsecnum,twocolumn,tightenlines,floats,floatfix,prc,
aps,showpacs]{revtex4-1}

\def\bea{\begin{eqnarray}}
\def\eea{\end{eqnarray}}

\newcommand{\ket}[1]{|#1\rangle}

\def\be{\begin{equation}}
\def\ee{\end{equation}}
\def\ba{\begin{eqnarray}}
\def\ea{\end{eqnarray}}
\def\bal{\begin{align}}

\usepackage{graphics}
\usepackage{graphicx}
\usepackage{epsf} 
\usepackage{amsmath}
\usepackage{amssymb}
\usepackage{slashed}
\usepackage{dcolumn}

\usepackage{bm}
\usepackage{color}
\usepackage{ulem}
\newcommand{\change}[2]{#2}

\def\sfrac#1#2{{\textstyle \frac{#1}{#2}}}

\begin{document}

\phantom{0}
\vspace{-0.3in}
\hspace{5.5in}\parbox{1.5in}{ \leftline{JLAB-THY-10-1193}
                \leftline{}\leftline{}\leftline{}\leftline{}
}

\title
{\bf Covariant spectator theory of $np$ scattering:\\  Effective range expansions and relativistic deuteron wave functions}

\author{Franz Gross\vspace{-0.15in}}
\email{gross@jlab.org} 
\affiliation{
College of William and Mary, Williamsburg, Virginia 23185\vspace{-0.15in}}
\affiliation{
Thomas Jefferson National Accelerator Facility, Newport News, VA 23606\vspace{-0.1in}}
\author{Alfred Stadler\vspace{-0.15in}}
\email{stadler@cii.fc.ul.pt}
\affiliation{
$^3$Centro de F\'\i sica Nuclear da Universidade de Lisboa, P-1649-003 Lisboa, Portugal\vspace{-0.15in}}
\affiliation{
$^4$Departamento de F\'\i sica da Universidade de \'Evora, P-7000-671 \'Evora, Portugal
}


\date{\today}

\begin{abstract} 

We present the effective range expansions for the $^1S_0$ and $^3S_1$ scattering phase shifts, and the relativistic deuteron wave functions that accompany our recent high precision fits (with $\chi^2/N_\mathrm{data} \simeq 1$) to the  2007 world $np$  data below 350 MeV.  The wave functions are expanded in a series of analytical functions (with the correct asymptotic behavior at both large and small arguments)  that can be Fourier-transformed from momentum to coordinate space and are convenient to use in any application.  A fortran subroutine to compute these wave functions can be obtained from the authors.
\end{abstract}


\preprint{JLAB-THY-10-???}

\maketitle


\section{Introduction}

This paper presents the effective range expansions and relativistic deuteron wave functions that accompany the recent high precision fits to the  2007 world $np$  data (containing 3788 data)  below 350 MeV \cite{Gross:2008ps}.  These fits, obtained using the covariant spectator theory (CST) \cite{Gross:1969rv,Gross:1972ye}, resulted in two new one boson exchange models of the $np$ interaction, both of which also reproduce the experimental triton binding energy without introducing additional irreducible three-nucleon forces \cite{Stadler:1997iu}.   One model (WJC-1) has 27 parameters and  fits with a $\chi^2/N_\mathrm{data} = 1.06$.  The other model  (WJC-2)  has only 15 parameters and fits with a $\chi^2/N_\mathrm{data} = 1.12$.    

The main body of the paper is divided into three sections.  In Sec.~II the 6th order effective range expansions for the $^1S_0$ and $^3S_1$ phase shifts are given; it is shown that these expansions provide an excellent description of the phase shifts up to 50 MeV lab kinetic energy.  Then, in Sec.~III, the relativistic deuteron wave functions that automatically emerge from these fits are discussed.  The deuteron binding energy was constrained during the fits, so both models have the correct binding energy.  

Section III begins with a review of the definitions of the wave functions and a description of how they are related to the relativistic $dnp$ vertex functions defined in the CST.   It is shown how the CST wave functions are decomposed into four independent amplitudes, and how these can be identified with the familiar $u$ (S-state) and $w$ (D-state) components, plus two P-state components of purely relativistic origin \cite{Buck:1979ff,BbC}.  Because these P-states are associated with the virtual antinucleon degrees of freedom they have positive parity.  One P-wave component has a symmetric total spin triplet structure (denoted $v_t$) and the other an antisymmetric total spin singlet structure (denoted $v_s$).  The antisymmetric $v_s$ wave function would be zero if both nucleons had the same energy, but because one is on-shell and the other off-shell, the relative energy is $E_p-\frac12M_d$ (where $E_p$ is the total relativistic energy of a nucleon with three momentum ${\bf p}$ and $M_d$ is the deuteron mass), allowing for the existence of this odd state.  The normalization of the wave functions is discussed \cite{Adam:1997rb}.  It turns out that the probability of the WJC-1 P-state components is only about 0.3\% while the WJC-2 P-state components are much smaller (with a combined probability of less than 0.02\%).  These components are retained because they are {\it required by the manifest covariance of the CST\/}, but they are not present in some other relativistic approaches \cite{Keister:1991sb}.  The wave functions are expanded in a series of analytic functions that can be conveniently Fourier-transformed to coordinate space, giving analytical wave functions in both momentum and coordinate space convenient to use in any application.  The section concludes with a discussion of the asymptotic $D/S$ ratio.  Calculations of the magnetic and quadrupole moments require the evaluation of interaction currents, which is beyond the scope of this paper and is deferred to future work.  

Our results are summarized in the conclusion, Sec.~IV.  There are also three Appendices. Appendix \ref{app:A} gives details of the construction and evaluation of the kernels omitted from Ref.~I \cite{Gross:2008ps}.   Appendix \ref{app:B} discusses the method used to evaluate the angular integrals, complicated by the presence of a cusp in the region of integration.  Finally, Appendix \ref{app:C} gives more details about the definitions, extraction, and normalization of deuteron wave functions.


\section{Effective range expansions}

The four-term effective range expansion for the $^1S_0$ and $^3S_1$ phases that we use is 
\bea
k\cot\delta=-\frac1a+r_0k^2\Big(\sfrac12 +(r_0k)^2\big[P+Q\,(r_0k)^2\big]\Big)\qquad\label{eq:2.1}
\eea
where $a$ is the scattering length, $r_0$ the effective range, and $P$ and $Q$ the dimensionless third and fourth order parameters.  Here 
\bea
k^2=\sfrac12 mE_{\rm lab} \label{eq:22}
\eea
where $k$ is the magnitude of the cm momentum of the nucleons, $E_{\rm lab}$ the laboratory kinetic energy, and we used $m=938.9$ MeV and $\hbar c$= 197.3288 MeV fm = 1.  Recall that Eq.~(\ref{eq:22}) holds for both relativistic and nonrelativistic kinematics, allowing us to compare our results directly with nonrelativistic calculations.

To fix these parameters we evaluated the phase shifts at four energies: 0.0001, 10, 25, and 50 MeV. The first energy of 0.0001 MeV is so low that it effectively fixes the scattering length (and hence the total cross section at the $np$ threshold).  The use of these four energies resulted in a small error between the phases calculated at 1 and 5 MeV, and those predicted using (\ref{eq:2.1}).  The effective range parameters and the errors at 1 and 5 MeV are summarized in Table \ref{tab:effective}.

\begin{table}[b]
\begin{minipage}{3in}
\caption{The effective range parameters defined in Eq.~(\ref{eq:2.1}) and the fractional errors $\Delta$ at 1 and 5 MeV for each model. The fractional error is defined to be $\Delta=(\cot\delta-\cot\delta_{{\rm eff}})/\cot\delta$, where $\delta_{\rm eff}$ is the phase predicted by Eq.~(\ref{eq:2.1}) and $\delta$ the phase emerging from solutions of the scattering equation.} 
\label{tab:effective}
\begin{ruledtabular}
\begin{tabular}{lrrrr}
& \multicolumn{2}{c}{$^1S_0$} & \multicolumn{2}{c}{$^3S_1$} \cr
parameter & WJC-1  & WJC-2  & WJC-1& WJC-2 \cr
\tableline
$a$ (fm) & $-$23.7494 & $-$23.7496   &  5.4295 & 5.4342  \cr
$r_0$ (fm) & 2.6261    &     2.6623    &  1.7601   & 1.7666       \cr
 $P$  &    0.0007   &   $-$0.0005      &   0.0172  & 0.0206       \cr
  $Q$ & 0.0033 &  0.0032 &     0.0060  & 0.0058 \cr
$\Delta$ (1 MeV) & 0.0044 & 0.0038 & $-$0.0011& $-$0.0009\cr 
 $\Delta$ (5 MeV) & 0.0033 & 0.0030 & $-$0.0012 & $-$0.0012
\end{tabular}
\end{ruledtabular}
\end{minipage}
\end{table}
%

\begin{figure}
\includegraphics[width=3in]{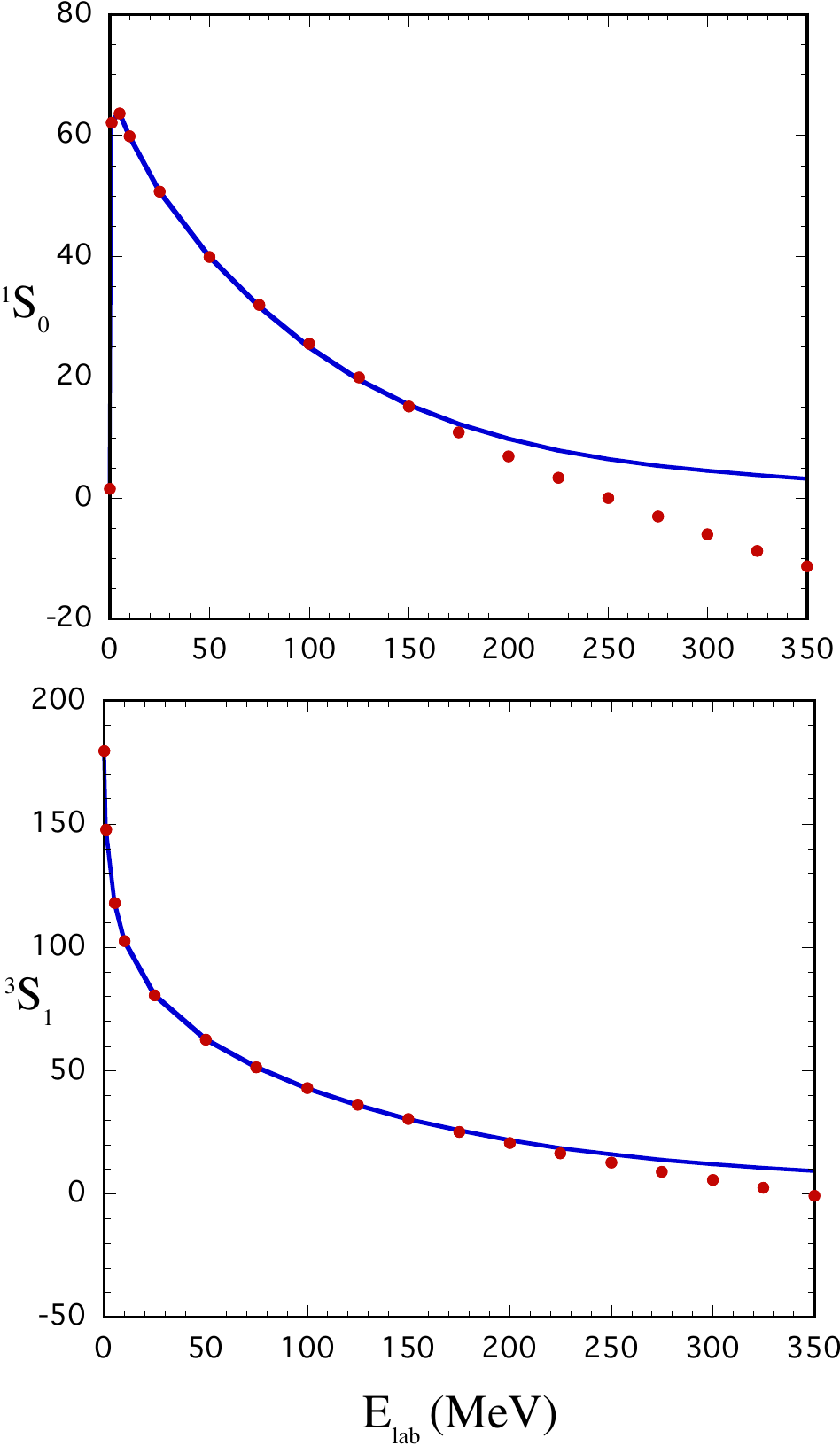}
\caption{\footnotesize\baselineskip=10pt (Color online) The effective range expansions for the S-waves (smooth line) compared to calculated phase shifts (solid dots).  The phases are in degrees. }
\label{fig:phases}
\vspace{0.5in}
\end{figure}

The effective range expansion is surprisingly accurate over a wide energy range.  As shown in Fig.~\ref{fig:phases}, the expansions are qualitatively accurate up to $E_{\rm lab}$ of about 150 MeV (the small errors shown in Table \ref{tab:effective} are completely invisible on the figure).  In the fitting we used the expansions instead of the phase shifts to calculate the $S$ wave contributions to the observables of all energies below 50 MeV.  In doing the final minimizations we constrained both the deuteron binding energy and the $^1S_0$ phase shift at 0.0001 MeV (which we fixed at $\delta=1.4937^{\rm o}$).  This was a very effective way to maintain an accurate fit to the low energy cross section data.  

\section{Deuteron wave functions}

The CST deuteron wave functions have been developed and defined in a number of references.  The earliest reference is the work of Buck and Gross \cite{Buck:1979ff}; more recent references include Refs.~\cite{GVOH} (referred to as Ref.~II) and Ref.~\cite{Adam:2002cn}, where many practical details for how to use  the wave functions in practical calculations are developed.  While all of these references use the same basic definitions, details of how to interpret and use the wave functions have improved with time.  In section we first present a brief review of the necessary definitions (relating them to the earlier references), and then present the numerical results for the wave functions of models WJC-1 and WJC-2.

\subsection{Definitions of the vertex function and the covariant  wave function}

The equation satisfied by the $np$ scattering amplitude is (in the notation of Eq.~(2.3) of Ref.~I)
\begin{align}
M_{12}(p, p'; P)=&
\overline {V}_{12}(p, p'; P)\cr
-\int_{k_1} &
\overline {V}_{12}(p,k;P)G_2(k_2)
{M}_{12}(k,p';P)\, ,\qquad \label{eq:spec}
\end{align}
where $P$ is the conserved total four-momentum, and $p, p'$, and $k$ are relative four-momenta related to the momenta of particles  1 and 2 by
$p_1=\sfrac12 P+p$, $p_2=\sfrac12 P-p$. $E_{k_1}=\sqrt{m^2+{\bf k}_1^2}$ is the energy of the on-shell particle 1, and the covariant integral is
\bea
\int_k\equiv\int \frac{d^3 k}{(2\pi)^3} \frac{m}{E_{k}} \, .
\eea
Note that these covariant operators can be written either in terms of the independent momenta $\{P,p,p'\}$ or $\{P, p_1,p_1'\}$.  The scattering amplitude,
\begin{align}
&M_{12}(p, p'; P)\equiv M_{\lambda_1\lambda'_1,\beta\beta'}(p, p'; P)\cr
&\qquad=\bar u_\alpha({\bf p}_1,\lambda_1){\cal M}_{\alpha\alpha',\beta\beta'}(p, p'; P)u_{\alpha'}({\bf p}'_1,\lambda_1'),\qquad
\end{align}
is the matrix element of the Feynman scattering amplitude ${\cal M}$ between positive-energy Dirac spinors of particle 1.  The definition of the nucleon spinors $u({\bf p}_1,\lambda_1)$ (with $\lambda_1$ the helicity of the nucleon) is given in Eq.~(\ref{A1});  the on-shell spinor has four-momentum $p_1$ with three momentum component ${\bf p}_1$, and $p_1^2=m^2$.
The propagator for the off-shell particle 2 is
\bea
G_2(k_2)\equiv G_{_{\beta\beta'}}(k_2)=\frac{\left(m+\slashed{k}_2\right)_{_{\beta\beta'}}}{m^2-k_2^2-i\epsilon}\,h^2(k_2)
\eea
with $k_2=P-k_1$, $k_1^2=m^2$, and $h(k_2)$ [denoted $H(k_2)$ in Ref.~I] the form factor of the off-shell nucleon (related to its self energy), normalized to unity when $k_2^2=m^2$
\bea
h(k_2)=\left[\frac{(\Lambda_N^2-m^2)^2}{(\Lambda^2_N-m^2)^2+(m^2-k_2^2)^2}\right]^2\, . \label{eq:nff}
\label{nuclff}
\eea
See Appendix \ref{app:C} for further discussion of the nucleon form factor $h$ and the role it plays in the deuteron wave functions.

\begin{figure}
\includegraphics[width=1.5in]{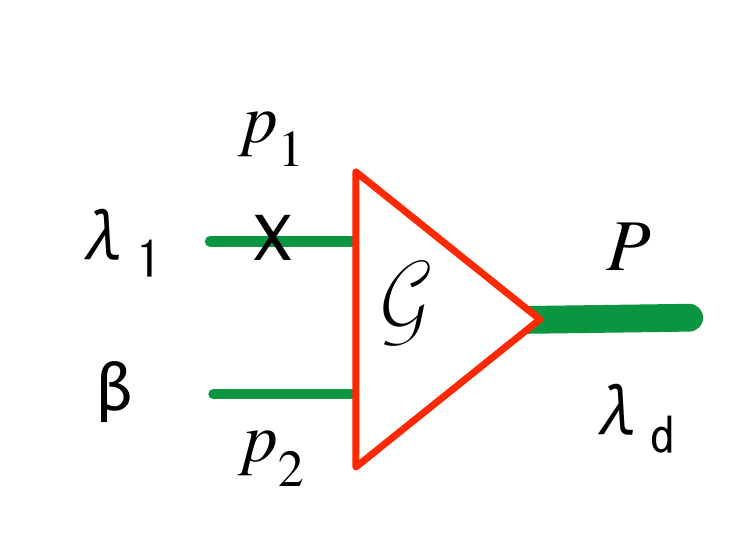}
\caption{\footnotesize\baselineskip=10pt (Color online) Diagrammatic representation of the vertex function (\ref{eq:32a}).  The on-shell particle is labeled by the $\times$, so that $p_1^2=m^2$, and the relative four-momentum is $p=\sfrac12(p_1-p_2)$.  The deuteron is also on-shell, with $P^2=M_d^2$.}
\label{fig:vertex}
\end{figure} 

Just below threshold, the $np$ scattering amplitude has a pole at the deuteron mass, $P^2=M_d^2$.  Near the pole this amplitude, in the CST, can be written 
\begin{align}
M_{\lambda_1\lambda'_1,\beta\beta'}&(p,p';P)=R_{\lambda_1\lambda'_1,\beta\beta'}(p,p';P)
\nonumber\\
&-\sum_{\lambda_d}\frac{{\cal G}_{\lambda_1\beta}(p,P,\lambda_d)\;\bar{ {\cal G}}_{\lambda'_1\beta'}(p',P,\lambda_d)}{M_d^2-P^2-i\epsilon} \label{eq:3.1}
\end{align}
where $\lambda_1, \lambda'_1$ are the helicities of the outgoing and incoming on-shell particle 1, $\beta, \beta'$ are the Dirac indices of the off-shell particle 2, $\lambda_d$ is the helicity of the deuteron, ${\cal G}$ is the {\it contracted\/} $dnp$ vertex functions describing the coupling of the deuteron to the neutron and proton, and $R$ is a remainder function finite at the pole.  
The equation for ${\cal G}$ can be found by substituting (\ref{eq:3.1}) into the scattering equation  (\ref{eq:spec}) and demanding that it hold at the pole (see Refs.~\cite{Gross:1972ye,Adam:1997rb} and II), giving
\bea
{\cal G}_{\lambda_1\beta}(p,P,\lambda_d)=&&-\int_{k_1'}  
\overline {V}_{\lambda_1\lambda_1',\beta\alpha}(p,k;P)
\nonumber\\
&&\qquad\times G_{\alpha\beta'}(k_2)\,
{\cal G}_{\lambda_1'\beta'}(k,P,\lambda_d) \, .\qquad
\label{eq:36}
\eea

The contracted vertex function is a Lorentz scalar product of a Dirac matrix element and the covariant polarization vector of the deuteron, $\xi_\mu(\lambda_d)$:
\bea
{\cal G}_{\lambda_1\beta}(p,P,\lambda_d)\equiv
{\cal O}^\mu_{\lambda_1\beta}(p,P)\,\xi_\mu(\lambda_d) \, , \label{eq:32a}
\eea
where ${\cal O}$ is the ``uncontracted'' $dnp$ vertex function for particle 2 off-shell [c.f. Eq.~(2.27) of Ref.~II]
\bea
{\cal O}^\mu_{\lambda_1\beta}(p,P)\equiv\Big[\Gamma^\mu(p,P)C\Big]_{\beta\beta'}\bar u^T_{\beta'}({\bf p}_1,\lambda_1) \, , \label{eq:32}
\eea
with $C$ the Dirac charge conjugation matrix. 
To simplify the language, ${\cal G}$ will always be called the contracted vertex function, and the functions ${\cal O}$ (or sometimes $\Gamma$) will be referred to simply as the ``vertex function''.  The contracted vertex function  is illustrated diagramatically in Fig.~\ref{fig:vertex}.

 The relativistic deuteron wave functions were defined previously in Eq.~(2.32) of Ref.~II  and Eq.~(2.29) of Ref.~\cite{Adam:2002cn}: 
 \bea
 \psi_{\lambda_1\beta,\lambda_d}(p,P)=N_d \,G_{\beta\beta'}(p_2) {\cal G}_{\lambda_1\beta'}({p},P,\lambda_d).\quad  \label{eq:d2}
 \eea
Here $N_d$ is a normalization constant,  chosen to be  
\bea
N_d=\frac1{\sqrt{(2\pi)^3\, 2M_d}}\, . \label{eq:nconst}
\eea
%
As discussed in Ref.~\cite{Adam:2002cn}, it is the vertex function that enters directly into any Feynman diagram involving an incoming or outgoing deuteron state, so if a Feynman amplitude is expressed in terms of the wave function instead of the vertex function, one must be careful to divide by the normalization constant (\ref{eq:nconst}).  The reason for this choice of normalization constant will be discussed further below.

The normalization of the wave function (\ref{eq:d2}) is discussed in detail in Refs.~\cite{Gross:1972ye,Adam:1997rb}, II, and Appendix \ref{app:C}.  The exact result, in the notation of Eq.~(\ref{eq:36}), is 
\bea
\delta_{\lambda_d\lambda'_d}=&&\frac{1}{2M_dN_d^2}\bigg\{\int_{p_1} {\psi}^\dagger_{\lambda_1\beta,\lambda_d}(p,P)\gamma^0_{\beta\beta'}\,{\psi}_{\lambda_1\beta',\lambda'_d}(p,P)
\nonumber\\
&&-\int_{p_1}\int_{p'_1}  {\psi}^\dagger_{\lambda_1\beta,\lambda_d}(p,P)\,
\Delta \overline {V}_{\lambda_1\lambda'_1,\beta\beta'}(p,p';P)
\nonumber\\
&& \quad\qquad\qquad\times{\psi}_{\lambda_1\beta,\lambda'_d}(p',P)\bigg\}\, ,\qquad
\label{eq:relnorm}
\eea
where summation over repeated indices is implied, and
\bea
\Delta  \overline {V}_{\lambda_1\lambda'_1,\beta\beta'}(p,p';P) = \frac{\partial}{\partial M_d}\overline {V}_{\lambda_1\lambda'_1,\beta\beta'}(p,p';P) \, ,\qquad
\eea
with the partial derivative holding $p_1$ and $p_1'$ constant.
This normalization integral will be simplified and related to the usual nonrelativistic normalization in Appendix~\ref{app:norm} below.

While it is often convenient to use the fully covariant expression (\ref{eq:d2}), the physical content of the wave function can be displayed if we expand the off-shell propagator into positive and negative energy pieces.  (Note that an alternative expansion given in Eq.~(A21) of Ref.~\cite{Adam:2002cn}  is sometimes more convenient for applications.)   Working in the rest frame, where we choose ${\bf p}_1=-{\bf p}_2={\bf p}$, the propagator can be decomposed 
\bea
G_{\alpha\beta}&&(p_2)=\frac{m}{E_p}\sum_{\lambda_2}\bigg\{
\frac{u_\alpha(-{\bf p},\lambda_2)\bar u_\beta(-{\bf p},\lambda_2)}{2E_p-M_d}
\nonumber\\ 
&&\qquad\qquad\qquad-\frac{v_\alpha({\bf p},\lambda_2)\bar v_\beta({\bf p},\lambda_2)}{M_d}\bigg\},\qquad
\nonumber\\
&&=\frac{m}{E_p}\sum_{\lambda_2\rho_2} G^{\rho_2}(p)
u^{\rho_2}_\alpha(-{\bf p},\lambda_2) \bar u^{\rho_2}_\beta(-{\bf p},\lambda_2)\qquad
\eea
where 
$\rho_2$ is the $\rho$-spin of the off-shell particle 2.  The $\rho$-spin propagators $G^\rho(p)=G^\pm(p)$ are
\bea
G^+(p)=\frac1{2E_p-M_d}\qquad G^-(p)=-\frac1{M_d},\label{eq:Groh}
\eea
and $u$ and $v$ are the positive- and negative-energy spinors of particle 2, with $u^\rho=u^\pm$ and $u^+(-{\bf p},\lambda_2)=u(-{\bf p},\lambda_2)$ and $u^-(-{\bf p},\lambda_2)=v({\bf p},\lambda_2)$ [see Eq.~(\ref{A1})].  Using this notation the wave function can be conveniently written
 \bea
 \psi_{\lambda_1\alpha,\lambda_d}(p,P)=\sum_{\rho_2\lambda_2}\psi^{\rho_2}_{\lambda_1\lambda_2,\lambda_d}({\bf p}) u_\alpha^{\rho_2}(-{\bf p},\lambda_2)\, ,\qquad\label{eq:34a}
  \eea
where the wave functions $\psi^{\rho_2}$ are
\bea
\psi^{\rho_2}_{\lambda_1\lambda_2,\lambda_d}({\bf p})&=&N_d\frac{m}{E_p} G^{\rho_2}(p)\, {\bf \Gamma}_{\lambda_1\lambda_2,\lambda_d}^{\rho_2}({\bf p}).\quad
\label{eq:d1}
\eea
Suppressing the Dirac indices in the matrix elements gives
\begin{align}
&{\bf \Gamma}_{\lambda_1\lambda_2,\lambda_d}^{\rho_2}({\bf p})=\bar u^{\rho_2}_\beta(-{\bf p},\lambda_2){\cal G}_{\lambda_1\beta}(p,P,\lambda_d)
\nonumber\\
&\qquad=\bar u^{\rho_2}(-{\bf p},\lambda_2)\Gamma^\mu({p},P)C\bar u^T({\bf p},\lambda_1)\,\xi_\mu(\lambda_d). \label{eq:37}
\end{align}
These definitions agree with Eq.~(2.67) of Ref.~II.  Note that ${\bf \Gamma}_{\lambda_1\lambda_2,\lambda_d}^{\rho_2}({\bf p})$ contains no additional factors of $m/E$; these are written explicitly in (\ref{eq:d1}), and will be identified whenever they appear below.

\subsection{Partial wave CST equations for the deuteron vertex functions} \label{sec:3C}

Equation (\ref{eq:spec}), and the companion equation (\ref{eq:36}) for the bound state are solved by expanding the amplitudes into partial waves.  The connection between the wave functions in momentum space and their partial waves will be briefly outlined in this subsection; for further details see Appendix E of Ref.~I, and, for the deuteron channel, Appendix \ref{app:A}.

The first step is to extract the $J=1$ partial waves from the momentum space amplitudes ${\bf \Gamma}_{\lambda_1\lambda_2,\lambda_d}^{\rho_2}({\bf p})$ defined in Eq.~(\ref{eq:37}).  These amplitudes have a simple angular dependence.  If ${\bf p}={\rm p}\{\sin\theta\cos\phi,\sin\theta\sin\phi,\cos\theta\}$ (where, from here up to section \ref{sec:3G} below and again in Appendix \ref{app:C}, we will use the notation ${\rm p}=|{\bf p}|$  in order to avoid confusion with the four-momentum $p$), the form of these functions  is
\bea
{\bf \Gamma}_{\lambda_1\lambda_2,\lambda_d}^{\rho_2}({\bf p})=\sqrt{\sfrac{3}{4\pi}}\, D^{1*}_{\lambda_d \lambda}&&(\phi,\theta,0)
{\bf \Gamma}_{\lambda_1\lambda_2,\lambda_d}^{\rho_2}({\rm p}),\qquad\label{eq:16a}
\eea
where $\lambda=\lambda_1-\lambda_2$,  $\lambda_d=M_J$ is the polarization of the deuteron, and $D^1$ are the $J=1$ rotation matrices. (This can be obtained from Eq.~(E29) of Ref.~I  by  recalling that the scattering amplitude at the deuteron pole is proportional  to the product of vertex functions for the incoming and outgoing nucleons.) In order to keep our notation succinct, we suppress the usual index $J$ on the partial-wave vertex function ${\bf \Gamma}_{\lambda_1\lambda_2,\lambda_d}^{J\,\rho_2}({\rm p})$. An argument $\rm p$ instead of $\bf p$ identifies it unambiguously as a partial-wave vertex function. If we choose ${\bf p}$ to lie in the $xz$ plane ($\phi=0$), we can use (\ref{eq:16a}) to find the connection between  ${\bf \Gamma}_{\lambda_1\lambda_2,\lambda_d}^{\rho_2}({\bf p})$ and ${\bf \Gamma}_{\lambda_1\lambda_2,\lambda_d}^{\rho_2}({\rm p})$.  Using the normalization condition for the $d^J$'s
\bea
1=\sfrac12(2J+1)\int_0^\infty \sin\theta d\theta \left[d^J_{\lambda'\lambda}(\theta)\right]^2, \label{eq:djnorm}
\eea
the relation is
\bea
{\bf \Gamma}_{\lambda_1\lambda_2,\lambda_d}^{\rho_2}({\rm p})&=&\sqrt{{3\pi}}\int_0^\pi \sin\theta\, d\theta\, 
\nonumber\\&&\quad\times\;
d^{1}_{\lambda_d,\lambda}(\theta)\, 
{\bf
\Gamma}_{\lambda_1\lambda_2,\lambda_d}^{\rho_2}({\bf p})\, ,\qquad \label{eq:320a}
\eea
and the deuteron wave functions corresponding to (\ref{eq:d1}) are
\bea
\psi^{\rho_2}_{\lambda_1\lambda_2,\lambda_d}({\rm p})&=&N_d\frac{m}{E_p} G^{\rho_2}(p)\, {\bf \Gamma}_{\lambda_1\lambda_2,\lambda_d}^{\rho_2}({\rm p}).\quad
\label{eq:d1a}
\eea

The next step is to use the properties of the partial wave amplitudes under parity and particle interchange (derived in Refs.~I and II) to show that there are only four independent vertex functions. We begin by restoring reference to the $\rho$-spin of particle 1 and the relative energy $p_0=(p_{10}-p_{20})/2$ (previously suppressed for convenience), which  transforms the partial wave vertex functions (\ref{eq:320a}) into
\bea
{\bf
\Gamma}_{\lambda_1\lambda_2,\lambda_d}^{\rho_2}({\rm p})\to {\bf
\Gamma}_{\lambda_1\lambda_2,\lambda_d}^{+\rho_2}({\rm p},p_0)\, ,
\eea
where, in the cm frame with particle 1 on shell, $p_0=E_p-M_d/2>0$.
Under parity (${\cal P}$) and particle interchange (${\cal P}_{12}$), it was shown in Appendix E of Ref.~I that these partial wave amplitudes (for $J=1$) satisfy the transformation properties
\bea
{\cal P}\,{\bf \Gamma}_{\lambda_1\lambda_2,\lambda_d}^{+\rho_2}({\rm p},p_0)&=&\rho_2{\bf \Gamma}_{-\!\lambda_1\,-\!\lambda_2,\lambda_d}^{+\rho_2}({\rm p},p_0)
\nonumber\\
{\cal P}_{12}{\bf \Gamma}_{\lambda_1\lambda_2,\lambda_d}^{+\rho_2}({\rm p},p_0)&=&{\bf \Gamma}_{\lambda_2\lambda_1,\lambda_d}^{\rho_2+}({\rm p}, -p_0)\, .\label{eq:321}
\eea
The parity relation shows that only two (of the four possible) helicity states are independent (which we chose to be $\lambda_1=+$; the states with $\lambda_1=-$ can be obtained using parity) and the exchange relation shows that the exchange amplitudes (with particle 2 on-shell and {\it negative\/} relative energy) can be obtained from the direct amplitudes (with particle 1 on-shell and {\it  positive\/} relative energy).  The detailed construction of the deuteron bound state equation is reviewed in Appendix \ref{app:C1}, where it is shown how these symmetry properties are derived and used to reduce the initial set of coupled equations to only four independent ones.

Using these results, the vertex functions (\ref{eq:320a}) can be organized into the column vector  
\bea
\ket{{\bf \Gamma}_{\lambda_d}(p)}=N_d\frac{m}{E_p}  \left(\begin{array}{c}
 {\bf \Gamma}_{++,\lambda_d}^+({\rm p})  \\[0.15in]
 {\bf \Gamma}_{++,\lambda_d}^-({\rm p}) \\[0.15in]
 {\bf \Gamma}_{+-,\lambda_d}^+({\rm p}) \\[0.15in]
 {\bf \Gamma}_{+-,\lambda_d}^-({\rm p}) 
\end{array}
\right)
\, , \label{eq:15}
\eea
where, in the matrix representations, we return to the notation $p$ to represent the momentum dependence of the vertex function.
With this notation, the original Eq.~(\ref{eq:36}) can now be written as four independent equations for these vertex functions.  In a  convenient matrix form the equations are
\bea
\ket{{\bf \Gamma}_{\lambda_d}(p)} =-\int_0^\infty k^2 dk\,{\bf V_{d}}(p,k)\;{\bf g}(k)\;\ket{{\bf \Gamma}_{\lambda_d}(k)} \, ,\qquad\label{eq:CST}
\eea
where ${\bf V_{d}}(p,k)$  (the kernel at the deuteron pole) and ${\bf g}(k)$ are 4$\times$4 matrices.  The equations are independent of the projection of the total angular momentum, which for the deuteron is its helicity ($m_J=\lambda_d$).  With the factors of $m/E$ explicitly included in the definition of the column vector (\ref{eq:15}) and in the kernel ${\bf V_{d}}(p,k)$, the propagator is the 4$\times$4 diagonal matrix:
\bea
{\bf g}(k)=
\left(
\begin{array}{cccc}
 G^+(k)  & 0 & 0 & 0 \\[0.15in]
 0 & G^-(k)& 0 & 0  \\[0.15in]
 0 & 0 &  G^+(k)  & 0 \\[0.15in]
0 & 0 & 0 &   G^-(k)  
\end{array} \label{eq:12a}
\right),\qquad
\eea
with $G^\rho(k)$ defined in Eq.~(\ref{eq:Groh}),  and the kernel is the 4$\times$4 matrix
\bea
{\bf V_{d}}(p,k)
&&=\left(\begin{array}{cc|cc} 
v_9^{++} & v_1^{+-}  &v^{++}_{11} &v_{3}^{+-}    \\[0.1in]
 v_{9}^{-+} &  v_{1}^{--} & v_{11}^{-+}  &v_{3}^{--} \\[0.05in]
\tableline\\[-0.1in]
 v^{++}_{12} & v_{8}^{+-}  & v_{10}^{++} & v_{6}^{+-} \\[0.1in]
 v_{16}^{-+} & v_{4}^{--}&v_{14}^{-+}& v_{2}^{--} 
\end{array}\right). \label{coupled}
\eea
where 
the $v_i^{\rho_1\rho_2}$ were previously defined in Eq.~(E45) of Ref.~I and also in Eq.~(\ref{eq:C10}).


\subsection{Decomposition into the $S$, $D$, and $P$ states}

The Dirac operator $\Gamma^\mu$ [from Eq.~(\ref{eq:32})] can be expanded in terms of four scalar invariant functions, $F, G, H,$ and $I$ as given in Eq.~(33) of Ref.~\cite{Buck:1979ff} and Eq.~(2.28) of Ref.~II,
\bea
\Gamma^\mu(p,P)&=&F\gamma^\mu + \frac Gm p^\mu
\nonumber\\
&&\qquad-\frac{(m-\not\!p)}{m}\Big(H\gamma^\mu + \frac Im p^\mu\Big).\qquad \label{eq:FGHI}
\eea
These functions differ by some factors of 2 from those originally introduced by Blankenbecler and Cook \cite{BbC}.  Since the deuteron is on mass-shell, so that $P^2=W^2=M_d^2$, they are functions of $p^2$ only.

Using the explicit definition of the spinors given in Appendix \ref{app:A} and substituting the expansion (\ref{eq:FGHI}) into the Dirac matrix elements (\ref{eq:37}), the wave functions (\ref{eq:d1}) can be reduced to 2-component matrix elements, with four independent scalar wave functions $u, w, v_t,$ and $v_s$ written as linear combinations of  the four invariant functions $F, G, H,$ and $I$ \cite{Buck:1979ff}:
\begin{align}
&\psi^+_{\lambda_1\lambda_2,\lambda_d}({\bf p}) =
{1\over\sqrt{4\pi}} \chi^\dagger_{_{-\!\lambda_2}}\bigg[u({\rm p})\,
{\bm\sigma\cdot \bm\xi}_{_{\lambda_d}} 
\nonumber\\
&\qquad+
{w({\rm p})\over\sqrt{2}}\left(3\,{\bf \hat p\cdot 
\bm\xi}_{_{\lambda_d}} {\bm{\sigma}\cdot \bf\hat
p}- {\bm{\sigma}\cdot\bm\xi}_{_{\lambda_d}}
\right)\bigg] {i\sigma_2\over\sqrt{2}}\,\chi_{_{\lambda_1}}
\nonumber\\
&\psi^-_{\lambda_1\lambda_2,\lambda_d}({\bf p}) =
\sqrt{{3\over 4\pi}} \chi^\dagger_{_{-\lambda_2}}\bigg[v_s({\rm p})\;
{\bf\hat p\cdot \bm\xi}_{_{\lambda_d}}
\nonumber\\
&\qquad-{v_t({\rm p})\over\sqrt{2}} \left({\bf \bm\sigma\cdot\hat p\,
\bm\sigma\cdot \bm\xi}_{_{\lambda_d}} -  {\bf\hat p}\cdot
{\bm\xi}_{_{\lambda_d}} \right)
\bigg]{i\sigma_2\over\sqrt{2}}\,\chi_{_{\lambda_1}} .
\label{B6}
\end{align}
For explicit formulae connecting the $F, G, H,$ and $I$ to $u, w, v_t,$ and $v_s$, see Eqs.~(45) and (46) of Ref.~\cite{Buck:1979ff} [note that the normalization factor (\ref{eq:nconst}) sets the scale of the  connection]. 
Here $\hat{\bf p}=\{\sin\theta,0,\cos\theta\}$ lies in the $x,z$ plane.  For a deuteron at rest with its polarization vector defined along the $z$ direction, the three-vector components of the polarization are
\bea
\xi_0=
\left(
\begin{array}{c}
 0   \\
 0   \\
 1   
\end{array}
\right), \qquad \xi_\pm=
\mp\frac1{\sqrt{2}}\left(
\begin{array}{c}
 1   \\
 \pm i   \\
 0  
\end{array}
\right).
\eea
The nucleon spinors can be either helicity spinors, or spinors with spin projections along the (fixed) $z$ direction.  In the latter case, we must replace $\lambda_1\to s_1$ and $-\lambda_2\to s_2$ (since, when $\theta=0$, the helicity and spin projection of particle 2 are in opposite directions).  In either case (by inspection) the wave functions are real.

The wave functions $u, w, v_t,$ and $v_s$ have a convenient physical interpretation that follows from the structure of the 2-component matrix elements that multiply them.  This follows most directly 
if the nucleon spin projections are fixed in the $z$ direction.  In this case the $\rho=+$ wave functions is symmetric in the spins,
\begin{align}
\Big(\psi^{+}_{s_1s_2,\lambda_d}({\bf p})\Big)^* &=\psi^+_{s_1s_2,\lambda_d}({\bf p}) =\psi^+_{s_2 s_1,\lambda_d}({\bf p}) \, ,
\end{align}
showing that the spin state is a spin-one triplet.  The $\rho=-$ term includes a symmetric piece (the spin triplet $v_t$) and an antisymmetric piece (corresponding to a spin singlet state, denoted by the $s$ subscript on $v_s$). 

In the CST, wave functions in coordinate space are {\it defined\/} to be the Fourier transforms of the 
momentum space wave functions (\ref{B6}) \cite{Buck:1979ff}
\bea
\psi^\rho_{s_1s_2,\lambda_d}({\bf r})=\frac{1}{(2\pi)^{3/2}}\int d^3p\; e^{i{\bf p}\cdot{\bf r}} \psi^\rho_{s_1s_2,\lambda_d}({\bf p})\, .\qquad
\eea
Using the expansion of the plane wave
\bea
e^{i{\bf p}\cdot{\bf r}}=4\pi\sum_\ell\sum_{m=-\ell}^\ell i^\ell  j_\ell({\rm p r}) Y_{\ell m}^*(\hat{\bf p})Y_{\ell m}(\hat{\bf r})
\eea
and noting that the coefficient of $w({\rm p})$ involves a linear combination of the $Y_{2m}$ spherical harmonics
\bea
3\hat{\bf p}^i\hat{\bf p}^j-\delta^{ij}=\sum_m c^{ij}_m \;Y_{2m}(\hat{\bf p})
\eea
(where the $c_{ij}$ are constants)
it follows that the coordinate space spin representation of the deuteron wave functions is
\begin{align}
{\rm r}\,\psi^+_{s_1s_2,\lambda_d}({\bf r}) &=
{1\over\sqrt{4\pi}} \chi^\dagger_{_{s_2}}\bigg[u({\rm r})\,
{\bm{\sigma}\cdot \bm\xi}_{_{\lambda_d}} 
\nonumber\\
&-
{w({\rm r})\over\sqrt{2}}\left(3\,{\bf \hat r\cdot 
\bm\xi}_{_{\lambda_d}} {\bm{\sigma}\cdot \bf\hat
r}- {\bm{\sigma}\cdot\bm\xi}_{_{\lambda_d}}
\right)\bigg] {i\sigma_2\over\sqrt{2}}\,\chi_{_{s_1}}
\nonumber\\
{\rm r}\,\psi^-_{s_1s_2,\lambda_d}({\bf r}) &=i\,
\sqrt{{3\over 4\pi}} \chi^\dagger_{_{s_2}}\bigg[v_s({\rm r})\;
{\bf\hat r\cdot \bm\xi}_{_{\lambda_d}}
\nonumber\\
&-{v_t({\rm r})\over\sqrt{2}} \left({\bm \sigma\cdot\bf\hat r\,
\bm\sigma\cdot \bm\xi}_{_{\lambda_d}} -  {\bf\hat r}\cdot
{\bm\xi}_{_{\lambda_d}} \right)
\bigg]{i\sigma_2\over\sqrt{2}}\,\chi_{_{s_1}}\, ,
\label{B6a}
\end{align}
where, denoting the typical wave function by $z_\ell$ (so that $z_0=u$, $z_2=w$, and $z_1=v_t$ or $v_s$),  the momentum and position space wave functions are related by the spherical Bessel transforms
\bea
z_\ell({\rm p})&=&\sqrt{\sfrac2{\pi}}\int_0^\infty {\rm r} d{\rm r}\,j_\ell({\rm p}{\rm r})\,z_\ell({\rm r})
\nonumber\\
\frac{z_\ell({\rm r})}{{\rm r}}&=&\sqrt{\sfrac2{\pi}}\int_0^\infty {\rm p}^2d{\rm p}\,j_\ell({\rm p}{\rm r})\,z_\ell({\rm p})
\label{eq:besseltrans}
\eea
where $j_\ell$ is the spherical Bessel function of order $\ell$ with the convenient recursion relation
\bea
j_\ell(z)=z^\ell\left(-\frac{1}{z}\frac{d}{dz}\right)^\ell\frac{\sin z}{z}. \label{eq:recursion}
\eea
{\it Note the appearance of the minus sign multiplying $w({\rm r})$ and the factor of $i$ multiplying both of the P-state terms\/}.  These come from the factor of $i^\ell$ in the plane wave expansion and can  be easily overlooked.

\subsection{Normalization of the wave functions} \label{sec:norm}

In Appendix \ref{app:norm} the normalization condition (\ref{eq:relnorm}) is reduced  
to the following simple form
\bea
1&=&\int_0^\infty {\rm p}^2 d{\rm p} \Big[ u^2({\rm p})+w^2({\rm p})+v_t^2({\rm p})+v_s^2({\rm p})\Big]
\nonumber\\&&
\qquad+\left<\frac{dV}{d M_d}\right>\, . \label{eq:norm1}
\eea
The derivative term is negative and of the order of a few percent (see Table \ref{tab:prob}).

The origin of the derivative term can be understood in two rather different ways.  First, it can be derived from the requirement that 
a nonlinear version of the equation, used to derive the unitarity relation for positive energies, also hold at the deuteron pole \cite{Gross:1972ye,Adam:1997rb}.  From this point of view, the normalization condition ensures that the strength of the pole is not altered by repeated interactions near the pole and shows that the normalization is a consequence of the equation itself.  But the normalization condition also follows from the requirement of current conservation, where it is seen to give precisely the correct factor to insure that the conservation of charge is an automatic consequence of the correct normalization of the deuteron wave functions \cite{Adam:1997rb}.

The second interpretation shows, indirectly, that interaction current contributions to the deuteron charge (which give rise to the derivative term) are necessarily of the order of a few percent.  Hence,  the calculation of any electromagnetic property  of the deuteron (including the magnetic and quadrupole moments) that does {\it not\/} include interaction currents can be expected {\it ab initio\/} to be in error by a few percent.   Since the famous discrepancy in the quadrupole moment is about 5\%,  it is quite possible that a careful CST calculation (including interaction currents) could explain it, but this calculation is, unfortunately, beyond the scope of this paper.  For this reason, we will not report deuteron moments here.

\begin{table}
\begin{minipage}{3.2in}
\caption{The deuteron probabilities (in percent).  Both the exact and scaled probabilities are shown.}
\label{tab:prob}
\begin{ruledtabular}
\begin{tabular}{lrrrr}
& \multicolumn{2}{c}{WJC-1} & \multicolumn{2}{c}{WJC-2} \cr
probability & exact  & scaled & exact & scaled \cr 
\tableline
$P_s$ & 97.3876 & 92.3330 & 95.7607  & 93.5985  \cr
$P_d$ & 7.7452 & 7.3432 & 6.5301 & 6.3827     \cr
$P_{v_t}$ &0.1180 & 0.1119 & 0.0103 &  0.0101     \cr
$P_{v_s}$ & 0.2234 & 0.2118 & 0.0090&   0.0088 \cr
\tableline
$\sum P$ & 105.4743  & 100.0000 & 102.3101 & 100.0000\cr
$\left<V'\right>$ &$-$5.4743 & ---$\quad$ &$-$2.3101 &  ---$\quad$\cr
\tableline
total & 100.0000 & ---$\quad$  & 100.0000 &  ---$\quad$  
\end{tabular}
\end{ruledtabular}
\end{minipage}
\end{table}
%

If these wave functions in are used in any calculation that neglects  interaction currents, it is probably a better approximation to use what we will refer to as the {\it scaled\/} normalization condition
\bea
1&=&\int_0^\infty {\rm p}^2 d{\rm p} \Big[ u^2({\rm p})+w^2({\rm p})+v_t^2({\rm p})+v_s^2({\rm p})\Big].\qquad \label{eq:norm2}
\eea
This is the normalization condition used in the earliest treatments of the CST deuteron wave functions \cite{Buck:1979ff}, and is also appropriate for electromagnetic calculations  using the relativistic impulse approximation (RIA) \cite{VanOrden:1995eg,Adam:2002cn}.

\begin{figure*}
\includegraphics[width=6.5in]{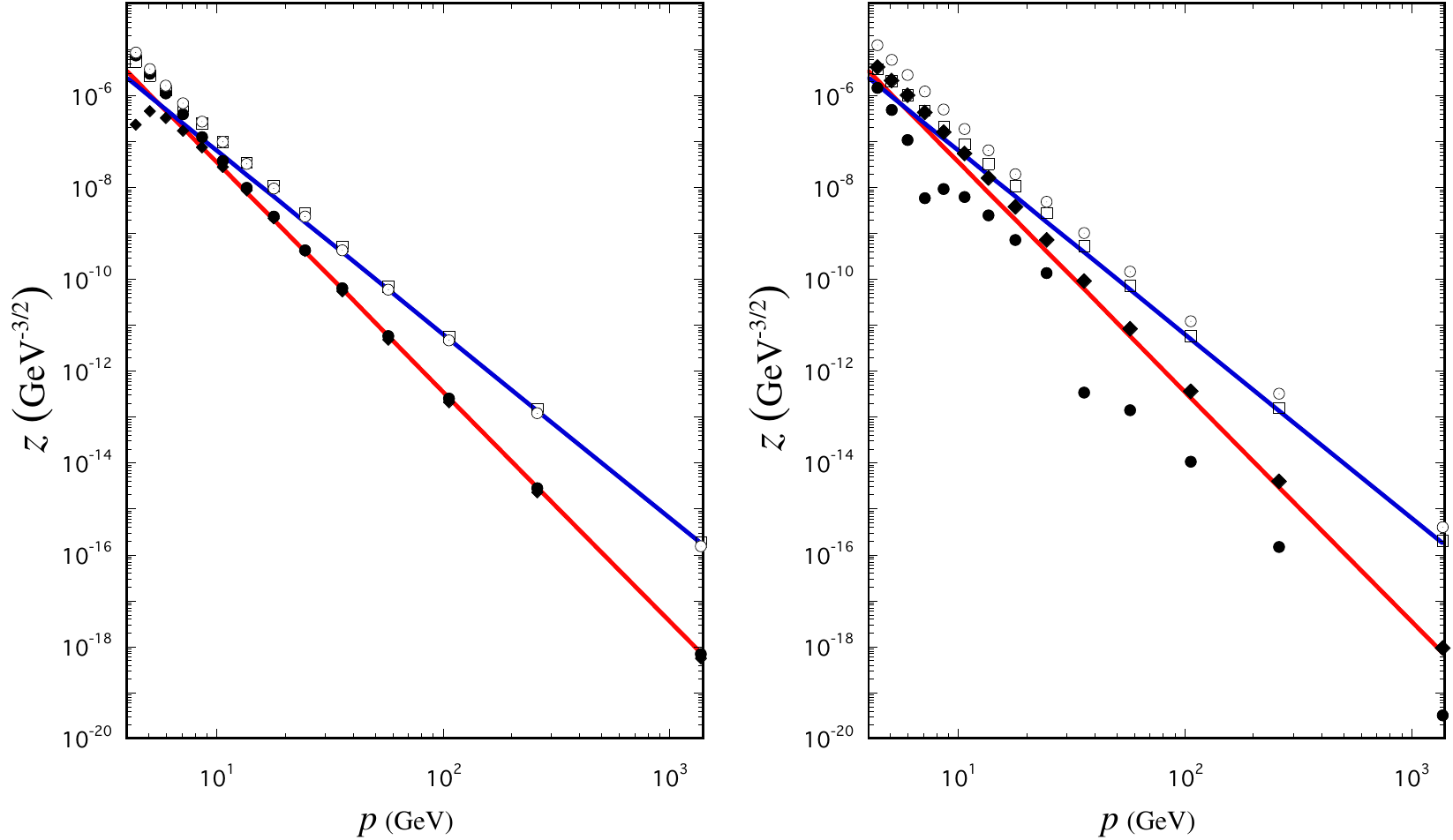}
\caption{\footnotesize\baselineskip=10pt (Color online) Log-log plot of the momentum space wave functions $z_\ell({\rm p})$, defined in Eq.~(\ref{B6}).  (Left panel): WJC-1 wave functions tabulated in Table \ref{tab:raw-wfs}, compared to the asymptotic limits given in Eq.~(\ref{eq:asymp}).  Solid circles: $u_0$; solid diamonds: $w_2$; open squares: $v_t$; open circles: $v_s$.  The solid lines are proportional to $p^{-5}$ and $p^{-4}$.  (Right panel):  WJC-2 wave functions tabulated in Table \ref{tab:raw-wfs-2}.   The solid lines are proportional to $p^{-5}$ and $p^{-4}$, and are {\it the same\/} as those shown in the left panel.}
\label{fig:loglog}
\end{figure*} 

\subsection{Asymptotic behavior of the wave functions}

Using the CST equations for the bound state it is possible to predict the asymptotic behavior of the wave functions. The derivation and discussion are given in Appendix \ref{app:C}.  The result is:
\bea
u({\rm p})&\to& \frac{1}{{\rm p}^5}\qquad w({\rm p})\to \frac{1}{{\rm p}^5}
\nonumber\\
v_t({\rm p})&\to& \frac{1}{{\rm p}^4}\qquad v_s({\rm p})\to \frac{1}{{\rm p}^4}. \label{eq:asymp}
\eea
Before the convergence of the angular integrals was improved (as discussed in Appendix \ref{app:B}) the numerical solutions had power law behaviors which differed from the integers predicted in (\ref{eq:asymp})  by about 0.1 to 0.2.  With the newly converged angular integrals, the actual solutions exhibit  the correct behavior.  

The odd powers for $u$ and $w$ (and even powers for the $v$'s) require special consideration when analytic representations of the wave functions are constructed in the following sections.

\subsection{Numerical results for the wave functions}

The solutions for the wave functions are tabulated in Tables \ref{tab:raw-wfs} and \ref{tab:raw-wfs-2}, with the high momentum behavior  shown graphically in Fig.~\ref{fig:loglog}.  The figure shows that the asymptotic estimates (\ref{eq:asymp}) hold for momenta larger than about 10 GeV, and that in this large momentum region the small P-state wave functions are larger than the dominant S and D-state wave functions.  The wave functions of model WJC-1 satisfy the approximate relations  $u\simeq w$, and the D-state wave functions of the two models are also comparable, but the S-state wave function of model WJC-2 has a zero at large momentum and is considerably smaller than its counterpart.  The P-state wave functions for both models are comparable to within a factor of 2, with $v_t\sim-v_s$.  

The probabilities for each of the components of the wave function are summarized in Table \ref{tab:prob}.  This Table reports both the exact probabilities (\ref{eq:norm1}) and the renormalized probabilities (\ref{eq:norm2}).

\begingroup
\squeezetable
\begin{table*}
\begin{minipage}{6in}
\squeezetable
\caption{Model WJC-1 momentum space wave functions $z_\ell({\rm p})$, defined in Eq.~(\ref{B6}).  The momenta are in GeV and the wave functions are in units of (GeV)$^{-3/2}$.  These wave functions are normalized to 105.4743\%.  The original value of $w$ at the lowest momentum point (-0.690345E-02), which resulted from the imprecise cancellation of two very large terms, can be replaced by the bold face number obtained using $p^2$ scaling.}
\label{tab:raw-wfs}
\begin{ruledtabular}
\begin{tabular}{ccccc}
p & $u$ & $w$ & $v_t$ & $v_s$ \cr
\tableline
    0.527313E-03  &    0.146227E+03  &   {\bf 0.495991E-03}
    &   -0.623843E-03  &   -0.358658E-03\cr
    0.277683E-02  &    0.145699E+03  &    0.137542E-01  &   -0.326777E-02  &   -0.192465E-02\cr
    0.681763E-02  &    0.142975E+03  &    0.813298E-01  &   -0.802292E-02  &   -0.471289E-02\cr
    0.126402E-01  &    0.135554E+03  &    0.264888E+00  &   -0.148513E-01  &   -0.871988E-02\cr
    0.202305E-01  &    0.121587E+03  &    0.607884E+00  &   -0.236810E-01  &   -0.139049E-01\cr
    0.295712E-01  &    0.101834E+03  &    0.108549E+01  &   -0.343738E-01  &   -0.201942E-01\cr
    0.406428E-01  &    0.798054E+02  &    0.160146E+01  &   -0.466979E-01  &   -0.274694E-01\cr
    0.534249E-01  &    0.594035E+02  &    0.204959E+01  &   -0.603055E-01  &   -0.355616E-01\cr
    0.678972E-01  &    0.428005E+02  &    0.236963E+01  &   -0.747258E-01  &   -0.442522E-01\cr
    0.840413E-01  &    0.303155E+02  &    0.255133E+01  &   -0.893826E-01  &   -0.532839E-01\cr
    0.101842E+00  &    0.213130E+02  &    0.261197E+01  &   -0.103638E+00  &   -0.623822E-01\cr
    0.121288E+00  &    0.149401E+02  &    0.257733E+01  &   -0.116855E+00  &   -0.712774E-01\cr
    0.142375E+00  &    0.104515E+02  &    0.247282E+01  &   -0.128471E+00  &   -0.796948E-01\cr
    0.165107E+00  &    0.728541E+01  &    0.232096E+01  &   -0.137990E+00  &   -0.876522E-01\cr
    0.189495E+00  &    0.504321E+01  &    0.213982E+01  &   -0.145164E+00  &   -0.948577E-01\cr
    0.215565E+00  &    0.344843E+01  &    0.194353E+01  &   -0.149827E+00  &   -0.101215E+00\cr
    0.243352E+00  &    0.231128E+01  &    0.174361E+01  &   -0.151964E+00  &   -0.106954E+00\cr
    0.272908E+00  &    0.149976E+01  &    0.154713E+01  &   -0.151653E+00  &   -0.111870E+00\cr
    0.304301E+00  &    0.922698E+00  &    0.135976E+01  &   -0.149087E+00  &   -0.115989E+00\cr
    0.337619E+00  &    0.515394E+00  &    0.118496E+01  &   -0.144410E+00  &   -0.119420E+00\cr
    0.372971E+00  &    0.231152E+00  &    0.102511E+01  &   -0.137607E+00  &   -0.122164E+00\cr
    0.410488E+00  &    0.350502E-01  &    0.881942E+00  &   -0.128414E+00  &   -0.123858E+00\cr
    0.450334E+00  &   -0.994438E-01  &    0.755982E+00  &   -0.116562E+00  &   -0.124047E+00\cr
    0.492699E+00  &   -0.189481E+00  &    0.645099E+00  &   -0.102598E+00  &   -0.121827E+00\cr
    0.537814E+00  &   -0.242417E+00  &    0.544069E+00  &   -0.878148E-01  &   -0.116934E+00\cr
    0.585950E+00  &   -0.262050E+00  &    0.448848E+00  &   -0.732091E-01  &   -0.109318E+00\cr
    0.637426E+00  &   -0.255246E+00  &    0.360144E+00  &   -0.592921E-01  &   -0.995041E-01\cr
    0.692620E+00  &   -0.230996E+00  &    0.280430E+00  &   -0.465163E-01  &   -0.878985E-01\cr
    0.751979E+00  &   -0.196819E+00  &    0.211737E+00  &   -0.352080E-01  &   -0.752255E-01\cr
    0.816027E+00  &   -0.159320E+00  &    0.154625E+00  &   -0.255953E-01  &   -0.621228E-01\cr
    0.885387E+00  &   -0.122585E+00  &    0.109110E+00  &   -0.177552E-01  &   -0.494042E-01\cr
    0.960798E+00  &   -0.900194E-01  &    0.742130E-01  &   -0.116628E-01  &   -0.377210E-01\cr
    0.104314E+01  &   -0.631224E-01  &    0.485750E-01  &   -0.718102E-02  &   -0.276183E-01\cr
    0.113346E+01  &   -0.421787E-01  &    0.305968E-01  &   -0.408644E-02  &   -0.193771E-01\cr
    0.123304E+01  &   -0.270876E-01  &    0.185189E-01  &   -0.208407E-02  &   -0.130319E-01\cr
    0.134340E+01  &   -0.166413E-01  &    0.107639E-01  &   -0.902849E-03  &   -0.841563E-02\cr
    0.146641E+01  &   -0.978419E-02  &    0.600927E-02  &   -0.275271E-03  &   -0.522831E-02\cr
    0.160437E+01  &   -0.557178E-02  &    0.322545E-02  &    0.208268E-04  &   -0.313370E-02\cr
    0.176010E+01  &   -0.306684E-02  &    0.165657E-02  &    0.125739E-03  &   -0.181763E-02\cr
    0.193715E+01  &   -0.161854E-02  &    0.809542E-03  &    0.135471E-03  &   -0.102257E-02\cr
    0.213998E+01  &   -0.824465E-03  &    0.375303E-03  &    0.110722E-03  &   -0.559203E-03\cr
    0.237429E+01  &   -0.413458E-03  &    0.166244E-03  &    0.808655E-04  &   -0.297772E-03\cr
    0.264741E+01  &   -0.199638E-03  &    0.674221E-04  &    0.532186E-04  &   -0.154652E-03\cr
    0.296895E+01  &   -0.923325E-04  &    0.240784E-04  &    0.325043E-04  &   -0.783631E-04\cr
    0.335166E+01  &   -0.410173E-04  &    0.695348E-05  &    0.187892E-04  &   -0.386958E-04\cr
    0.381276E+01  &   -0.175134E-04  &    0.111592E-05  &    0.103540E-04  &   -0.185709E-04\cr
    0.437596E+01  &   -0.741984E-05  &   -0.231960E-06  &    0.551835E-05  &   -0.861492E-05\cr
    0.507462E+01  &   -0.295514E-05  &   -0.452993E-06  &    0.276257E-05  &   -0.386774E-05\cr
    0.595690E+01  &   -0.110492E-05  &   -0.326876E-06  &    0.131169E-05  &   -0.167290E-05\cr
    0.709434E+01  &   -0.387509E-06  &   -0.173383E-06  &    0.590317E-06  &   -0.691226E-06\cr
    0.859696E+01  &   -0.126207E-06  &   -0.755500E-07  &    0.249574E-06  &   -0.269811E-06\cr
    0.106411E+02  &   -0.375810E-07  &   -0.277414E-07  &    0.976773E-07  &   -0.979770E-07\cr
    0.135237E+02  &   -0.999262E-08  &   -0.854233E-08  &    0.346150E-07  &   -0.323958E-07\cr
    0.177768E+02  &   -0.228855E-08  &   -0.215203E-08  &    0.107312E-07  &   -0.945155E-08\cr
    0.244325E+02  &   -0.428701E-09  &   -0.419613E-09  &    0.272700E-08  &   -0.230841E-08\cr
    0.357133E+02  &   -0.637757E-10  &   -0.539018E-10  &    0.510267E-09  &   -0.433381E-09\cr
    0.571587E+02  &   -0.579367E-11  &   -0.486092E-11  &    0.700683E-10  &   -0.590489E-10\cr
    0.105975E+03  &   -0.256335E-12  &   -0.211897E-12  &    0.562191E-11  &   -0.463011E-11\cr
    0.260189E+03  &   -0.281688E-14  &   -0.229780E-14  &    0.150575E-12  &   -0.121213E-12\cr
    0.137016E+04  &   -0.682890E-18  &   -0.553955E-18  &    0.193220E-15  &   -0.153367E-15\cr
\end{tabular}
\end{ruledtabular}
\end{minipage}
\end{table*}
\endgroup
%

\begingroup
\squeezetable
\begin{table*}
\begin{minipage}{6in}
\squeezetable
\caption{Model WJC-2 momentum space wave functions $z_\ell({\rm p})$, defined in Eq.~(\ref{B6}).  The momenta are in GeV and the wave functions are in units of (GeV)$^{-3/2}$.   These wave functions are normalized to 102.3101\%.  The original value of $w$ at the lowest momentum point (-0.743127E-02), which resulted from the imprecise cancellation of two very large terms, can be replaced by the bold face number obtained using $p^2$ scaling.}
\label{tab:raw-wfs-2}
\begin{ruledtabular}
\begin{tabular}{ccccc}
p & $u$ & $w$ & $v_t$ & $v_s$ \cr
\tableline
    0.527313E-03  &    0.146161E+03  &   {\bf 0.497127E-03}  
    &    0.673653E-04  &    0.139978E-04\cr
    0.277683E-02  &    0.145632E+03  &    0.137857E-01  &    0.372102E-03  &    0.368387E-04\cr
    0.681763E-02  &    0.142908E+03  &    0.815168E-01  &    0.908775E-03  &    0.100679E-03\cr
    0.126402E-01  &    0.135483E+03  &    0.265495E+00  &    0.168223E-02  &    0.192525E-03\cr
    0.202305E-01  &    0.121510E+03  &    0.609254E+00  &    0.268867E-02  &    0.316519E-03\cr
    0.295712E-01  &    0.101747E+03  &    0.108784E+01  &    0.392312E-02  &    0.478611E-03\cr
    0.406428E-01  &    0.797033E+02  &    0.160461E+01  &    0.537988E-02  &    0.685611E-03\cr
    0.534249E-01  &    0.592857E+02  &    0.205290E+01  &    0.705316E-02  &    0.942644E-03\cr
    0.678972E-01  &    0.426693E+02  &    0.237212E+01  &    0.893675E-02  &    0.125069E-02\cr
    0.840413E-01  &    0.301750E+02  &    0.255185E+01  &    0.110216E-01  &    0.160505E-02\cr
    0.101842E+00  &    0.211677E+02  &    0.260934E+01  &    0.132917E-01  &    0.199444E-02\cr
    0.121288E+00  &    0.147943E+02  &    0.257044E+01  &    0.157173E-01  &    0.240633E-02\cr
    0.142375E+00  &    0.103090E+02  &    0.246071E+01  &    0.182443E-01  &    0.286055E-02\cr
    0.165107E+00  &    0.714947E+01  &    0.230262E+01  &    0.208563E-01  &    0.313824E-02\cr
    0.189495E+00  &    0.491677E+01  &    0.211456E+01  &    0.234179E-01  &    0.337366E-02\cr
    0.215565E+00  &    0.333407E+01  &    0.191087E+01  &    0.258522E-01  &    0.353615E-02\cr
    0.243352E+00  &    0.221125E+01  &    0.170294E+01  &    0.280836E-01  &    0.333024E-02\cr
    0.272908E+00  &    0.141604E+01  &    0.149824E+01  &    0.300543E-01  &    0.284267E-02\cr
    0.304301E+00  &    0.856967E+00  &    0.130257E+01  &    0.317477E-01  &    0.198415E-02\cr
    0.337619E+00  &    0.468983E+00  &    0.111930E+01  &    0.332767E-01  &    0.574512E-03\cr
    0.372971E+00  &    0.205168E+00  &    0.950678E+00  &    0.348229E-01  &   -0.146236E-02\cr
    0.410488E+00  &    0.303889E-01  &    0.798954E+00  &    0.363889E-01  &   -0.389499E-02\cr
    0.450334E+00  &   -0.828321E-01  &    0.665687E+00  &    0.373820E-01  &   -0.658983E-02\cr
    0.492699E+00  &   -0.154049E+00  &    0.551266E+00  &    0.364870E-01  &   -0.910693E-02\cr
    0.537814E+00  &   -0.193834E+00  &    0.452057E+00  &    0.328528E-01  &   -0.113767E-01\cr
    0.585950E+00  &   -0.206848E+00  &    0.363715E+00  &    0.273257E-01  &   -0.131172E-01\cr
    0.637426E+00  &   -0.198813E+00  &    0.284931E+00  &    0.211844E-01  &   -0.143788E-01\cr
    0.692620E+00  &   -0.177269E+00  &    0.216739E+00  &    0.152062E-01  &   -0.150297E-01\cr
    0.751979E+00  &   -0.148546E+00  &    0.159772E+00  &    0.980525E-02  &   -0.151078E-01\cr
    0.816027E+00  &   -0.118123E+00  &    0.113878E+00  &    0.537109E-02  &   -0.145747E-01\cr
    0.885387E+00  &   -0.890513E-01  &    0.783167E-01  &    0.202329E-02  &   -0.134838E-01\cr
    0.960798E+00  &   -0.639660E-01  &    0.518825E-01  &   -0.137523E-03  &   -0.119400E-01\cr
    0.104314E+01  &   -0.437558E-01  &    0.329899E-01  &   -0.128366E-02  &   -0.101160E-01\cr
    0.113346E+01  &   -0.283328E-01  &    0.201385E-01  &   -0.167665E-02  &   -0.815911E-02\cr
    0.123304E+01  &   -0.176467E-01  &    0.117937E-01  &   -0.152450E-02  &   -0.629427E-02\cr
    0.134340E+01  &   -0.104205E-01  &    0.658279E-02  &   -0.117834E-02  &   -0.462869E-02\cr
    0.146641E+01  &   -0.581003E-02  &    0.349531E-02  &   -0.802520E-03  &   -0.324538E-02\cr
    0.160437E+01  &   -0.315563E-02  &    0.176854E-02  &   -0.460812E-03  &   -0.218651E-02\cr
    0.176010E+01  &   -0.164951E-02  &    0.831201E-03  &   -0.227186E-03  &   -0.141637E-02\cr
    0.193715E+01  &   -0.800130E-03  &    0.349702E-03  &   -0.970097E-04  &   -0.879905E-03\cr
    0.213998E+01  &   -0.366387E-03  &    0.123929E-03  &   -0.288285E-04  &   -0.527587E-03\cr
    0.237429E+01  &   -0.175371E-03  &    0.333862E-04  &    0.684399E-05  &   -0.308052E-03\cr
    0.264741E+01  &   -0.775657E-04  &   -0.229350E-05  &    0.151418E-04  &   -0.173993E-03\cr
    0.296895E+01  &   -0.303299E-04  &   -0.119914E-04  &    0.132282E-04  &   -0.949967E-04\cr
    0.335166E+01  &   -0.104213E-04  &   -0.111922E-04  &    0.932916E-05  &   -0.502335E-04\cr
    0.381276E+01  &   -0.309564E-05  &   -0.763178E-05  &    0.592149E-05  &   -0.257338E-04\cr
    0.437596E+01  &   -0.144689E-05  &   -0.413223E-05  &    0.374297E-05  &   -0.128285E-04\cr
    0.507462E+01  &   -0.488220E-06  &   -0.212835E-05  &    0.203145E-05  &   -0.613856E-05\cr
    0.595690E+01  &   -0.107614E-06  &   -0.100778E-05  &    0.101279E-05  &   -0.280394E-05\cr
    0.709434E+01  &   -0.588336E-08  &   -0.430027E-06  &    0.476792E-06  &   -0.121466E-05\cr
    0.859696E+01  &    0.931326E-08  &   -0.163878E-06  &    0.211653E-06  &   -0.493916E-06\cr
    0.106411E+02  &    0.613471E-08  &   -0.550345E-07  &    0.873897E-07  &   -0.185900E-06\cr
    0.135237E+02  &    0.246297E-08  &   -0.159505E-07  &    0.327267E-07  &   -0.635345E-07\cr
    0.177768E+02  &    0.718045E-09  &   -0.386029E-08  &    0.106738E-07  &   -0.191935E-07\cr
    0.244325E+02  &    0.135807E-09  &   -0.732832E-09  &    0.281290E-08  &   -0.493279E-08\cr
    0.357133E+02  &    0.340558E-12  &   -0.932273E-10  &    0.532783E-09  &   -0.101591E-08\cr
    0.571587E+02  &   -0.140690E-12  &   -0.832796E-11  &    0.721192E-10  &   -0.146914E-09\cr
    0.105975E+03  &   -0.105532E-13  &   -0.362792E-12  &    0.579471E-11  &   -0.119175E-10\cr
    0.260189E+03  &   -0.149527E-15  &   -0.394161E-14  &    0.157413E-12  &   -0.317980E-12\cr
    0.137016E+04  &   -0.319205E-19  &   -0.952160E-18  &    0.204258E-15  &   -0.406280E-15\cr
\end{tabular}
\end{ruledtabular}
\end{minipage}
\end{table*}
\endgroup
%

\subsection{Fitted wave functions} \label{sec:3G}

It is convenient to fit the deuteron wave functions to a series of simple functions that can be analytically integrated to obtain the wave functions in $r$ space, and to interpolate for any value of the momentum.  This subsection will describe in detail how this is done.

\subsubsection{Scaling the wave functions}

The first step in the fitting process is to scale out the rapid dependence of each wave function on the momentum, so that it may be studied on a linear plot.  Constructing these scaling functions took some care, but in the end we found that the functions
\begin{align}
u_{\rm scale}(p)&=\frac{N}{(m_{s1}^2+p^2)^{3/2}}\bigg[\frac{1}{(\alpha_0^2+p^2)} +\frac{R}{(m_{s1}^2+p^2)}\bigg]
\nonumber\\
w_{\rm scale}(p)&=\frac{N\,p^2}{(m_{s1}^2+p^2)^{5/2}}\bigg[\frac{1}{(\alpha_0^2+p^2)} +\frac{R}{(m_{s1}^2+p^2)}\bigg]\qquad
\nonumber\\
v_{\rm scale}(p)&=N\,p\bigg[\frac{1}{(m_{s0}^2+p^2)^{5/2}}+\frac{R}{(m_{s1}^2+p^2)^{5/2}}\bigg]\,\qquad
\label{eq:356}
\end{align}
(where the functional form of $v_{\rm scale}$ is used to scale both $v_t$ and $v_s$, and we return to denoting the magnitude of the three-momentum by $p$ [instead of p, as we did in the previous sections]) work very well.  Note that these scaling functions have the desired (and observed) $p^\ell$ behavior at small $p$ and the correct asymptotic $p$ dependence for all the states.  The parameter $\alpha_0$ is determined from the asymptotic relation
\bea
2\sqrt{m^2+\alpha_0^2}-2m=\epsilon
\eea
where $\epsilon=2.2246$ MeV is the deuteron binding energy.  This gives
\bea
\alpha_0=\sqrt{m\epsilon+\sfrac14\epsilon^2}=45.7159\;{\rm MeV} \label{eq:alpha}
\eea
which differs slightly from the nonrelativistic $\alpha_0=\sqrt{m\epsilon}$, as has been emphasized in the literature \cite{Keister:1991sb}.  This parameter is fixed by the deuteron binding energy, but since the P-states vanish outside of the range of the potential, their leading mass, $m_{s0}$ is treated as a free parameter. This and the other parameters used in the scale functions are given in Tables \ref{tab:scale} and \ref{tab:scale-2}.

\begin{figure*}
\includegraphics[width=6in]{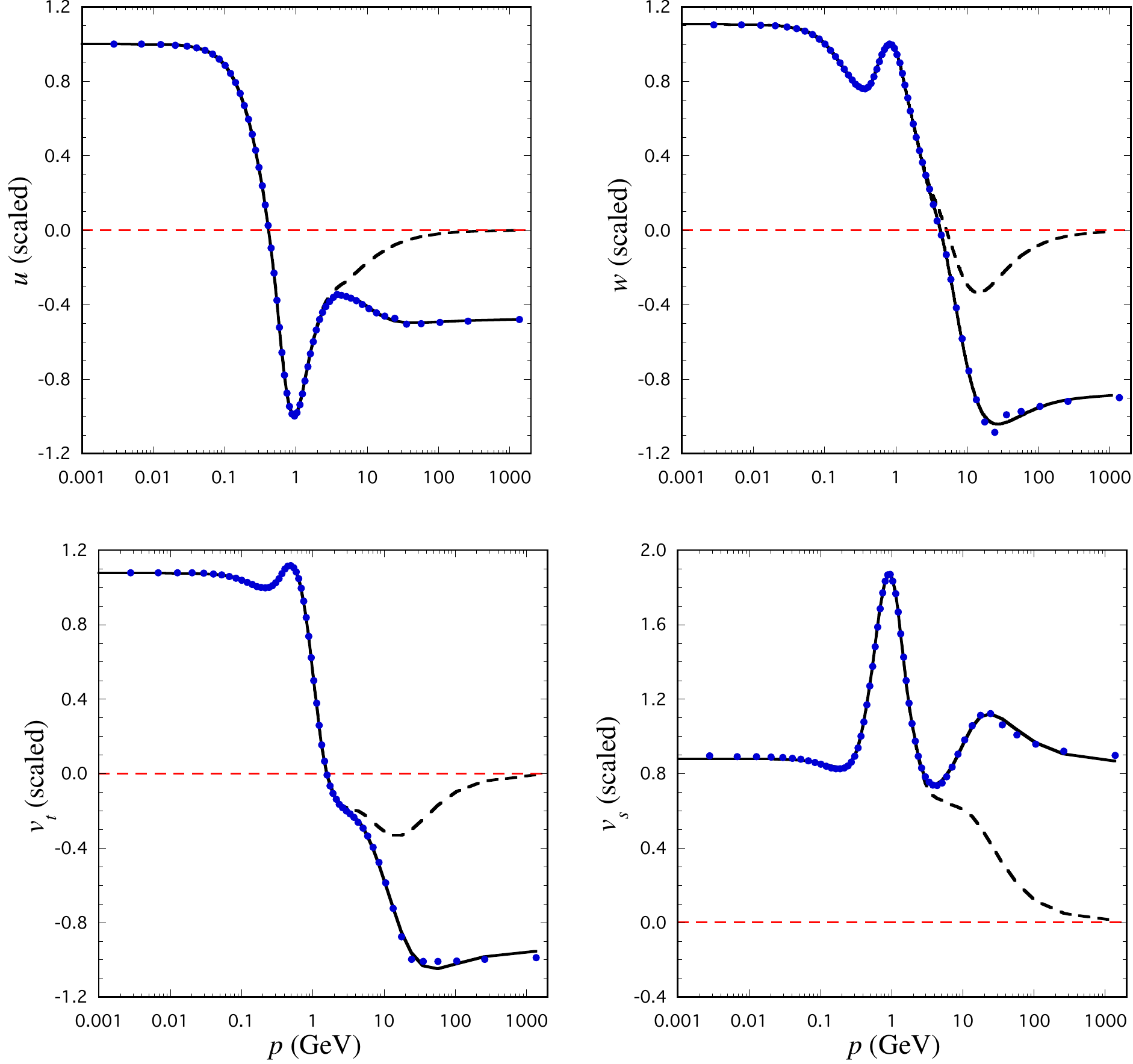}
\caption{\footnotesize\baselineskip=10pt (Color online) Ratio of the raw wave functions for model WJC-1 (taken from Table \ref{tab:raw-wfs} with the probabilities ``scaled'' to 100\% as outlined in Table \ref{tab:prob}) to the scale functions from Eq.~(\ref{eq:356}).  In each panel the dots are the tabulated wave functions for 60 Gauss points, the solid curve the fit, and the dashed curve the fit {\it without\/} the tail wave function of Eq.~(\ref{eq:tail}). }
\label{Four-I}
\end{figure*} 

\begin{figure*}
\includegraphics[width=6in]{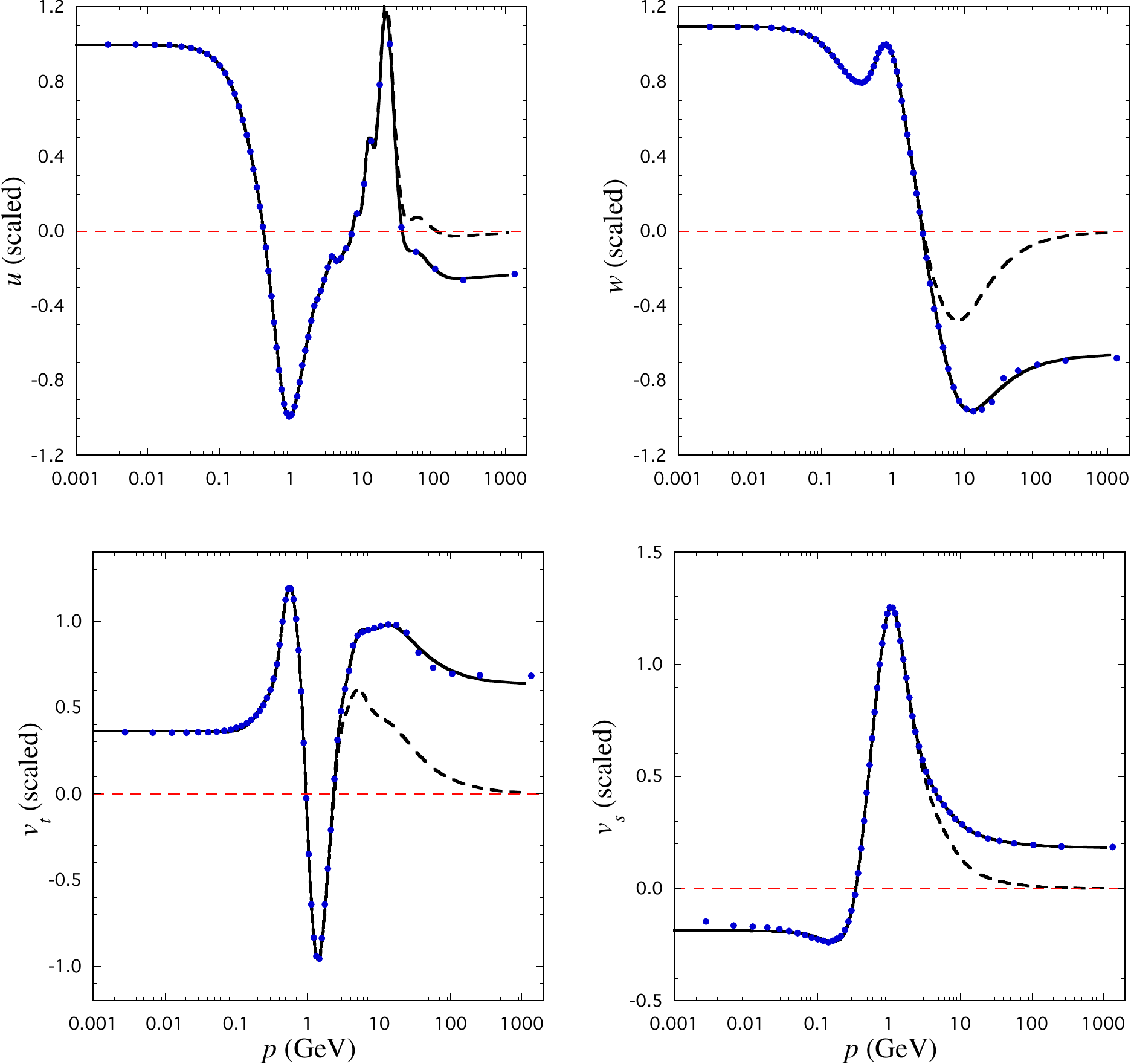}
\caption{\footnotesize\baselineskip=10pt (Color online) Ratio of the raw wave functions for model WJC-2 (taken from Table \ref{tab:raw-wfs-2} with the probabilities ``scaled'' to 100\% as outlined in Table \ref{tab:prob}) to the scale functions from Eq.~(\ref{eq:356}).  In each panel the dots are the tabulated wave functions for 60 Gauss points, the solid curve the fit, and the dashed curve the fit {\it without\/} the tail wave function of Eq.~(\ref{eq:tail}).   The origin of the many small wiggles in $u$ is not clear, but $u$ is {\it very\/} small in this region, as illustrated in Fig.~\ref{fig:loglog}.  The first $v_s$ point is probably inaccurate because of numerical cancellations.}
\label{Four-II}
\end{figure*} 

\subsubsection{Fitting the momentum space wave functions}

In order to obtain simple and accurate fits to the wave functions, we found it convenient to use two types of expansion functions.  For all but the high momentum tail, we use functions that go asymptotically like an {\it even\/} power of $p$ for the S and D-states (in practice, $p^{-6}$), and an {\it odd\/} power of $p$ for the P-states ( $p^{-5}$).  These functions go  to zero precisely one power of $p$ faster than the observed asymptotic behavior and are very conveniently transformed to coordinate space giving a superposition of simple exponentials.  However, the asymptotic  tail cannot be accurately described by such functions, and it is best to use one special function with the correct asymptotic behavior in order to describe the wave functions at very large $p$.  This function has a Fourier transform that can be written in terms of the modified Bessel functions of the second kind, $K_n(z)$.  

\begin{figure}
\includegraphics[width=3.5in]{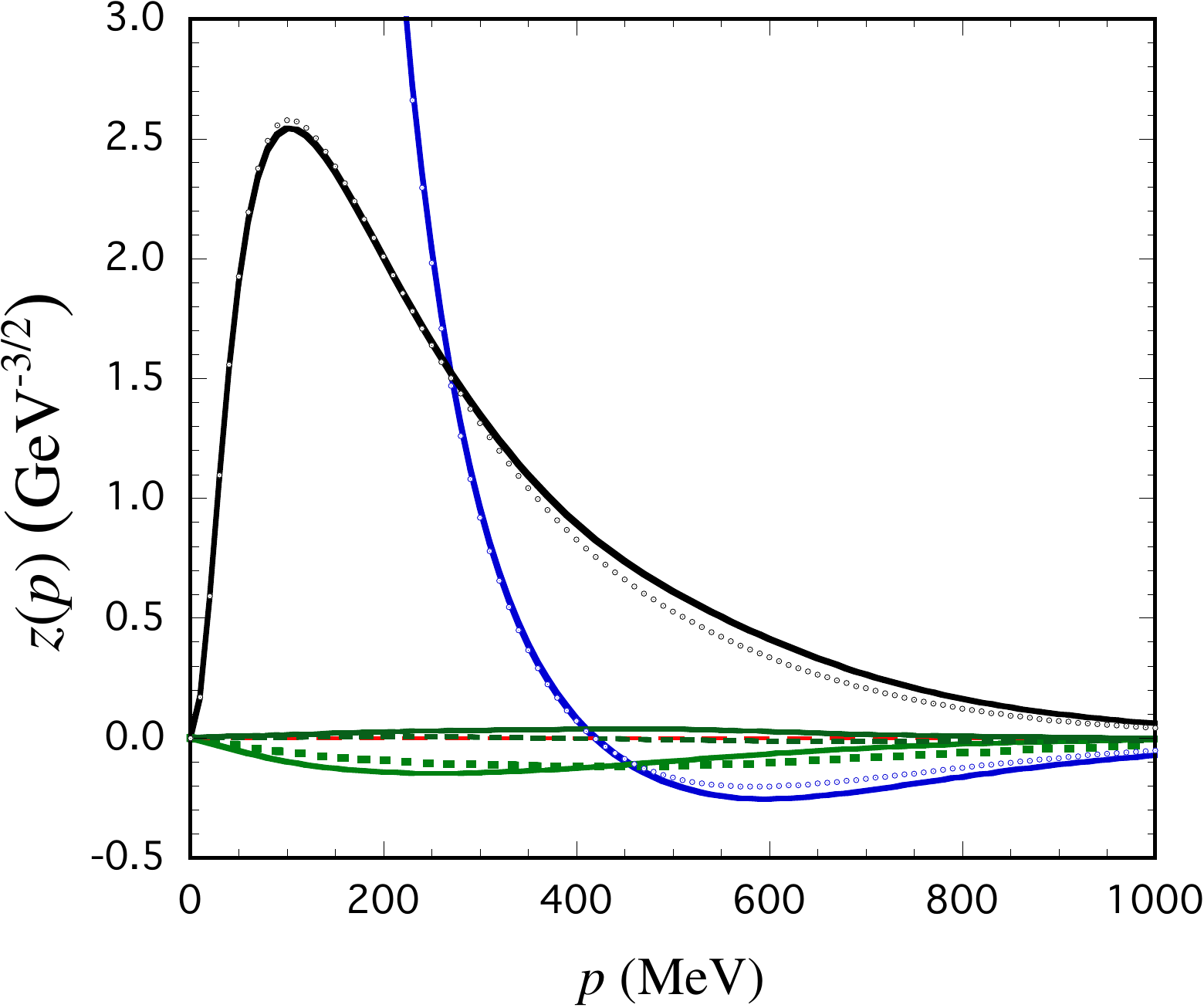}
\caption{\footnotesize\baselineskip=10pt (Color online) Momentum space wave functions for both models.  The largest two wave functions are $u$ (off-scale at small momenta) and $w$ (zero at small momenta),  with WJC-1 wave functions solid and WJC-2 dotted.  The P-state wave functions are very small, with {\it both\/} $v_t$ wave functions solid ($v_t(1)$ is negative and $v_t(2)$ positive) and both $v_s$ wave functions dotted (with ($v_s(1)$ negative and($v_s(2)$ very near zero).  Note that the S-state wave function has a zero around 400 MeV, and that above this momentum the P-state wave functions for WJC-1 are no longer completely negligible, as illustrated in Fig.~\ref{fig:All-P(fine)}.}
\label{fig:All-P}
\end{figure} 

\begin{table}
\begin{minipage}{3.2in}
\caption{The parameters used in the scaling functions (\ref{eq:356}) for model WJC-1.}
\label{tab:scale}
\begin{ruledtabular}
\begin{tabular}{lcccc}
 & $N$ (GeV$^2$)  & $R$ & $m_{s1}$ (GeV) & $m_{s0}$ (GeV) \cr 
\tableline
$u$ & 0.4775 & $-$0.9856 & 1.160  & --  \cr
$w$ & 0.5131 & $-$0.9942 & 0.685  & --      \cr
$v_t$ &-0.03024 & $-$0.9772 & 1.600 &  0.488     \cr
$v_s$ & -0.1038 & $-$0.9942 & 0.840  &   0.633 \cr
\end{tabular}
\end{ruledtabular}
\end{minipage}
\end{table}
%

\begin{table}
\begin{minipage}{3.2in}
\caption{The parameters used in the scaling functions (\ref{eq:356}) for model WJC-2.}
\label{tab:scale-2}
\begin{ruledtabular}
\begin{tabular}{lcccc}
 & $N$ (GeV$^2$)  & $R$ & $m_{s1}$ (GeV) & $m_{s0}$ (GeV) \cr 
\tableline
$u$ & 0.3060 & $-$0.9978 & 1.000  & --  \cr
$w$ & 0.3474 & $-$0.9805 & 0.632  & --      \cr
$v_t$ &0.02700 & $-$0.9610 & 0.700 &  0.551     \cr
$v_s$ & -0.2356 & $-$0.9671 & 1.054  &   0.977 \cr
\end{tabular}
\end{ruledtabular}
\end{minipage}
\end{table}
%

\begin{figure}
\includegraphics[width=3.5in]{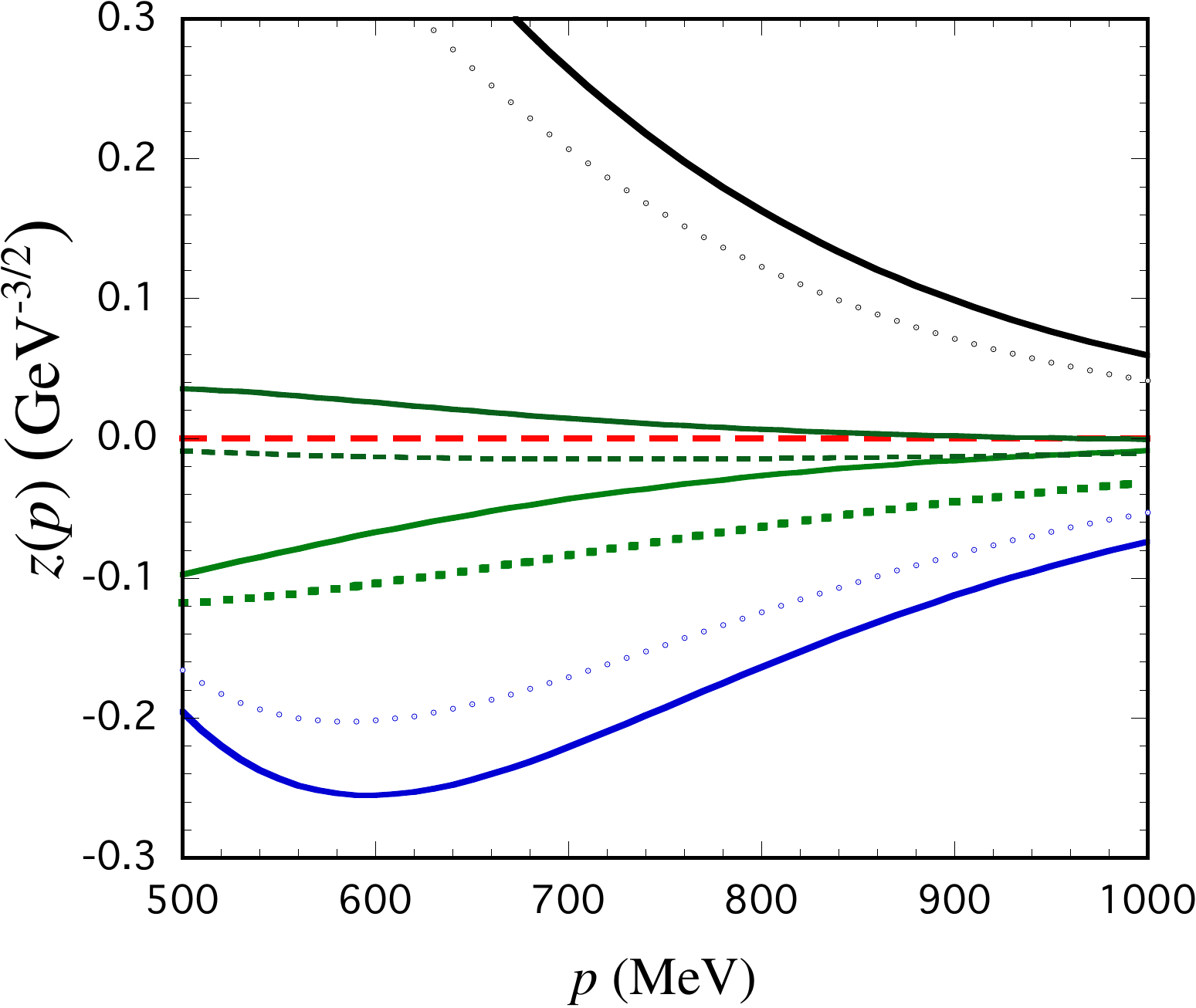}
\caption{\footnotesize\baselineskip=10pt (Color online) Enlarged view of the momentum space wave functions for both models. The curves are labeled as in Fig.~\ref{fig:All-P}, with a dashed zero reference line.  The large positive states are $w$; the large negative ones $u$, with WJC-1 (solid) and WJC-2 (dotted).  Both of the $v_t$ wave functions are solid, and both $v_s$ wave functions are dotted, with WJC-1 negative in both cases and WJC-2 both close to zero.}
\label{fig:All-P(fine)}
\end{figure} 

\begingroup
\squeezetable
\begin{table}
\begin{minipage}{3.5in}
\caption{The expansion parameters for model WJC-1 wave functions.  First line is the total number of terms $n$ in the sum (\ref{eq:354a}), second is the step mass $m_x$ (in MeV), the next lines are the coefficients $b_i^z, \{i=1,n-1\}$, with the last line the coefficient $b_n^z$, written in the notation $125.(-6)=125\times 10^{-6}$, all in units of GeV$^{-3/2}$.  The ``tail'' mass $M_n=2$ GeV in all cases.  We have given the $b_i^z$ coefficients to three decimal places, sufficient to reproduce the fits to better than 1\% accuracy; if greater accuracy is needed use the routine supplied by the authors. }
\label{tab:parameters-1}
\begin{ruledtabular}
\begin{tabular}{rrrr}
 \multicolumn{1}{c}{$u$} & \multicolumn{1}{c}{$w$} & \multicolumn{1}{c}{$v_t$} & \multicolumn{1}{c}{$v_s$} \\
 \tableline
                   17 &              16 &              12 &               12 \cr
                  75.00 &             80.00 &            109.09 &             109.09  \cr
\tableline
              134.963 &          23.813 &         -27.120 &          -78.763  \cr
               52.871 &          32.709 &         233.155 &          688.895  \cr
             -217.709 &        -111.381 &        -998.893 &        -2910.298  \cr
             1876.699 &         844.861 &        2679.013 &         7690.084  \cr
           -11369.449 &       -4376.965 &       -4894.348 &       -13861.596  \cr
            49427.176 &       16489.297 &        6277.553 &        17571.825  \cr
          -156695.247 &      -45122.587 &       -5674.481 &       -15720.926  \cr
           369322.468 &       91061.493 &        3550.952 &         9747.293  \cr
          -655367.189 &     -136672.748 &       -1467.429 &        -3994.747  \cr
           879178.453 &      152411.033 &         360.796 &          975.086  \cr
          -887103.150 &     -124633.670 &         -40.010 &         -107.489  \cr
           662786.733 &       72617.712 &           0.000 &            0.000  \cr
          -355720.583 &      -28549.974 &           0.000 &            0.000  \cr
           129738.381 &        6790.953 &           0.000 &            0.000  \cr
           -28805.739 &        -738.436 &           0.000 &            0.000  \cr
             2939.797 &           0.000 &           0.000 &            0.000  \cr
-125. (-6)&     -100.(-6) &      498.(-7) &      -394.(-7)  \cr
\end{tabular}
\end{ruledtabular}
\end{minipage}
\end{table}
\endgroup
%

\begingroup
\squeezetable
\begin{table}
\begin{minipage}{3.5in}
\caption{The expansion parameters for each of the model WJC-2 wave functions.  For an explanation, see the caption to Table \ref{tab:parameters-1}.  Here the ``tail'' mass $M_n=11$ GeV for $u$ and 2 GeV for all other wave functions. }
\label{tab:parameters-2}
\begin{ruledtabular}
\begin{tabular}{rrrr}
 \multicolumn{1}{c}{$u$} & \multicolumn{1}{c}{$w$} & \multicolumn{1}{c}{$v_t$} & \multicolumn{1}{c}{$v_s$} \\
 \tableline
                   30 &              18 &              21 &               18 \cr
                 344.83 &             70.59 &             60.00 &              70.59  \cr
\tableline
              181.569 &          19.228 &       33002.947 &       -74492.403  \cr
               -0.817 &          30.600 &     -580810.178 &      1001357.141  \cr
               11.779 &        -107.606 &     4678729.666 &     -6147239.139  \cr
              -64.782 &         897.969 &   -22669425.846 &     22510592.546  \cr
              247.517 &       -5434.481 &    72755203.760 &    -53617329.732  \cr
             -772.915 &       25662.551 &  -159377601.760 &     83750225.006  \cr
             1948.950 &      -93771.705 &   232305154.041 &    -77842609.919  \cr
            -3813.276 &      265664.025 &  -193963168.786 &     20071103.929  \cr
             5397.722 &     -582644.240 &    15154785.470 &     48069105.480  \cr
            -4637.491 &      986160.259 &   160238368.984 &    -61037252.628  \cr
              533.832 &    -1280768.137 &  -139567635.494 &      8837981.631  \cr
             3806.187 &     1262888.436 &   -52643827.848 &     50533251.186  \cr
            -3119.875 &     -927821.391 &   170318699.570 &    -65739754.705  \cr
            -2058.321 &      491712.394 &   -69098137.722 &     43130588.287  \cr
             3773.554 &     -177560.697 &  -117233319.092 &    -16799889.802  \cr
             1092.507 &       39095.262 &   192985378.690 &      3717624.361  \cr
            -3919.342 &       -3959.947 &  -138452165.069 &      -363261.256  \cr
             -736.912 &           0.000 &    56855870.564 &            0.000  \cr
             4018.336 &           0.000 &   -13049893.573 &            0.000  \cr
              612.635 &           0.000 &     1310791.818 &            0.000  \cr
            -4213.343 &           0.000 &           0.000 &            0.000  \cr
             -260.753 &           0.000 &           0.000 &            0.000  \cr
             4581.603 &           0.000 &           0.000 &            0.000  \cr
            -1171.113 &           0.000 &           0.000 &            0.000  \cr
            -4559.802 &           0.000 &           0.000 &            0.000  \cr
             5566.714 &           0.000 &           0.000 &            0.000  \cr
            -2984.961 &           0.000 &           0.000 &            0.000  \cr
              814.407 &           0.000 &           0.000 &            0.000  \cr
              -92.509 &           0.000 &           0.000 &            0.000  \cr
          -118.(-11) &     -172.(-6) &      514.(-7) &      -110.(-6)  \cr
\end{tabular}
\end{ruledtabular}
\end{minipage}
\end{table}
\endgroup
%

For the first set of functions we choose 
\bea
G^i_\ell(p)=\sqrt{\frac2\pi}\frac{p^\ell m_i^2\,M_{i}^{2 n_\ell-\ell}}{(m_i^2+p^2)(M_{i}^2+p^2)^{n_\ell}},\qquad \label{eq:317}
\eea
where the factor of $\sqrt{2/\pi}$ is introduced for convenience, $\ell$ is the angular momentum of the state, $n_\ell$=2 for the S and P states and $n_\ell=3$ for the D-state.  Note that, near $p=0$, the expansion functions have the normalization
\bea
G^i_\ell(0)\to\sqrt{\frac2\pi}\,\left(\frac{p}{M_i}\right)^\ell. \label{eq:317a}
\eea

The tail wave function, chosen to have the correct fall-off as $p\to\infty$ and the same normalization at $p=0$, is
\bea
G^n_\ell(p)= \sqrt{\frac2\pi}\frac{p^\ell\,M_n^{2n_\ell+1-\ell}}{(M_n^2+p^2)^{n_\ell+\frac12}} \label{eq:tail}
\eea

Denoting the typical wave function by $z_\ell$ (as we did above) the full momentum space wave functions are expanded in terms of $G_\ell$
\bea
z_\ell(p)&=&\sum_{i=1}^n b_i^z G^i_\ell(p)\, .
\label{eq:354a}
\eea
As this notation implies, the last function in the sum (when $i=n$) is the ``tail''  function (\ref{eq:tail}), while the first $n-1$ are of the type (\ref{eq:317}).  Since all of the functions have the same normalization at $p=0$, the relative size of the expansion coefficients is a measure of the relative size of each function (at least at small momenta).

The masses $m_i$ and $M_i$ (for $i=1$ to $n-1$) used in each of the wave functions (\ref{eq:317}) were defined by the relations
\bea
m_i&=&\alpha_\ell +(i-1)m_x  
\nonumber\\
M_i&=&m_i+m_x
\eea
where the ``step'' mass $m_x$ was chosen to depend on the number of functions $n$ used in the expansion, so that
\bea
m_x=\frac{M_0}{n-1}\, .
\eea
Hence, as $n$ increases, $m_x$ decreases, giving a finer ``grid'' of mass scales.  It was found that the precise value of $M_0$ was not critical, except that fitting the complicated structure of $u(2)$ at large $p$  (to save writing in this section, we will sometimes use the notation $z(i)$ to denote the generic wave function for model WJC-$i$,  with $i=1, 2$) required a large value of $M_0$ which we chose to be 10 GeV.  For all other wave functions we chose $M_0=1.2$ GeV.  The leading mass $\alpha_\ell$ determines the asymptotic behavior of the wave function at large distances in coordinate space, and was therefore chosen to be $\alpha_0$ of Eq.~(\ref{eq:alpha}) for the S and D states, and $m_{s0}$ from Tables \ref{tab:scale} and \ref{tab:scale-2} for the P states.  The mass $M_n$ used in the tail wave functions was fixed at 2 GeV for all $z_i$ except $u(2)$, where it was fixed at 11 GeV.

Using these choices, excellent fits to the momentum space wave functions were found.  These are shown in Figs.~\ref{Four-I} and \ref{Four-II}.  The expansion parameters are given in Tables \ref{tab:parameters-1} and \ref{tab:parameters-2}.  

Note that the coefficients of the tail wave functions are very small.  The tail makes a negligible contribution at low momenta, but is very important for the description of the high momentum components (for $p\agt 3$ GeV).  Remember that the figures show {\it ratios\/}; the actual values of the wave functions are quite small above $p\agt 3$ GeV, as already shown in Fig.~\ref{fig:loglog}.  In any case it will be possible to study the sensitivity of any observable to the very high momentum components simply by setting this last coefficient to zero.

The components $z(i)$ are compared in Fig.~\ref{fig:All-P}, which shows the wave functions for momenta up to 1 GeV, and in Fig.~\ref{fig:All-P(fine)}, which  gives an exploded view of the wave functions from 500 MeV to 1 GeV, a region where all of the components might be important, depending on whether or not the larger $u$ and $w$ components interfere in a given matrix element.  

In general we observe that the P-state components are very small, particularly so for WJC-2.  However, in the region shown in Fig.~\ref{fig:All-P(fine)} all of the WJC-1 components are larger than the WJC-2 ones, and it is possible that, in some observables, P-state components might compensate for the differences between the larger S and D-state components.

\subsubsection{Transformations to coordinate space}\label{sec:Rspace}

The coordinate space wave functions are constructed from the spherical Bessel transforms (\ref{eq:besseltrans}).  The details are given in Appendix \ref{app:Rspace}.   The coordinate space expansion functions for the terms $i<n$ are
\bea
G^i_0(r)&=&
A_i\Bigg\{e^{-z_i}
-e^{-Z_i}\left[1+\sfrac12 Z_i\left(1-R_i^2\right)\right]\Bigg\}
\nonumber\\
G^i_1(r)&=&
A_i\Bigg\{R_ie^{-z_i}\Big[1+\frac1{z_i}\Big]
\nonumber\\
&&\qquad-e^{-Z_i}\left[1+\frac1{Z_i}+\sfrac12 Z_i\left(1-R_i^2\right)\right]\Bigg\},\qquad
\nonumber
\eea
\bea
G^i_2(r)&=&
B_i\Bigg\{R_i^2e^{-z_i}\Big[1+\frac3{z_i}+\frac3{z_i^2}\Big]
\nonumber\\
&&\qquad-\,e^{-Z_i}\Big[1+\frac3{Z_i}+\frac3{Z_1^2}+\sfrac12(1-R_i^2)(1+Z_i)
\nonumber\\
&&\qquad\qquad\quad+\sfrac18 Z_i^2\left(1-R_i^2\right)^2\Big]\Bigg\},
\label{eq:343}
\eea
where
\bea
z_i&=&m_i r\, ,\qquad Z_i=M_i r\, ,\qquad R_i=\frac{m_i}{M_i}\, ,
\nonumber\\
A_i&=&\frac{m_i^2 M_i^4}{(M_i^2-m_i^2)^2}\, ,\quad B_i=A_i\frac{M_i^2}{M_i^2-m_i^2}\, . \label{eq:343a}
\eea
The last terms with $i=n$ are
\bea
G^n_0(r)
&=&\frac2{3\pi} M_n^2\,Z_n^2 K_1(Z_n)
\nonumber\\
G^n_1(r)&=&\frac2{3\pi} M_n^2 Z_n^2 K_0(Z_n)
\nonumber\\
G^n_2(r)&=&\frac2{15\pi} M_n^2 Z_n^3 K_0(Z_n),
\eea
where $K_n(z)$ are the modified Bessel functions of the second kind (see Appendix \ref{app:C}).

\begin{figure}
\includegraphics[width=3.5in]{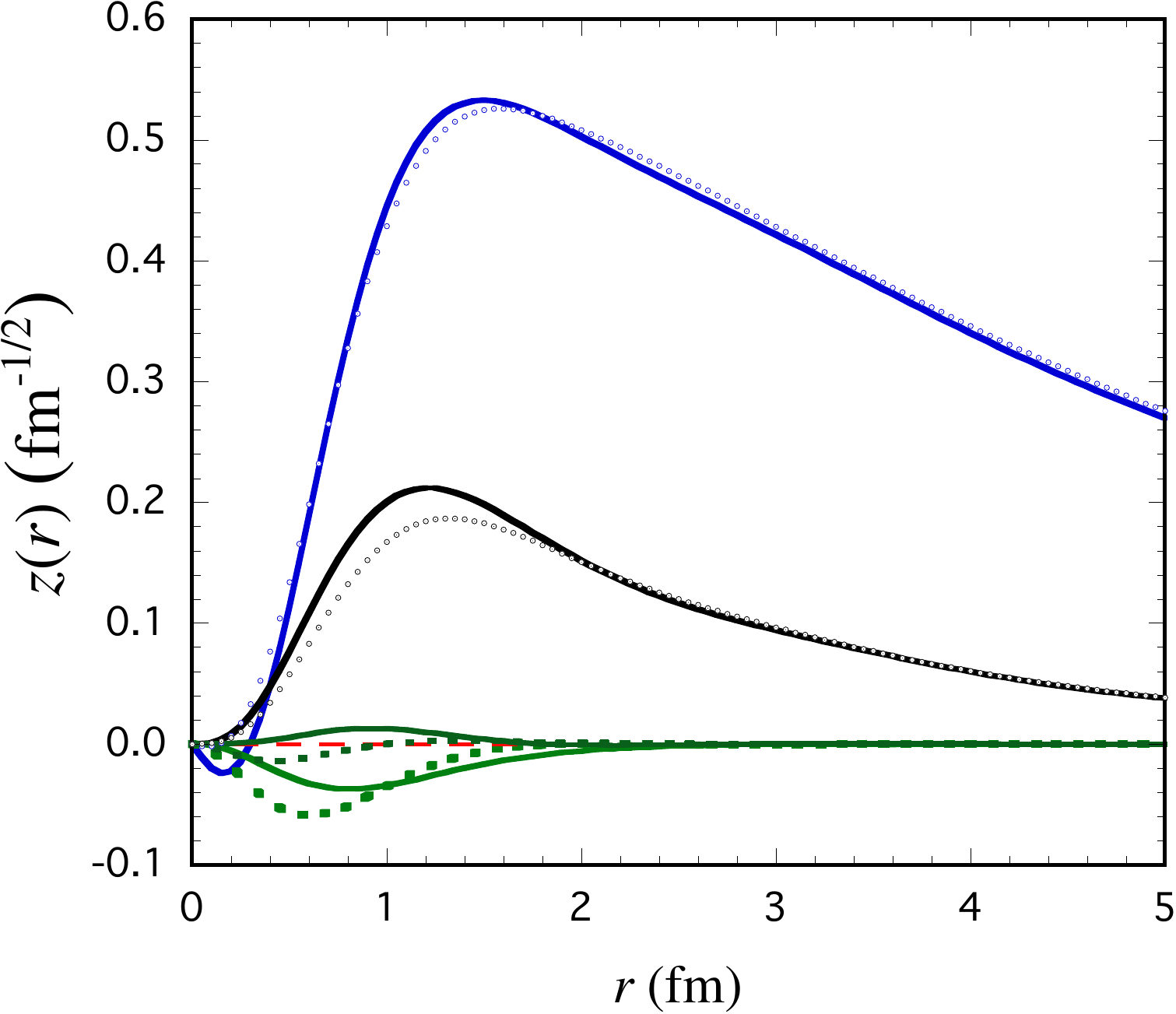}
\caption{\footnotesize\baselineskip=10pt (Color online) The coordinate space wave functions for both models. The curves are labeled as in Fig.~\ref{fig:All-P}.  The largest two wave functions are the familiar $u$ and $w$ wave functions, with WJC-1 solid and WJC-2 dotted.  The $v_t$ wave functions are both solid, with WJC-1 negative and WJC-2 small and positive.  The $v_s$ wave functions are both dotted, with WJC-1 negative and WJC-2 close to zero.   }
\label{fig:All-R}
\end{figure} 

\begin{figure}
\includegraphics[width=3.5in]{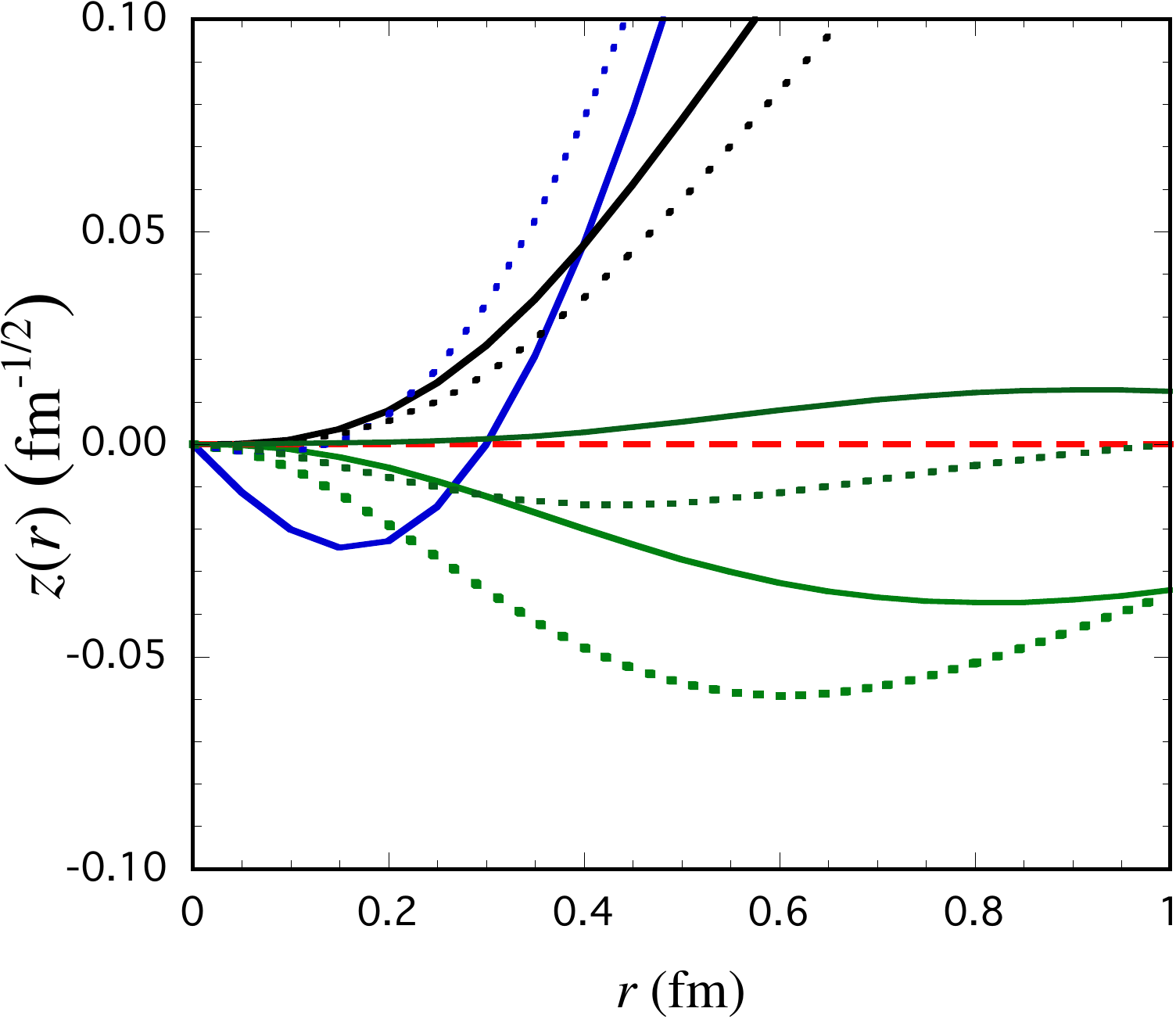}
\caption{\footnotesize\baselineskip=10pt (Color online) Expanded view of the coordinate space wave functions for both models at small $r$. The curves are labeled as in Fig.~\ref{fig:All-P}.  }
\label{fig:All-R(fine)}
\end{figure} 

At large $r$ the functions $G_\ell^i  (i<n)$ have the asymptotic behavior expected for solutions of the Schr\"odinger equation with orbital angular momentum $\ell$.   The tail functions fall-off like exponentials multiplied by a fractional power of $r$, but, because of their large mass (short range), do not contribute to the overall asymptotic behavior of the wave functions in coordinate space.  
At small $r$ the functions $G_\ell^i (i<n)$ have the expected $r^{\ell+1}\sim Z_i^{\ell+1}$ behavior.  However, the tail functions contribute some nonanalytic logarithmic behavior at small $r$, as described in Appendix \ref{app:C}.  As it turns out, this behavior is too small to be seen at the level of one percent.

\subsubsection{Wave functions in coordinate space}

The coordinate-space wave functions are shown in Figs.~\ref{fig:All-R} and \ref{fig:All-R(fine)}.  Note that the P-state wave functions vanish beyond 2 fm, and that they are larger for WJC-1, as already seen in momentum space.  Particularly notable is the zero in $u(1)$ near 0.35 fm, leading to a dip at very small $r$.  This dip is not an artifact of the fits; it is related to the deeper dip in the momentum space $u(1)$ near 600 MeV, clearly seen in Figs.~\ref{fig:All-P} and \ref{fig:All-P(fine)}.  A similar (but much smaller and not visible in the figures) dip is also present in $u(2)$ inside of 0.15 fm.  The fact that the dip in WJC-1 is larger may be a consequence of its  larger P-state components.

Before any physical conclusions can be drawn from the differences in the wave functions for WJC-1 and WJC-2, observables must be calculated from these models.  Electromagnetic observables (such as the quadrupole moment) will be sensitive to interaction currents, which will be different in each case, and might very well compensate for any difference arising from the wave functions.  However, these calculations, which will be a subject for a future paper, are beyond the scope of this work.

\subsubsection{Asymptotic normalization and the $D/S$ ratio}

One observable that can be calculated immediately is the ratio of the asymptotic normalization constants for the S and D-state wave functions.  The results for the two models, together with the asymptotic normalization $A_s$, are summarized in Table \ref{tab:DtoS}.  These quantities can be extracted from the expansions (\ref{eq:343}):
\bea
A_s&=&b_1^uA_1, \qquad A_d=b_1^wB_1 R_1^2
\nonumber\\
\eta&=&A_d/A_s\, .
\eea
Note that $\eta$ is more accurately determined than $A_s$, and that the values of $\eta$ predicted by the two models are in agreement with each other (within errors), and also agree (to less than 2 standard deviations) with the results of Ref.~\cite{Rodning:1990zz}. The agreement is even better with most other experimental and theoretical results for $\eta$ quoted in Ref.~\cite{Rodning:1990zz}, which are all situated in the range between 0.0259 and 0.0272.

\begin{table}
\begin{minipage}{3.2in}
\caption{The asymptotic normalization $A_s$ and $D/S$ ratio $\eta$ for the two models.  Results are shown for variations of $n$ around the central (best) values of 17 and 16 for WJC-1 and 30 and  18 for WJC-2, with errors based on the fluctuations.  These may be compared to the recently determined $\eta=0.0256(4)$ \cite{Rodning:1990zz}.}
\label{tab:DtoS}
\begin{ruledtabular}
\begin{tabular}{cllll}
& \multicolumn{2}{c}{WJC-1} & \multicolumn{2}{c}{WJC-2} \cr
$n$ & $\quad\eta$  & $\quad A_s$ & $\quad\eta$ & $\quad A_s$ \cr 
\tableline
$n_0-2$ & 0.02625  &   0.86585 & 0.02635  &   0.87471    \cr
$n_0-1$ & 0.02623   &  0.86837 & 0.02639   & 0.87693      \cr
$n_0$ & 0.02619    &  0.86416   & 0.02630 &    0.87768    \cr
$n_0+1$ & 0.02623   &  0.86604  & 0.02643  &   0.87829 \cr
$n_0+2$ & 0.02622   &  0.86436 & 0.02614  &   0.87879 \cr
\tableline
$n_0$(error) & 0.02619(4)    &  0.864(2) & 0.0263(1) &    0.8777(15)  \cr
\end{tabular}
\end{ruledtabular}
\end{minipage}
\end{table}
%

\section{Conclusions}

In this paper we present (Table \ref{tab:effective}) effective range expansions for the $^1S_0$ and $^3S_1$ phase shifts and relativistic deuteron wave functions based on the precision models WJC-1 and WJC-2 of Ref.~I.

Since the effective range expansions emerge directly from the new high precision phase shift analysis described in Ref.~I, they can be regarded as an up-to-date, precision determination of the effective range parameters that should constrain any modern theory of the nuclear force, such as effective (or chiral effective) field theories.  It does not matter that they were determined using the CST.

The covariant deuteron wave functions presented here  have four components:\ two that have a nonrelativistic analogue, and two (the P-states) of purely relativistic origin.  Convenient analytic representations of these wave functions are presented, with expansion coefficients given in Tables \ref{tab:parameters-1} and \ref{tab:parameters-2}. These expansions (and subroutines that are available from the authors) make it easy and convenient to use these wave functions with any calculation.  When these wave functions are used in a covariant theory with one nucleon off-shell they provide a precise description of the non-perturbative interactions that lead to the deuteron bound state.  However, even in this case a completely consistent description requires that we include other contributions, such as final state interactions or interaction currents, that are also generated by the one-boson-exchange dynamics.  That these effects cannot be ignored, even in the low-energy or momentum limit, is indicated by the size of the derivative of the kernel that contributes to the normalization [recall Eq.~(\ref{eq:norm1})].  This is about $-$5\% for WJC-1 and $-$2\% for WJC-2, and could, of course, be much larger at higher energies for some observables.  In particular,  the calculation of any electromagnetic property  of the deuteron (including the magnetic and quadrupole moments) that does {\it not\/} include interaction currents can be expected {\it ab initio\/} to be in error by a few percent. It is therefore quite possible that a careful CST calculation of the quadrupole moment could resolve the current $\sim5\%$ discrepancy between theory and experiment, but this calculation must wait until the exchange currents have been accurately calculated.

For this reason it is also unclear how to incorporate the wave functions given in this paper into a calculation which uses dynamics different from the CST (nonrelativistic or another form of relativistic dynamics -- including light-front).  In such cases we are inclined to suggest that the best procedure is to renormalize the wave functions as shown in Eq.~(\ref{eq:norm2}), which is also why our numerical wave functions are presented here in this normalization.

One observable that does {\it not\/} depend on the normalization of the wave functions is the asymptotic $D/S$ ratio.  Our results (Table \ref{tab:DtoS}) determine this ratio to an accuracy about 10 times smaller than the experimental accuracy and are in good agreement with measured values.  Still, it would be interesting to review the theory that has gone into the analysis of the experiments that determine the $D/S$ ratio and to see if there are any relativistic corrections previously overlooked.

\acknowledgments

This work is the continuation of an effort extending over more that a decade, supported initially by the DOE  through grant  No.~DE-FG02-97ER41032, and recently supported by Jefferson Science Associates, LLC under U.S. DOE Contract No.~DE-AC05-06OR23177. A.\ S.\ was supported by FEDER and Funda\c c\~ao para a Ci\^encia e a Tecnologia (FCT) under grant No.~POCTI/ISFL/2/275 and thanks the Jefferson Lab Theory Group for the hospitality extended to him during several visits while this work was performed.  We also acknowledge prior work by R. Machleidt and J.W. Van Orden, who wrote some earlier versions of parts of the $NN$ code.  The data analysis used parts of the SAID code supplied to us by R.\ A.\ Arndt.  Dick was a good friend and we will miss him very much.  Helpful conversations with the the Nijmegen group (J. J. de Swart, M. C. M. Rentmeester, and R.G.E. Timmermans) and with R.\ Schiavilla are gratefully acknowledged.  Finally, we thank  J.~J.~Adam, Jr.\ for a review of the code which lead to a realization that the angular integrals originally used were slow to converge; the new mappings described in Appendix \ref{app:B} have improved the convergence while not altering the results previously published in Ref.~I.



\appendix



\section{CONSTRUCTION OF THE KERNEL}\label{app:A}

In this appendix we present details, omitted from Ref.~I  \cite{Gross:2008ps}, of the construction of the $NN$ kernel directly from the sum of Feynman amplitudes that define it.  This method is completely different from that used in Ref.~II  \cite{GVOH}; it is simpler and more transparent allowing changes in the kernel to be made more easily.

\subsection{Review of definitions from Ref.~I}

In Ref.~I the kernels are defined to be
\bal
&{\bf  V}^{J\,\rho_1\rho_2,\rho'_1\rho'_2}_{\lambda_1\lambda_2,\lambda'_1\lambda'_2}(\delta_{_{p_0}},\delta_S)
\nonumber\\
&\quad=
\sfrac{1}{2}\Big\{V^{J\,\rho_1\rho_2,\rho'_1\rho'_2}_{{\rm dir}\;\lambda_1\lambda_2,\lambda'_1\lambda'_2}(p_0) + \delta_S V^{J\,\rho_1\rho_2,\rho'_1\rho'_2}_{{\rm dir}\;\lambda_1\lambda_2,\,-\!\lambda'_1\,-\!\lambda'_2}(p_0) 
\nonumber\\
&+\delta_{_{p_0}}V^{J\,\rho_2\rho_1,\rho'_1\rho'_2}_{{\rm dir}\;\lambda_1\lambda_2,\lambda'_1\lambda'_2}(-p_0)
+\delta_{_{p_0}}\delta_S V^{J\,\rho_2\rho_1,\rho'_1\rho'_2}_{{\rm dir}\;\lambda_1\lambda_2,\,-\!\lambda'_1\,-\!\lambda'_2}(-p_0)\Big\}
, \qquad\label{A31a}
\end{align}
where $\delta_{_{p_0}}=(-1)^I\eta(\rho_T\delta_S)^\lambda$ (with $\rho_T=\rho_1\rho_2\rho'_1\rho'_2$, $\lambda=\lambda_1-\lambda_2$, and $I$ the isospin) is the phase of ${\bf V}$ under $p_0\to -p_0$ and $\rho_1\leftrightarrow \rho_2$, and $\delta_S=\delta_P\,\rho'_1\rho'_2\,\eta$ (with $\delta_P=\pm1$ the parity), and in both experssions $\eta=(-1)^{J-1}$ is a ubiquitous phase.  {\it These linear combinations are very similar to (but different from) those used referred to as Ref.~II}.

If particle 1 is on-shell,  $\rho_1=\rho'_1=+$.   Using the parity relation,
\bal
V^{J\,\rho_1\rho_2,\rho'_1\rho'_2}_{{\rm dir}\,\lambda_1\lambda_2,\lambda'_1\lambda'_2}(\pm p_0)&=\rho_{\rm T}
V^{J\,\rho_1\rho_2,\rho'_1\rho'_2}_{{\rm dir}\,-\!\lambda_1\,-\!\lambda_2,\,-\!\lambda'_1\,-\!\lambda'_2}(\pm p_0), \label{A30}
\end{align}
%
the amplitudes (\ref{A31a}) can always be organized so that $\lambda_1=\lambda'_1=+1/2$, leaving the helicities of particle 2 and the phases $\delta_S$ and $\delta_{_{p_0}}$ unconstrained. Hence there are $2^4=16$  independent kernels for {\it each\/} $J, \rho_2$ and $\rho'_2$. Our particular choice of independent kernels is denoted $v_i$ and defined in Table XI of Ref.~I. Their behavior  under parity and interchange symmetry for each combination of $\{\rho_2,\,\rho'_2\}$ is given in Table XII of Ref.~I.

The evaluation of the matrix elements $V^J$ is simplified by the fact that one can write the nucleon spinors as direct products of two two-component spinors, one of which contains all angle-dependence, while the other describes the dependence on momentum and $\rho$-spin.  Recall that the nucleon spinors (from Eq.~(E1) of Ref.~I) are:
\begin{align}
u^\rho_1({\bf p},\lambda)&=N_\rho(p\lambda)\otimes \chi_{_\lambda}(\theta) = \begin{cases} u({\bf p}, \lambda) & \rho=+  \cr v(-{\bf p}, \lambda) & \rho=- \end{cases}
\nonumber\\
u^\rho_2({\bf p},\lambda)&=N_\rho(p\lambda)\otimes \chi_{_{-\lambda}}(\theta) = \begin{cases} u(-{\bf p}, -\lambda) & \rho=+  \cr v({\bf p}, -\lambda) & \rho=- \end{cases}
\label{A1}
\end{align}
where the \change{rho}{$\rho$}-space spinors are 
\bal
N_+(p\lambda)&=\left(\begin{array}{c} \cosh\sfrac12 \zeta \\ [0.1in]
2\lambda\sinh \sfrac12\zeta
\end{array}\right) \nonumber\\
N_-(p\lambda)&=\left(\begin{array}{c} -2\lambda\sinh \sfrac12\zeta \\ [0.1in]
\cosh\sfrac12 \zeta
\end{array}\right), \label{eq:14}
\end{align}
with $p=|{\bf p}|$ and $\tanh \zeta = p/E_p$.  Note that, at large $p$,
\bea
&&\cosh\sfrac12 \zeta=\sqrt{\frac{E_p+m}{2m}}\to \sqrt{\frac{p}{2m}}
\nonumber\\
&&\sinh\sfrac12\zeta=\frac{p}{\sqrt{2m(E_p+m)}}\to \sqrt{\frac{p}{2m}}\, . \label{eq:A5}
\eea  
For momenta limited to the $\hat x\hat z$ plane, the spin 1/2 spinors are
\begin{align}
\chi_{_{1/2}}(\theta)&=R_y(\theta)\left(\begin{array}{c} 1 \\ 0\end{array}\right)=
\left(\begin{array}{c} \cos\sfrac12\theta \\ [0.1in]
\sin \sfrac12\theta
\end{array}\right) \nonumber\\
\chi_{_{-1/2}}(\theta)&=R_y(\theta)\left(\begin{array}{c} 0 \\ 1\end{array}\right)=\left(\begin{array}{c} -\sin\sfrac12\theta \\ [0.1in]
\cos \sfrac12\theta
\end{array}\right). \label{A3}
\end{align}
%


\subsection{Extracting the angular integrals}

Each term in the Feynman meson-exchange operator can be written in the generic form
\bea
{\cal V}_b(\theta,p_0)&=&\sum_{ij} g_i s_j {\cal D}_b(\theta, p_0) 
\nonumber\\
&&\times [{\cal O}^i(p_0)\otimes{\cal S}^j]_1^b\; [{\cal O}^{i}(p_0)\otimes{\cal S}^{j}]_2^b\, , \;\;
\eea
 where ${\cal O}^i$ are operators in the $\rho$-spin space, ${\cal S}^j$ operators in the spin-space,  $g_i$ and $s_j$ are constants, and the sums are over all operators needed to describe the meson exchange interaction. 
Here ${\cal D}_b$ is the scalar part of the propagator
\bea
{\cal D}_b(\theta, p_0)=\frac{f_b(\Lambda_b,q)}{m_b^2+|q(\theta,p_0)|^2}, \label{eq:A8}
\eea
with $q(\theta, p_0)$ the four-momentum transferred by the exchanged meson and $f_b$ the meson form factor.  The operator ${\cal O}(p_0)$ acts in the $2\times2$  $\rho$-spin space, and ${\cal S}$ operates in the $2\times2$ spin space, so that ${\cal O}\otimes{\cal S}$ is a 4$\times$4 matrix.    Using this notation, the matrix elements of ${\cal V}$ reduce to (suppressing the index $b$ for simplicity)
%
\bal
V^{\rho_1\rho_2,\rho'_1\rho'_2}_{{\rm dir}\;\lambda_1\lambda_2,\lambda'_1\lambda'_2}(\theta, \pm p_0)
=&{\cal O}^{\rho_1\rho'_1,\rho_2\rho'_2}_{\lambda_1\lambda_2,\lambda'_1\lambda'_2,\pm}
\nonumber\\&
\left<\lambda_1\lambda_2|\lambda'_1\lambda'_2\right>{\cal D}(\theta, \pm p_0).
\end{align}
where
\bal
{\cal O}^{\rho_1\rho'_1,\rho_2\rho'_2}_{\lambda_1\lambda_2,\lambda'_1\lambda'_2,\pm}=&\sum_ig_i \,\Big[\bar N_{\rho_1}(p\lambda_1)\, {\cal O}^i(\pm p_0) N_{\rho'_1}(p' \lambda'_1)\Big]
\nonumber\\&\qquad\times 
\Big[\bar N_{\rho_2}(p\lambda_2)\, {\cal O}^i(\pm p_0)N_{\rho'_2}(p' \lambda'_2)\Big] 
\nonumber\\
\left<\lambda_1\lambda_2|\lambda'_1\lambda'_2\right>=& \sum_j s_j\,\Big[\chi^\dagger_{_{\lambda_1}}(\theta)\, {\cal S}^j \chi_{_{\lambda'_1}}(0)\Big]
\nonumber\\&\qquad\times
\Big[\chi^\dagger_{_-{\lambda_2}}(\theta)\, {\cal S}^j \chi_{_{-\lambda'_2}}(0)\Big]\, , \label{eq:18}
\end{align}
%
with $\bar N=N^\dagger\tau_3$ the Dirac conjugation in $\rho$-spin space, and use is made of the fact that  the partial wave amplitude may be calculated by aligning the initial momentum ${\bf p}'$ in the $+z$ direction.
Note that the sign of $p_0$ in the operators ${\cal O}(\pm p_0)$ is captured in the last subscript of the matrix elements ${\cal O}$, \change{and}{} that  the superscripts for ${\cal O}$ group together the $\rho$-spin indices for particle 1, followed by the  $\rho$-spin indices for particle 2, and that the \change{${\cal S}$ matrix elements}{matrix elements of ${\cal S}$} do not depend on the \change{rho}{$\rho$}-spin of the states.  We keep $\lambda_1$ unspecified in these formulae even though only the case $\lambda_1=+$ needs to be considered.  The linear combinations (\ref{A31a}) restrict $\lambda'_1=+$, but the individual terms in the sum use both signs of $\lambda'_1$, so it cannot be restricted in (\ref{eq:18}).  Recalling that the partial wave projections are of the form
\bal
V^{J\,\rho_1\rho_2,\rho'_1\rho'_2}_{{\rm dir}\;\lambda_1\lambda_2,\lambda'_1\lambda'_2}(p_0)=& 
2\pi \int_0^\pi \sin\theta\,d\theta\,d^J_{\lambda'\lambda}(\theta)\nonumber\\
&\times
V^{J\,\rho_1\rho_2,\rho'_1\rho'_2}_{{\rm dir}\;\lambda_1\lambda_2,\lambda'_1\lambda'_2}(p,p';P),
 \label{A26}
\end{align}
we see that each of the matrix elements $\left<\lambda_1\lambda_2|\lambda'_1\lambda'_2\right>$ will be multiplied by the rotation function $d^J_{\lambda'\lambda}(\theta)$, where $\lambda=\lambda_1-\lambda_2$ and $\lambda'=\lambda'_1-\lambda'_2$.  It turns out that the $\rho$-spin matrix elements  generate an additional factor of $z=\cos\theta$, so matrix elements with an additional factor of $z$ are needed; these are handled by multiplying each of the matrix elements by $z^i$, where $i=0$ or 1.  The 16 integrals we require are defined in Table \ref{tab:5}, with explicit expressions for these 16  matrix elements in terms of 5 independent integrals summarized in Table \ref{tab:3}.   The 5 independent integrals over Legendre polynomials and their derivatives are given in Table \ref{tab:VIIa}, with convenient identities showing how to reduce 5 other integrals to these 5 given in Table \ref{tab:VII},

\begin{table*}
\begin{minipage}{6in}
\caption{The 16 matrix elements $A^i_{\lambda_1\lambda_2,\lambda'_1\lambda'_2}=z^i\left<+\lambda_2|\lambda'_1\lambda'_2\right>d^J_{\lambda'\lambda}$ for $\lambda_1=+$ and $i=0$ or 1.  Here $\lambda=\pm\sfrac12$ is written as $\lambda=\pm$ and $z=\cos\theta$.}
\label{tab:5}
\begin{ruledtabular}
\begin{tabular}{ll} 
$A^i_{++,++}=z^i\left<++|++\right>d^J_{00}=z^i\sfrac12(1+z)d^J_{00}$& 
$A^i_{++,--}=z^i\left<++|--\right>d^J_{00}=-z^i\sfrac12(1-z)d^J_{00}$   \cr
$A^i_{+-,++}=z^i\left<+-|++\right>d^J_{01}=z^i\sfrac12\sin\theta \,d^J_{01}$& 
$A^i_{+-,--}=z^i\left<+-|--\right>d^J_{01}=z^i\sfrac12\sin\theta \,d^J_{01}$   \cr
$A^i_{++,+-}=z^i\left<++|+-\right>d^J_{10}=-z^i\sfrac12\sin\theta\,d^J_{10}$& 
$A^i_{++,-+}=z^i\left<++|-+\right>d^J_{-\!1,0}=z^i\sfrac12\sin\theta\,d^J_{-\!1,0}$   \cr
$A^i_{+-,+-}=z^i\left<+-|+-\right>d^J_{11}=z^i\sfrac12(1+z)d^J_{11}$& 
$A^i_{+-,-+}=z^i\left<+-|-+\right>d^J_{-\!1,1}=z^i\sfrac12(1-z)\,d^J_{-\!1,1}$   \cr
\end{tabular}
\end{ruledtabular}
\end{minipage}
\end{table*}
%

\begin{table*}
\begin{minipage}{6.5in}
\caption{Matrix elements of Table \ref{tab:5} expressed in terms of the angular functions from Table \ref{tab:VIIa} and $a_6$, $a_7$, $a_8$ and $a_9$ from Table \ref{tab:VII}.  Note that $d^J_{-\!1,0}=-d^J_{1,0}$ and $d^J_{01}=-d^J_{10}$.  Here $a_i=a_i^J(\theta)$.  The integrals $a_4$ and $a_5$ are needed only for the $A^1$ and hence only for vector mesons.  The matrix elements ${\cal A}_{ij,k}$ have $i,j=\{1,4\}$ according to the code $1=++, 2=+-, 3=-+, 4=--$ and $k=1$ for $A^0$ and $k=2$ for $A^1$.}
\label{tab:3}
\begin{ruledtabular}
\begin{tabular}{ll} 
$A^0_{++,++}={\cal A}_{11,1}=\sfrac12(1+z)d^J_{00}=\sfrac12(a_1+a_2)$ & 
$A^0_{++,--}={\cal A}_{14,1}=-\sfrac12(1-z)d^J_{00}=-\sfrac12(a_1-a_2)$   \cr
$A^1_{++,++}={\cal A}_{11,2}=\sfrac12z(1+z)d^J_{00}=\sfrac12(a_2+a_4)$ & 
$A^1_{++,--}={\cal A}_{14,2}=-\sfrac12z(1-z)d^J_{00}=-\sfrac12(a_2-a_4)$   \cr
$A^0_{+-,++}={\cal A}_{21,1}=\sfrac12\sin\theta \,d^J_{01}=-\sfrac12 a_7$& 
$A^0_{+-,--}={\cal A}_{24,1}=\sfrac12\sin\theta \,d^J_{01}=-\sfrac12 a_7$   \cr
$A^1_{+-,++}={\cal A}_{21,2}=\sfrac12z\sin\theta \,d^J_{01}=-\sfrac12 a_8$& 
$A^1_{+-,--}={\cal A}_{24,2}=\sfrac12z\sin\theta \,d^J_{01}=-\sfrac12 a_8$   \cr
$A^0_{++,+-}={\cal A}_{12,1}=-\sfrac12\sin\theta\,d^J_{10}=-\sfrac12 a_7$&
$A^0_{++,-+}={\cal A}_{13,1}=\sfrac12\sin\theta\,d^J_{-\!1,0}=-\sfrac12 a_7$   \cr
$A^1_{++,+-}={\cal A}_{12,2}=-\sfrac12 z\sin\theta\,d^J_{10}=-\sfrac12 a_8$&
$A^1_{++,-+}={\cal A}_{13,2}=\sfrac12 z\sin\theta\,d^J_{-\!1,0}=-\sfrac12 a_8$   \cr
$A^0_{+-,+-}={\cal A}_{22,1}=\sfrac12(1+z)d^J_{11}=\sfrac12\left(a_6+a_1\right)$& 
$A^0_{+-,-+}={\cal A}_{23,1}=\sfrac12(1-z)\,d^J_{-\!1,1}=\sfrac12\left(a_6-a_1\right)$   \cr
$A^1_{+-,+-}={\cal A}_{22,2}=\sfrac12 z(1+z)d^J_{11}=\sfrac12\left(a_9+a_2\right)$& 
$A^1_{+-,-+}={\cal A}_{23,2}=\sfrac12 z(1-z)\,d^J_{-\!1,1}=\sfrac12\left(a_9-a_2\right)$  
\end{tabular}
\end{ruledtabular}
\end{minipage}
\end{table*}


\begin{table}
\begin{minipage}{2.5in}
\caption{The five independent angular integrals, $a_i^J(\theta)$, needed for this calculation.  Four more frequently used linear combinations of these, denoted $a_j$, with $j=\{6,9\}$, are given in Table \ref{tab:VII}.}
\label{tab:VIIa}
\begin{ruledtabular}
\begin{tabular}{cc}
$a^J_1(\theta)$ &  
$P_J(z)$\\[0.05in]
$a^J_2(\theta)$ & $zP_J(z)$ \\[0.05in]
$a^J_3(\theta)$ & $P_{J-1}(z)$\\[0.05in] 
$a^J_4(\theta)$ &  $z^2 P_{J}(z)$ \\[0.05in] 
$a^J_5(\theta)$ & $zP_{J-1}(z)$ \\[0.05in]  
\end{tabular}
\end{ruledtabular}
\end{minipage}
\end{table}


\begin{table}
\begin{minipage}{2.5in}
\caption{Useful identities involving the $d$ functions.  Here $P_J=P_J(z)$, $a_i=a_i^J(\theta)$, and the new combinations, $a_6, a_7, a_8,$ and $a_9$ are defined.}
\label{tab:VII}
\begin{ruledtabular}
\begin{tabular}{l}
$\sfrac12\big[(1+z)d^J_{11}-(1-z)d^J_{-1,1}\big] =d^J_{00}=P_J=a_1$
\\[0.2in]
$\sfrac12\left[(1+z)d^J_{11}+(1-z)d^J_{-1,1}\right]=a_6$
\\[0.05in] 
$=\sfrac1{J+1}\left[J zP_{J} + P_{J-1}\right]
=\sfrac1{J+1}\left[J a_2 +  a_3\right] $
\\[0.2in]
$\sfrac12z\left[(1+z)d^J_{11}+(1-z)d^J_{-1,1}\right]=a_9$
\\[0.05in] 
$=\sfrac1{J+1}\left[J z^2P_{J} + zP_{J-1}\right]
=\sfrac1{J+1}\left[ J a_4 + a_5\right] $
\\[0.2in] 
$\sin\theta d^J_{10}=\sqrt{\sfrac{J}{J+1}}\left[zP_{J}- P_{J-1}\right]$\\[0.1in] 
$\qquad\qquad=\sqrt{\sfrac{J}{J+1}}\left[a_2- a_3\right]\equiv a_7$
\\[0.2in] 
$z\sin\theta d^J_{10}=\sqrt{\sfrac{J}{J+1}}\left[a_4- a_5\right]\equiv a_8$\\[0.1in] 
\end{tabular}
\end{ruledtabular}
\end{minipage}
\end{table}


The 16 generic integrals (for {\it each\/} value of $\rho_1$) defined in Tables XI and XII of Ref.~I can then be written as products of the 16 angular-dependent terms given in Table \ref{tab:5} and Dirac matrix elements (defined below) as follows:
\begin{widetext}
 \bal
& v_{1\atop 5}^{\rho_2\rho'_2}=\sum_i\Bigg\{ \int_{+}\Big(
{\cal O}^{\rho_1+,\rho_2\rho'_2,i}_{++,++,+}A^i_{++,++}- 
{\cal O}^{\rho_1+,\rho_2\rho'_2,i}_{++,--,+}A^i_{++,--}\Big)
\pm\int_{-}\Big(
{\cal O}^{\rho_2+,\rho_1\rho'_2}_{++,++,-}A^i_{++,++}-
{\cal O}^{\rho_2+,\rho_1\rho'_2}_{++,--,-}A^i_{++,--}\Big)\Bigg\}
 \nonumber\\
 & v_{2\atop6}^{\rho_2\rho'_2}=\sum_i\Bigg\{ \int_{+}\Big(
{\cal O}^{\rho_1+,\rho_2\rho'_2,i}_{+-,+-,+}A^i_{+-,+-}-
{\cal O}^{\rho_1+,\rho_2\rho'_2,i}_{+-,-+,+}A^i_{+-,-+}\Big)
\mp\int_{-}\Big(
{\cal O}^{\rho_2+,\rho_1\rho'_2}_{+-,+-,-}A^i_{+-,+-}-
{\cal O}^{\rho_2+,\rho_1\rho'_2}_{+-,-+,-}A^i_{+-,-+}\Big)\Bigg\}
 \nonumber\\
& v_{3\atop7}^{\rho_2\rho'_2}=\sum_i\Bigg\{ \int_{+}\Big(
{\cal O}^{\rho_1+,\rho_2\rho'_2,i}_{++,+-,+}A^i_{++,+-}-
{\cal O}^{\rho_1+,\rho_2\rho'_2,i}_{++,-+,+}A^i_{++,-+}\Big)
\pm\int_{-}\Big(
{\cal O}^{\rho_2+,\rho_1\rho'_2}_{++,+-,-}A^i_{++,+-}-
{\cal O}^{\rho_2+,\rho_1\rho'_2}_{++,-+,-}A^i_{++,-+}\Big)\Bigg\}
\nonumber\\
& v_{4\atop8}^{\rho_2\rho'_2}=\sum_i\Bigg\{ \int_{+}\Big(
{\cal O}^{\rho_1+,\rho_2\rho'_2,i}_{+-,++,+}A^i_{+-,++}-
{\cal O}^{\rho_1+,\rho_2\rho'_2,i}_{+-,--,+}A^i_{+-,--}\Big)
\mp\int_{-}\Big(
{\cal O}^{\rho_2+,\rho_1\rho'_2}_{+-,++,-}A^i_{+-,++}-
{\cal O}^{\rho_2+,\rho_1\rho'_2}_{+-,--,-}A^i_{+-,--}\Big)\Bigg\}
\nonumber\\
& v_{9\atop13}^{\rho_2\rho'_2}=\sum_i\Bigg\{ \int_{+}\Big(
{\cal O}^{\rho_1+,\rho_2\rho'_2,i}_{++,++,+}A^i_{++,++}+
{\cal O}^{\rho_1+,\rho_2\rho'_2,i}_{++,--,+}A^i_{++,--}\Big)
\pm\int_{-}\Big(
{\cal O}^{\rho_2+,\rho_1\rho'_2}_{++,++,-}A^i_{++,++}+
{\cal O}^{\rho_2+,\rho_1\rho'_2}_{++,--,-}A^i_{++,--}\Big)\Bigg\}\nonumber\\
& v_{10\atop14}^{\rho_2\rho'_2}=\sum_i\Bigg\{ \int_{+}\Big(
{\cal O}^{\rho_1+,\rho_2\rho'_2,i}_{+-,+-,+}A^i_{+-,+-}+
{\cal O}^{\rho_1+,\rho_2\rho'_2,i}_{+-,-+,+}A^i_{+-,-+}\Big)
\pm\int_{-}\Big(
{\cal O}^{\rho_2+,\rho_1\rho'_2}_{+-,+-,-}A^i_{+-,+-}+
{\cal O}^{\rho_2+,\rho_1\rho'_2}_{+-,-+,-}A^i_{+-,-+}\Big)\Bigg\}
\nonumber\\
& v_{11\atop15}^{\rho_2\rho'_2}=\sum_i\Bigg\{ \int_{+}\Big(
{\cal O}^{\rho_1+,\rho_2\rho'_2,i}_{++,+-,+}A^i_{++,+-}+
{\cal O}^{\rho_1+,\rho_2\rho'_2,i}_{++,-+,+}A^i_{++,-+}\Big)
\pm\int_{-}\Big(
{\cal O}^{\rho_2+,\rho_1\rho'_2}_{++,+-,-}A^i_{++,+-}+
{\cal O}^{\rho_2+,\rho_1\rho'_2}_{++,-+,-}A^i_{++,-+}\Big)\Bigg\}
\nonumber\\
&v_{12\atop16}^{\rho_2\rho'_2}=\sum_i\Bigg\{ \int_{+}\Big(
{\cal O}^{\rho_1+,\rho_2\rho'_2,i}_{+-,++,+}A^i_{+-,++}+
{\cal O}^{\rho_1+,\rho_2\rho'_2,i}_{+-,--,+}A^i_{+-,--}\Big)
\pm\int_{-}\Big(
{\cal O}^{\rho_2+,\rho_1\rho'_2}_{+-,++,-}A^i_{+-,++}+
{\cal O}^{\rho_2+,\rho_1\rho'_2}_{+-,--,-}A^i_{+-,--}\Big)\Bigg\},
 \label{eq:110}
 \end{align}
%
\end{widetext}
where the $+$ and $-$ integrals include the propagator
\bea
\int_{\pm}=\pi\int_0^\pi \sin\theta \,d\theta\,{\cal D}(\theta, \pm p_0)
\eea
and, in each case, the upper sign goes with the upper index (of the two lower indices) and the lower sign with the lower index (of the two lower indices), so that, for example, $v_1$ is symmetric under $p_0\to-p_0$ and $v_5$ is antisymmetric.  The 16 $\rho$-spin matrix elements that enter into these expressions, all independent of $\theta$, result from the fact that the general $\rho$-spin matrix elements ${\cal O}$ given in Eq.~(\ref{eq:18}) are linear in $z$, and can therefore be separated into two terms using the notation
\bea
{\cal O}^{\rho_1\rho'_1,\rho_2\rho'_2}_{\lambda_1\lambda_2,\lambda'_1\lambda'_2,\pm}=
{\cal O}^{\rho_1\rho'_1,\rho_2\rho'_2,0}_{\lambda_1\lambda_2,\lambda'_1\lambda'_2, \pm}
+z\,{\cal O}^{\rho_1\rho'_1,\rho_2\rho'_2,1}_{\lambda_1\lambda_2,\lambda'_1\lambda'_2, \pm}. \qquad\label{eq:113}
\eea
The ${\cal O}^1$'s arise only from the exchange of vector mesons; all of the ${\cal O}$'s arising from the exchange of each meson will be computed below.

\subsection{Particle momenta}

In some applications there is the possibility that {\it both\/} particles in the final state could be off-shell (the possibility that both particles in the initial state are off-shell  is not considered at this time).  To allow for this possibility, we will write the final state momenta in the cm system of the two-nucleon system in the general form
\bea
p_1&=&\{\sfrac12 W+p_0, {\bf p}\}=\{x_0E_p+(1-x_0)\sfrac12W,{\bf p}\}
\nonumber\\
p_2&=&\{\sfrac12 W-p_0, -{\bf p}\}
\nonumber\\
&=&\{-x_0E_p+(1+x_0)\sfrac12W,-{\bf p}\} \label{eq:147}
\eea
with
\bea
p_0&=&x_0(E_p-\sfrac12W) \, ,
 \label{eq:148}
\eea
where 
$x_0$ is a dimensionless number varying between $-\infty$ and $\infty$\change{}{, and $W$ is the total two-body energy}.  Note that when $x_0=1$, particle 1 is on-shell (with energy $E_p$), while when $x_0=-1$, particle 2 is on-shell.  Hence changing the sign of $x_0$ is a convenient way to interchange the energies of particles 1 and 2 in the final state, and we may construct the $\pm p_0$ combinations of Eq.~(\ref{A31a}) merely by changing the sign of $x_0$.  (A similar variable, $y_0$, could be used for the initial state.)

\subsection{Single particle $\rho$-spin matrix elements}

We now turn our attention to the form of the $\rho$-spin matrix elements ${\cal O}^{\rho_1\rho'_1,\rho_2\rho'_2,i}_{\lambda_1\lambda_2,\lambda'_1\lambda'_2}$.  In all but the simplest cases, these are best calculated by first constructing matrix elements on {\it each\/} nucleon line and then multiplying these together to get the total ${\cal O}$. The one-nucleon matrix elements are evaluated numerically by matrix multiplication, best described separately for each type of meson exchange.   The total $\rho$-spin matrix elements for each meson will be summarized in the next subsection.

\subsubsection{Scalar mesons}

The calculation of the {\it on-shell\/} scalar exchange  is very straightforward, with
\bea
{\cal O}^{\rho_1\rho'_1,\rho_2\rho'_2,0,s}_{\lambda_1\lambda_2,\lambda'_1\lambda'_2,\pm}&=&-\Big[\bar N_{\rho_1}(p\lambda_1)\, {\bf 1} N_{\rho'_1}(p' \lambda'_1)\Big]
\nonumber\\
 &&\times\Big[\bar N_{\rho_2}(p\lambda_2)\,{\bf 1} N_{\rho'_2}(p' \lambda'_2)\Big] 
 \nonumber\\
 {\cal O}^{\rho_1\rho'_1,\rho_2\rho'_2,1,s}_{\lambda_1\lambda_2,\lambda'_1\lambda'_2,\pm}&=&0 \label{eq:scalar}
\eea
where ${\bf 1}$ is the unit matrix in 2$\times$2 space, and the additional superscript $s$ labels these ${\cal O}$'s as the contributions from scalar mesons.  It is convenient to calculate the general matrix element 
\bea
{\bf 1}_i\equiv \Big[\bar N_{\rho_i}(p\lambda_i)\, {\bf 1} N_{\rho'_i}(p' \lambda'_i)\Big]\label{eq:121}
\eea
and then construct all of the products (\ref{eq:scalar}), avoiding as much algebra as possible.  This also allows the off-shell matrix elements to be calculated without any extra work.  In what follows we will generalize the notation of (\ref{eq:121}):
\bea
{\cal O}_i\equiv  \Big[\bar N_{\rho_i}(p\lambda_i)\, {\bf {\cal O}} N_{\rho'_i}(p' \lambda'_i)\Big]
\eea
where ${\cal O}$ is any 2$\times$2 operator in the $\rho$-spin space.  Explicitly, the Dirac matrices are written as a direct product of two 2$\times$2 matrices (where the first matrix operates in the 2$\times$2 $\rho$-spin space and the second in the spin space):
\bea
{\bf 1}={\bf 1}\otimes{\bf 1}&\qquad&\gamma^0=\tau^3\otimes{\bf 1}
\nonumber\\
\gamma^5=\tau^1\otimes{\bf 1}& &\gamma^i=i\tau^2\otimes\sigma^i.
\eea
Hence, the $\rho$-spin matrices in the 2$\times$2 spin independent part of the Dirac space are the familiar Pauli matrices.

We frequently encounter off-shell couplings, which in every case give factors of
\bea
\left(m-\not\!p\right) u^\rho({\bf p},\lambda)&=&\left(m+{\bm \gamma}\cdot{\bf p}-\gamma^0p_0\right)u_i^\rho({\bf p},\lambda)
\nonumber\\
&=&(\rho E_p-p_0)\gamma^0 u^\rho({\bf p},\lambda)
\eea
where 
we have used the Dirac equation for the ({\it always\/}) on-shell spinor $u^\rho({\bf p},\lambda)$ and recalled that $u^-({\bf p},\lambda)=v(-{\bf p},\lambda)$ so that its Dirac equation is 
$$(m+\not\!p)v(-{\bf p},\lambda)=(m+\gamma^0E_p +\bm\gamma\cdot{\bf p})v(-{\bf p},\lambda)=0.$$ 
This use of the Dirac equations allows us to replace the angular dependent ${\bm \gamma}\cdot{\bf p}$ term by an angular independent  factor.  (Similar steps work off-shell couplings in the final state.)  Off-shell couplings therefore do not involve any new angular integrals, but they do require evaluation of a new 2$\times$2 Dirac matrix element.  Since the most general scalar Feynman operator is
\bea
\Lambda_s(p,p')=g_s\,{\bf 1} -\nu_s\left[\frac{m-\not\!p}{2m}+\frac{m-\not\!p'}{2m}\right]  \label{B28}
\eea
the {\it spin independent\/} part of the matrix element in the general scalar case is then
\begin{align}
&\Lambda_{s\,i}(p_i,p'_i)\equiv\bar u^{\rho_i}({\bf p},\lambda_i) \Lambda_s(p_i,p'_i)  u^{\rho'_i}({\bf p}',\lambda'_i) 
\nonumber\\
&\qquad=g_s\,{\bf 1}_i-\frac{\nu_s}{2m}\tau^3_i\left[\rho_iE_p-p_{i0}+\rho'_iE_{p'}-p'_{i0} \right]
\nonumber\\
&\qquad\equiv R^{\rho_i\rho'_i}_{\lambda_i\lambda'_i,s}(\pm p_{0}, \pm p'_0),\label{eq:126}
\end{align}
where, in the last line, we {\it define\/} the first of the one-particle $\rho$-spin matrix elements needed in this calculation.  The definition of $R_s$ restores explicit reference to the helicities $\lambda_i$ and $\lambda'_i$ implicitly contained in the matrix elements ${\bf 1}_i$ and $\tau^3_i$, and replaces $p_{i0}$ by $\pm p_0$, with $p_{10}=W/2+p_0$ and $p_{20}=W/2-p_0$ (so that sign of $p_0$ is positive for particle 1 and negative for particle 2). 
We emphasize that the spin dependent angular part is unchanged by the off-shell couplings.  

We encounter the off shell factors numerous times, so it is convenient to denote
\bea
\Delta E_i&\equiv& \rho_i E_p-p_{i0}\nonumber\\
\Delta E'_i&\equiv& \rho'_i E_{p'}-p'_{i0}. \label{eq:A24}
\eea
 However, note that 
\bea
\Delta E_1 \Delta E_2=\Delta E'_1 \Delta E'_2=0
\label{eq:A25}
\eea
because one of the two particles is always in shell.

\begin{table*}
\begin{minipage}{6in}
\caption{Matrix elements $\left<\rho,\lambda|{\cal O}_n|\rho',\lambda'\right>=\bar N_\rho(p\lambda) {\cal O}_n N_{\rho'}(p'\lambda')\equiv D^{\rho\rho'}_{\lambda\lambda',n}$ for various operators ${\cal O}_n$, where  $n=\{1,4\}$ with the correspondence $\tau^1\to 1$, $i\tau^2\to 2$, $\tau^3\to 3$, and  ${\bf 1}\to4$.  The two-component $\rho$-spin spinors $N_\rho$ were defined in Eq.~(\ref{eq:14}).  In this table $c=\cosh\sfrac12\zeta$, $c'=\cosh\sfrac12\zeta'$,  $s=2\lambda \sinh\sfrac12\zeta$, and $s'=2\lambda'\sinh\sfrac12\zeta'$, where $\zeta$ was defined below Eq.~(\ref{eq:14}).}
\label{tab:Dirac}
\begin{ruledtabular}
\begin{tabular}{ll} 
$\left<+,\lambda|{\bf 1}|+,\lambda'\right>=-\left<-,\lambda|{\bf 1}|-,\lambda'\right>=cc'-ss'$& 
$\left<+,\lambda|{\bf 1}|-,\lambda'\right>=\left<-,\lambda|{\bf 1}|+,\lambda'\right>=-(cs'+sc')$   \cr
$\left<+,\lambda|\tau^3|+,\lambda'\right>=\left<-,\lambda|\tau^3|-,\lambda'\right>=cc'+ss'$& 
$\left<+,\lambda|\tau^3|-,\lambda'\right>=-\left<-,\lambda|\tau^3|+,\lambda'\right>=-(cs'-sc')$    \cr
$\left<+,\lambda|\tau^1|+,\lambda'\right>=\left<-,\lambda|\tau^1,\lambda|-,\lambda'\right>=cs'-sc'$& 
$\left<+,\lambda|\tau^1|-,\lambda'\right>=-\left<-,\lambda|\tau^1|+,\lambda'\right>=cc'+ss'$     \cr
$\left<+,\lambda|i\tau^2|+,\lambda'\right>=-\left<-,\lambda|i\tau^2|-,\lambda'\right>=cs'+sc'$& 
$\left<+,\lambda|i\tau^2|-,\lambda'\right>=\left<-,\lambda|i\tau^2,\lambda|+,\lambda'\right>=cc'-ss'$       \cr
\end{tabular}
\end{ruledtabular}
\end{minipage}
\end{table*}


\subsubsection{Pseudoscalar mesons}

The pseudoscalar Feynman operator is
\bal
\Lambda_p&(p,p')=g_p\,\gamma^5 
\nonumber\\
&-g_p(1-\lambda_p)\left\{\left[\frac{m-\not\!p}{2m}\right]\gamma^5+\gamma^5\left[\frac{m-\not\!p'}{2m}\right]\right\} \, .
\end{align}
Using the same arguments, the pseudoscalar matrix elements become
\bal
\Lambda_{p\,i}(p_i,p'_i) 
=& g_p\bigg\{\tau^1_i -\frac{(1-\lambda_p)}{2m}\,i\tau^2_i \Big[\Delta E_i-\Delta E'_i\Big]
\bigg\} \nonumber\\
\equiv& R^{\rho_i\rho'_i}_{\lambda_i\lambda'_i, p}(\pm p_{0}, \pm p'_0) , \label{eq:128}
\end{align}
where we introduced the new matrix element $R_p$.
This requires evaluation of the matrix elements of $\tau^1$ and $i\tau^2$, but the spin dependent matrix elements are the same as in the scalar case.  For the evaluation of the matrix elements, see Table \ref{tab:Dirac}.

\subsubsection{Vector mesons} 

The vector matrix elements introduce new spin dependent factors, and the terms ${\cal O}^1$ linear in $z$.

The first step in the reduction of the vector-meson exchange terms is to reduce the Pauli interaction term using the generalized Gordon decomposition:
%
\bea
\frac{i\sigma^{\mu\nu}(p-p')_\nu}{2m}&=&\gamma^\mu-\frac{(p+p')^\mu}{2m}
\nonumber\\&-&
\left(\frac{m-\not\!p}{2m}\right)\gamma^\mu -\gamma^\mu\left(\frac{m-\not\!p'}{2m}\right).\qquad\quad
\eea
Combining this result with the general definition of the vector Feynman operator
\begin{widetext}
\bea
\Lambda_v^\mu(p,p') &=& g_v\Bigg\{\gamma^\mu +\nu_v\left(\left[\frac{m-\not\!p}{2m}\right]\gamma^\mu +\gamma^\mu\left[\frac{m-\not\!p'}{2m}\right]\right)
+\frac{\kappa_v}{2m} \,i\sigma^{\mu\nu}(p-p')_\nu\Bigg\}
\eea
gives
\bal
\Lambda_v^\mu(p,p') &= g_v\Bigg\{\gamma^\mu(1+\kappa_v) -\frac{\kappa_v}{2m}(p+p')^\mu
+(\nu_v-\kappa_v)\left(\left[\frac{m-\not\!p}{2m}\right]\gamma^\mu +\gamma^\mu\left[\frac{m-\not\!p'}{2m}\right]\right)\Bigg\}
\nonumber\\
& \to g_v\Bigg\{\gamma^\mu(1+\kappa_v) -\frac{\kappa_v}{2m}(p+p')^\mu
+\frac{(\nu_v-\kappa_v)}{2m}\Big[\Delta E\gamma^0\gamma^\mu +\gamma^\mu\gamma^0\Delta E'\Big]\Bigg\},
\end{align}
where the last line anticipates later use of the Dirac equation to reduce the operators $m-\not\!p$.

These two vertex operators are contracted with the spin-one meson propagator, which gives
\begin{align}
\Lambda_{v\,1}^\mu(p_1,p_1')\Lambda_{v\,2}^\nu(p_2,p_2')\Delta_{\mu\nu}
=&\Lambda_{v\,1}^0(p_1,p_1')\Lambda_{v\,2}^0(p_2,p_2')
-\Lambda_{v\,1}^j(p_1,p_1')\Lambda_{v\,2}^j(p_2,p_2')
\nonumber\\
&+\frac{\eta}{m_v^2}\Big\{\Lambda_{v\,1}^\mu(p_1,p_1')(p_1-p'_1)_\mu
\Lambda_{v\,2}^\nu(p_2,p_2')(p_2-p'_2)_\nu\Big\}
\end{align}
where the factor $\eta=1$ is included solely to keep track of the effect of the $q^\mu q^\nu/m_v^2$ in the meson propagator (note that the sign follows from $-q^\mu q^\nu=(p_1-p'_1)^\mu(p_2-p'_2)^\nu$).  To simplify, note that
\begin{align}
&B_i\equiv\Lambda_{v\,i}^\nu(p_i,p_i')(p_i-p'_i)_\nu
=g_v\Bigg\{{\bf 1}_i\,\nu_v\,\frac{(p'^2_i-p^2_i)}{2m}
-\tau_i^3(1+\nu_v)(\Delta E_i-\Delta E'_i)\Bigg\}
\equiv R^{\rho_i\rho'_i}_{\lambda_i\lambda'_i,v2}(\pm p_{0}, \pm p'_0)
\nonumber\\
&\Lambda_{v\,i}^0(p_i,p_i')=g_v\Bigg\{\tau_i^3(1+\kappa_v)
+{\bf 1}_i\bigg[\frac{\nu_v}{2m}(\Delta E_i+\Delta E'_i)
-\frac{\kappa_v}{2m}\,(\rho_i E_p+\rho'_i E_{p'})\bigg]
\Bigg\}
\equiv R^{\rho_i\rho'_i}_{\lambda_i\lambda'_i,v1}(\pm p_{0}, \pm p'_0),\label{eq:133}
\end{align}
\end{widetext}
where we introduced the new matrix elements $R_{v1}$ and $R_{v2}$.

The three-vector part introduces the spin-dependent operators $\sigma^i$ and angle-dependent terms from the factor $(p+p')^i$.  These are first separated into two terms,
\bea
\Lambda_{v\,i}^j(p_i,p_i')=A_i\times \sigma_i^j-{\bf 1}_i\;\frac{g_v\,\kappa_v}{2m}\,(p_i+p'_i)^j,\qquad\label{eq:134}
\eea
where  we remind the reader that the subscript $i$ labels the particle number (1 or 2), and $A_i$ is the spin-independent coefficient of the spin-dependent operator $\sigma^j$, and the spin-dependent part of the second term is the identity, suppressed in the expression.  The coefficient $A_i$ is
\bal
A_i&=g_v\Bigg\{i\tau^2_i (1+\kappa_v)
+\tau^1_i\frac{(\nu_v-\kappa_v)}{2m}(\Delta E_i-\Delta E'_i) \Bigg\}
\nonumber\\
&\equiv R^{\rho_i\rho'_i}_{\lambda_i\lambda'_i,v3}(\pm p_{0}, \pm p'_0), \label{eq:135}
\end{align}
which defines the matrix element $R_{v3}$.

In calculating the operator $\Lambda_{v\,1}^j(p_1,p_1')\Lambda_{v\,2 }^j(p_2,p_2')$ we encounter the cross terms
\bea
\bm\sigma_1\cdot ({\bf p}_2+{\bf p}'_2)&=&-\bm\sigma_1\cdot ({\bf p}_1+{\bf p}'_1)
\nonumber\\
&=&-2(\lambda_1p+\lambda'_{1}p'),\qquad
\eea
where the evaluation has been carried out in the center of mass system and use was made of the fact that the helicity eigenstates satisfy the eigenvalue conditions
\bea
\bm\sigma\cdot {\bf p}\left|\lambda\right> &=& 2\lambda\, p\left| \lambda\right>\, .
\eea
(For incoming particle 2, for example, $\left|\sfrac12\right>$ is a state with spin {\it down\/} in the $\hat z$ direction, and ${\bf p}'_2$ is in the $-\hat z$ direction, so that $\bm\sigma\cdot {\bf p}'_2\left|\frac12\right> =-p\,\sigma^3 \left|\frac12\right> = p\left|\sfrac12\right>$.)
Finally, note that 
\bea
({\bf p}_1+{\bf p}'_1)\cdot ({\bf p}_2+{\bf p}'_2)&=&-({\bf p}_1+{\bf p}'_1)^2
\nonumber\\
&=&-({\bf p}^2+{\bf p}'^2+2pp'z),\qquad \label{eq:138}
\eea
where we use ${\bf p}^2$, $p^2$, and $p$ to denote the square of the three-vector, the four-vector, and the magnitude of the three-vector, respectively.
Combining (\ref{eq:134}) - (\ref{eq:138}) gives the following:
\bal
&\Lambda_{v1}^j(p_1,p_1')\Lambda_{v2}^j(p_2,p_2')=A_1A_2\left<\lambda_1\lambda_2|\bm\sigma_1\cdot\bm\sigma_2|\lambda'_1\lambda'_2\right>
\nonumber\\
&+\Bigg\{\frac{g_v\,\kappa_v}{m}\Big[A_1{\bf 1}_2(\lambda_1p+\lambda'_1p')+A_2{\bf 1}_1(\lambda_2p+\lambda'_2p')\Big]
\nonumber\\
&-\left(\frac{g_v\,\kappa_v}{2m}\right)^2{\bf 1}_1{\bf 1}_2({\bf p}^2+{\bf p}'^2+2pp'z)\Bigg\}\left<\lambda_1\lambda_2|\lambda'_1\lambda'_2\right>, \label{eq:139}
\end{align}
where, for clarity,  we have included the term $\left<\lambda_1\lambda_2|\lambda'_1\lambda'_2\right>$ that multiplies the terms in the $\{\}$;  this 
term is not part of the $\rho$-spin matrix element and will be removed once the effect of the operator $\bm\sigma_1\cdot\bm\sigma_2$ has been expressed using the relation
\bea
\left<\lambda_1\lambda_2|\bm\sigma_1\cdot\bm\sigma_2|\lambda'_1\lambda'_2\right>\to{\cal T} \left<\lambda_1\lambda_2|\lambda'_1\lambda'_2\right>,
\label{eq:140}
\eea
where ${\cal T}$ will be determined shortly.   The extra $z=\cos\theta$ dependence in the last term of (\ref{eq:139}) is the origin of the special linear $z$ dependent terms described in Eq.~(\ref{eq:113}) above.  In our model, only vector-meson exchanges contribute such terms.

To determine ${\cal T}$ correctly note that the helicity states $\left|+-\right>$ or $\left|-+\right>$ correspond to the {\it spin\/} states $\left|\uparrow \uparrow\right>$ or $\left|\downarrow\downarrow\right>$, respectively, and hence are spin triplet states with ${\cal T}=1$.  The same argument works for both initial or final states.  This leaves only the combinations with equal helicities in {\it both\/} the initial and final states: $\lambda_1=\lambda_2=\pm$ {\it and\/} $\lambda'_1=\lambda'_2=\pm$.   Looking at Eq.~(\ref{eq:110}), the only potentials affected are $v_1, v_5, v_9,$ and $v_{13}$, and using the results
\bea
\bm\sigma_1\cdot\bm\sigma_2\left|++\right>&=&-\left|++\right>+2\left|--\right>
\nonumber\\
\bm\sigma_1\cdot\bm\sigma_2\left|--\right>&=&-\left|--\right>+2\left|++\right>
\eea
the ${\cal T}$ term in these kernels leads to the decomposition
\bal
 &v_{i}^{\rho_2\rho'_2}\sim\Big(
{\cal O}^{++,\rho_2\rho'_2,i}_{++,++,+}{\cal T} A^{\;i}_{++,++}\pm
{\cal O}^{++,\rho_2\rho'_2,i}_{++,--,+} {\cal T} A^{\;i}_{++,--}\Big)
\nonumber\\
&=\sfrac12\Big({\cal O}^{++,\rho_2\rho'_2,i}_{++,++,+}\pm {\cal O}^{++,\rho_2\rho'_2,i}_{++,--,+}\Big)\Big(A^{i}_{++,++}+A^{i}_{++,--}\Big)\nonumber\\
&+\sfrac12S\Big({\cal O}^{++,\rho_2\rho'_2,i}_{++,++,+}\mp {\cal O}^{++,\rho_2\rho'_2,i}_{++,--,+}\Big)\Big(A^{i}_{++,++}-A^{i}_{++,--}\Big)\, ,
\end{align}
where $S=-3$.   If the ${\cal T}$ operator is absent, the same expression holds, but with $S=1$, so this decomposition can be used for both cases. 
With this relation it is not necessary to explicitly calculate the matrix elements $\left<\lambda_1\lambda_2|\bm\sigma_1\cdot\bm\sigma_2|\lambda'_1\lambda'_2\right>$.   

\subsubsection{Axial vector mesons}

Only the simplest possible coupling was used for the axial vector mesons, namely
\bea
\Lambda_{a\,i}^\mu(p_i,p'_i)=g_a\gamma^5_i\gamma^\mu_i \, ,
\eea
and this was contracted with  the $g_{\mu\nu}$ tensor, giving
\bea
&&\Lambda_{a1}^\mu(p_1,p'_1)\Lambda_{a2}^\nu(p_2,p'_2)g_{\mu\nu}=\Lambda^0_{a1}\Lambda^0_{a2}\left<\lambda_1\lambda_2|\lambda'_1\lambda'_2\right>
\nonumber\\
&&\qquad\qquad-C_1C_2\left<\lambda_1\lambda_2|\bm\sigma_1\cdot\bm\sigma_2|\lambda'_1\lambda'_2\right>,
\label{eq:146a}
\eea
with 
\bea
&&\Lambda^0_{ai}=-ig_a\tau^2_i \equiv R^{\rho_i\rho'_i}_{\lambda_i\lambda'_i,a1}(\pm p_{0}, \pm p'_0)
\nonumber\\
&&C_i=-g_a\tau_i^3 \equiv R^{\rho_i\rho'_i}_{\lambda_i\lambda'_i,a2}(\pm p_{0}, \pm p'_0),\label{eq:146}
\eea
which defines the last two $\rho$-spin matrix elements, $R_{a1}$ and $R_{a2}$.

\subsection{Two-particle $\rho$-spin matrix elements} \label{App:rho}

The two-particle $\rho$-spin matrix elements 
are the ones defined in Eq.~(\ref{eq:113}).  They are constructed from the helicity matrix elements $D_{\lambda\lambda',\,n}^{\rho\rho'}$  given in Table \ref{tab:Dirac}, and the one-nucleon matrix elements $R_b$ defined in Eqs.~(\ref{eq:126}), (\ref{eq:128}), (\ref{eq:133}), (\ref{eq:135}), and (\ref{eq:146}) above.
In constructing the matrix elements, note that $p_0\to -p_0$ corresponds to the interchange of $p_{10} \leftrightarrow p_{20}$, and similarily for $p'_0$.  The results, with all indices included, are:

\vskip0.2in

\noindent {\bf Scalar}, from Eq.~(\ref{eq:126}):
\bea
{\cal O}^{\rho_1\rho'_1,\rho_2\rho'_2,0,s}_{\lambda_1\lambda_2,\lambda'_1\lambda'_2,\pm}&=&-R_{\lambda_1\lambda'_1,s}^{\rho_1\rho'_1}(\pm p_0,p'_0)R_{\lambda_2\lambda'_2,s}^{\rho_2\rho'_2} (\mp p_0,-p'_0)\nonumber\\
 {\cal O}^{\rho_1\rho'_1,\rho_2\rho'_2,1,s}_{\lambda_1\lambda_2,\lambda'_1\lambda'_2,\pm}&=&0 \label{eq:A44}
\eea

\vskip0.2in

\noindent {\bf Pseudocalar}, from Eq.~(\ref{eq:128}):
\bea
{\cal O}^{\rho_1\rho'_1,\rho_2\rho'_2,0,p}_{\lambda_1\lambda_2,\lambda'_1\lambda'_2,\pm}
&=&R_{\lambda_1\lambda'_1,p}^{\rho_1\rho'_1}(\pm p_0,p'_0)
R_{\lambda_2\lambda'_2,p}^{\rho_2\rho'_2}(\mp p_0,-p'_0)
\nonumber\\
 {\cal O}^{\rho_1\rho'_1,\rho_2\rho'_2,1,p}_{\lambda_1\lambda_2,\lambda'_1\lambda'_2,\pm}&=&0 \label{eq:A45}
\eea

\vskip0.2in
\noindent {\bf Vector}, from Eqs.~(\ref{eq:133}), (\ref{eq:135}), and (\ref{eq:139}):
\begin{align}
{\cal O}^{\rho_1\rho'_1,\rho_2\rho'_2,0,v}_{\lambda_1\lambda_2,\lambda'_1\lambda'_2,\pm}&=
R_{\lambda_1\lambda'_1,v1}^{\rho_1\rho'_1}(\pm p_0, p'_0)
R_{\lambda_2\lambda'_2,v1}^{\rho_2\rho'_2}(\mp p_0,-p'_0)
\nonumber\\ 
+&\frac{\eta}{m_v^2}
R_{\lambda_1\lambda'_1,v2}^{\rho_1\rho'_1}(\pm p_0,p'_0)
R_{\lambda_2\lambda'_2,v2}^{\rho_2\rho'_2}(\mp p_0,-p'_0) \qquad
\nonumber\\
-&{\cal T} \,
R_{\lambda_1\lambda'_1,v3}^{\rho_1\rho'_1}(\pm p_0,p'_0)
R_{\lambda_2\lambda'_2,v3}^{\rho_2\rho'_2}(\mp p_0,-p'_0)
\nonumber\\
+&\left(\frac{g_v\,\kappa_v}{2m}\right)^2\left({\bf p}^2+{\bf p}'^2\right)D^{\rho_1\rho'_1}_{\lambda_1\lambda'_1,4} D^{\rho_2\rho'_2}_{\lambda_2\lambda'_2,4}
\nonumber\\
-&\frac{g_v\kappa_v}{m}\Big[(\lambda_1p+\lambda'_1p')R_{\lambda_1\lambda'_1,v3}^{\rho_1\rho'_1}(\pm p_0, p'_0)
D^{\rho_2\rho'_2}_{\lambda_2\lambda'_2,4}
\nonumber\\
&
+(\lambda_2p+\lambda'_2p')R_{\lambda_2\lambda'_2,v3}^{\rho_2\rho'_2}(\mp p_0, -p'_0)
D^{\rho_1\rho'_1}_{\lambda_1\lambda'_1,4}\Big]
\nonumber\\
{\cal O}^{\rho_1\rho'_1,\rho_2\rho'_2,1,v}_{\lambda_1\lambda_2,\lambda'_1\lambda'_2,\pm}&=2pp'\left(\frac{g_v\,\kappa_v}{2m}\right)^2D^{\rho_1\rho'_1}_{\lambda_1\lambda'_1,4}D^{\rho_2\rho'_2}_{\lambda_2\lambda'_2,4}.  \label{eq:A46}
\end{align}
In these expressions we used the fact that $D_4$ does not depend on $p_0$.  

\vskip0.2in
\noindent {\bf Axial Vector}, from Eqs.~(\ref{eq:146a}) and (\ref{eq:146}):
\begin{align}
{\cal O}^{\rho_1\rho'_1,\rho_2\rho'_2,0,a}_{\lambda_1\lambda_2,\lambda'_1\lambda'_2,\pm}&=g_b^2\Big\{ 
D^{\rho_1\rho'_1}_{\lambda_1\lambda'_1,2}D^{\rho_2\rho'_2}_{\lambda_2\lambda'_2,2}-{\cal T}
D^{\rho_1\rho'_1}_{\lambda_1\lambda'_1,3}D^{\rho_2\rho'_2}_{\lambda_2\lambda'_2,3}\Big\}
\nonumber\\
{\cal O}^{\rho_1\rho'_1,\rho_2\rho'_2,1,a}_{\lambda_1\lambda_2,\lambda'_1\lambda'_2,\pm}&=0\, . \label{eq:A47}
\end{align}
Once again we note that the $D$'s do not depend on $p_0$.

\vskip0.2in
\noindent{\bf Treatment of the photon}

For the photon we assume that $F_{1n}=0$ and that all other form factors are equal: $F_{2n}=F_{1p}=F_{2p}=F_D$.  The photon is an isovector-vector exchange, obtained using the following couplings (with $\alpha\simeq0.007297)$:
\bea
g^2&\to& 0\qquad {\rm no} \; e^2\;{\rm term}
\nonumber\\
g^2\kappa&\to& e^2 \kappa_n \to -\alpha(1.913)=-0.01396\equiv G_1
\nonumber\\
g^2\kappa^2&\to& e^2\kappa_p\kappa_n\to -\alpha(1.913) (1.793) 
\nonumber\\
&=& -0.02503\equiv G_2\, .
\eea
These couplings require a special construction and cannot easily be incorporated into the general vector formulae above.

As a first step, define artificial couplings designed to reproduce the ratio $G_2/G_1=\kappa_{\rm eff}=\kappa_p=1.7930$ and $g^2_{\rm eff}=G_1/\kappa_p=G_1^2/G_2=-0.00779$.  The upshot of this is to use the proton anomalous moment for $\kappa$ and a fictitious charge of $g^2_{\rm eff}=-0.00779$ so that $g^2_{\rm eff}\kappa_p=G_1$ and $g^2_{\rm eff}\kappa^2_p=G_2$ are reproduced.  Then, we will factor out the coupling $g_{\rm eff}$ from the $R$'s so that it occurs as an overall multiplicative factor with the correct (negative) sign.

Next, note that all terms in the photon exchange must involve at least one factor of $\kappa$, so that $B\sim R_{v2}$ does not contribute.  The $R_{v1}$ and $R_{v3}$ squared terms must have the $g^2_{\rm eff}$ terms subtracted.  Noting that $\nu=0$ for the photon, and {\it redefining the photon $R$'s so that $g_v=1$ and $\kappa_v=\kappa_p$\/}, this gives

\vskip0.2in
\noindent {\bf Photon}:
%
\begin{align}
&{\cal O}^{\rho_1\rho'_1,\rho_2\rho'_2,0,\gamma}_{\lambda_1\lambda_2,\lambda'_1\lambda'_2,\pm}=
g^2_{\rm eff}\Big(R_{\lambda_1\lambda'_1,\gamma1}^{\rho_1\rho'_1}(\pm p_0, p'_0)
R_{\lambda_2\lambda'_2,\gamma1}^{\rho_2\rho'_2}(\mp p_0,-p'_0)
\nonumber\\ &\qquad\qquad
-D^{\rho_1\rho'_1}_{\lambda_1\lambda'_1,3} D^{\rho_2\rho'_2}_{\lambda_2\lambda'_2,3}  \Big)
\nonumber\\
&\qquad-{\cal T}g^2_{\rm eff} \,\Big(
R_{\lambda_1\lambda'_1,\gamma3}^{\rho_1\rho'_1}(\pm p_0,p'_0)
R_{\lambda_2\lambda'_2,\gamma3}^{\rho_2\rho'_2}(\mp p_0,-p'_0)
\nonumber\\
&\qquad\qquad -D^{\rho_1\rho'_1}_{\lambda_1\lambda'_1,2} D^{\rho_2\rho'_2}_{\lambda_2\lambda'_2,2} \Big) 
\nonumber\\
&\qquad+G_2\frac{\left({\bf p}^2+{\bf p}'^2\right)}{4m^2}D^{\rho_1\rho'_1}_{\lambda_1\lambda'_1,4} D^{\rho_2\rho'_2}_{\lambda_2\lambda'_2,4}
\nonumber\\
&\qquad-\frac{G_1}{m}\Big[(\lambda_1p+\lambda'_1p')R_{\lambda_1\lambda'_1,\gamma3}^{\rho_1\rho'_1}(\pm p_0, p'_0)
D^{\rho_2\rho'_2}_{\lambda_2\lambda'_2,4}
\nonumber\\
&\qquad\qquad+(\lambda_2p+\lambda'_2p')R_{\lambda_2\lambda'_2,\gamma3}^{\rho_2\rho'_2}(\mp p_0, -p'_0)
D^{\rho_1\rho'_1}_{\lambda_1\lambda'_1,4}\Big]
\nonumber\\
&{\cal O}^{\rho_1\rho'_1,\rho_2\rho'_2,1,\gamma}_{\lambda_1\lambda_2,\lambda'_1\lambda'_2,\pm}=G_2 \frac{pp'}{2m^2}D^{\rho_1\rho'_1}_{\lambda_1\lambda'_1,4}D^{\rho_2\rho'_2}_{\lambda_2\lambda'_2,4}, \label{eq:162}
\end{align}
where the redefined $R$'s are [recalling Eqs.~(\ref{eq:133}) for $R_{v1}$ and (\ref{eq:135}) for $R_{v3}$ with $\nu=0$ and $g_v=1$]
\begin{align}
&R^{\rho\rho'}_{\lambda\lambda',\gamma1}(x_0,y_0)=(1+\kappa_p)D^{\rho_i\rho'_i}_{\lambda_i\lambda'_i,3}
\nonumber\\
&\qquad\qquad -
\frac{\kappa_p}{2m}D^{\rho\rho'}_{\lambda\lambda',4} \,(\rho E_p+\rho' E_{p'})
\nonumber\\
&R^{\rho\rho'}_{\lambda\lambda',\gamma3}(x_0,y_0)
=(1+\kappa_p)D^{\rho\rho'}_{\lambda\lambda',2}
\nonumber\\
&\qquad\qquad-\frac{\kappa_p}{2m}D^{\rho\rho'}_{\lambda\lambda',1}\Big[\Delta E_\rho(x_0)-\Delta E'_{\rho'}(y_0)\Big] \, .
\end{align}


\subsection{Isospin factor}

For the exchange of isovector mesons, one must evaluate the factor $\bm\tau_1\cdot\bm\tau_2$. This operator preserves the isospin symmetry of the state; therefore its value is the same for each of the 16 kernels (or 4 when $J=0$) that make up the coupled array describing the singlet (S), triplet (T), or coupled  (C) channels: namely it must equal unity on isovector $np$ states and $-3$ on isoscalar $np$ states.  To determine the correct value look at the leading amplitudes (those which are symmetric under the change in sign of $p_0$, and hence survive when both particles are on shell), and note that, for these amplitudes, the Pauli principle requires that $L+S+I$ be odd.  For singlet and triplet states, $J=L$ while for coupled states $J=L\pm1$.  Hence we have the identification:
\bea
{\rm singlet\,\&\,coupled}: &&\begin{cases}\;J\,{\rm even}&\to\;  I=1 \cr
\;J\,{\rm odd}&\to\: I=0 \end{cases} \nonumber\\
{\rm triplet}: &&\begin{cases}\;J\,{\rm even}&\to\;  I=0 \cr
\;J\,{\rm odd}&\to\: I=1\,. \end{cases}
\eea
These are summarized by: 
\bea
{\rm singlet\,\&\,coupled}: &&\;\bm\tau_1\cdot\bm\tau_2= -1+2(-1)^J\nonumber\\
{\rm triplet}: &&\;\bm\tau_1\cdot\bm\tau_2= -1-2(-1)^J.\qquad \label{eq:150}
\eea

To determine the correct isospin factor for each $v_i^{\rho_2\rho'_2}$ the channel to which it contributes must first be determined by examining the matrices Eqs.~(E43)-(E48) of Ref.~I.  
Inspection shows that the 
kernels $v_1, v_3, v_9,$ and $v_{11}$ do not contribute to the triplet channels, while kernels $v_5, v_7, v_{13},$ and $v_{15}$ contribute {\it only\/} to triplet channels.  The rest contribute to either, depending on their $\rho$-spin.  In particular, contributions to the triplet channels come from
\begin{itemize}
\item the $++$ sector of $v_6$ and $v_8$,
\item the $+-$ sector of $v_{10}$ and $v_{12}$,
\item the $-+$ sector of $v_2$ and $v_4$, and 
\item the $--$ sector of $v_{14}$ and $v_{16}$,
\end{itemize}
with the other sectors of these kernels contributing to either singlet or coupled.


\subsection{Charge symmetry breaking induced by pion exchange}

In order to include important violations of charge symmetry and to better describe the important pion exchange, both models treat the charged $\pi^\pm$ and the neutral $\pi^0$ as independent mesons, with their masses fixed to their experimental values.  Model WJC-1 allows the couplings of the pions to be varied independently while model WJC-2 assumes that the couplings (written in terms of $g$ and not $f$) of both are the same.

In addition to the couplings and masses, the correct isospin factor for the exchange of each pion must be worked out.  There are two ways to do this.  The operator $\bm\tau_1\cdot\bm\tau_2$ can be decomposed into
\bea
\bm\tau_1\cdot\bm\tau_2=\tau^3_1\tau^3_2 + 2\Big(\tau^+_1\tau^-_2 + \tau^-_1\tau^+_2\Big)
\eea
where $\tau^\pm=(\tau^1\pm\tau^2)/2$ are the raising and lowering operators (normalized to 1 and not $\sqrt{2}$ as is more commonly done) and the first term arrises from $\pi^0$ exchange and the second from $\pi^\pm$ exchange.  Noting that
\bea
\tau_1^3\tau_2^3\left|pn\right>&=&-\left|pn\right>
\nonumber\\
\tau_1^3\tau_2^3\left|np\right>&=&-\left|np\right>
\nonumber\\
2(\tau_1^+\tau_2^-+\tau_1^+\tau_2^-)\left|pn\right>&=&2\left|np\right>
\nonumber\\
2(\tau_1^+\tau_2^-+\tau_1^+\tau_2^-)\left|np\right>&=&2\left|pn\right>
\eea
we see that 
\bea
\bm\tau_1\cdot\bm\tau_2|_{\pi^0}&\to& -1 
\nonumber\\
\bm\tau_1\cdot\bm\tau_2|_{\pi^\pm}&\to&\begin{cases} \phantom{-}2 & I=1\;{\rm states} \cr -2 & I=0  \;{\rm states}\, . \label{eq:153}
\end{cases}
\eea
These factors can also be obtained by using the fact that isospin invariance gives couplings for $\{p\to p\pi^0, n\to n\pi^0, p\to n\pi^+, n\to p\pi^-\}$ in the ratios $\{1,-1,\sqrt{2},\sqrt{2}\}$.  

The implication of (\ref{eq:153}) for $\pi^\pm$ couplings is to replace the isospin factors from (\ref{eq:150}) by
\bea
\bm\tau_1\cdot\bm\tau_2|_{\pi^\pm}\to\begin{cases} \phantom{-}2(-1)^J & {\rm S\;and\;C\; states} \cr -2(-1)^J & {\rm T\; states}\end{cases}
\eea


\section{EVALUATION OF ANGULAR INTEGRALS} \label{app:B}

\subsection{Angular integrals}

The angular integrals needed for this problem are defined to be 
\bal
{\cal A}^{i,a,\pm}_{\lambda_1\lambda_2,\lambda'_1\lambda'_2}=\int_{-1}^1 dz &\;A^i_{\lambda_1\lambda_2,\lambda'_1\lambda'_2}
{\cal D}_b\Big(\theta, x_0(E_p-\sfrac12 W)\Big)
\end{align}
where ${\cal D}_b$ is the dressed propagator (including form factor) for meson $b$, defined in Eq.~(\ref{eq:A8}), and the four-momentum transfer for the direct $(+)$ and alternating $(-)$ terms is
\bea
q^2(x_0)&=&\Big[x_0(E_p-\sfrac12W)-y_0(E_{p'}-\sfrac12W)\Big]^2
\nonumber\\
&&-p^2-p'^2+2pp'z,
\eea
where in this paper the initial state always has particle 1 on-shell, so that $y_0=1$.
These integrals are assembled from the matrix elements of Table \ref{tab:5} and the functions of Table \ref{tab:VII}, leading to the identities of Table \ref{tab:3}.  All of the integrals ${\cal A}$ are recalculated for {\it each\/} $J, p, x_0,p',$ and $y_0$, so these arguments are not included in the list of subscripts denoting the dependences of ${\cal A}$.

\begin{figure}
\includegraphics[width=3in]{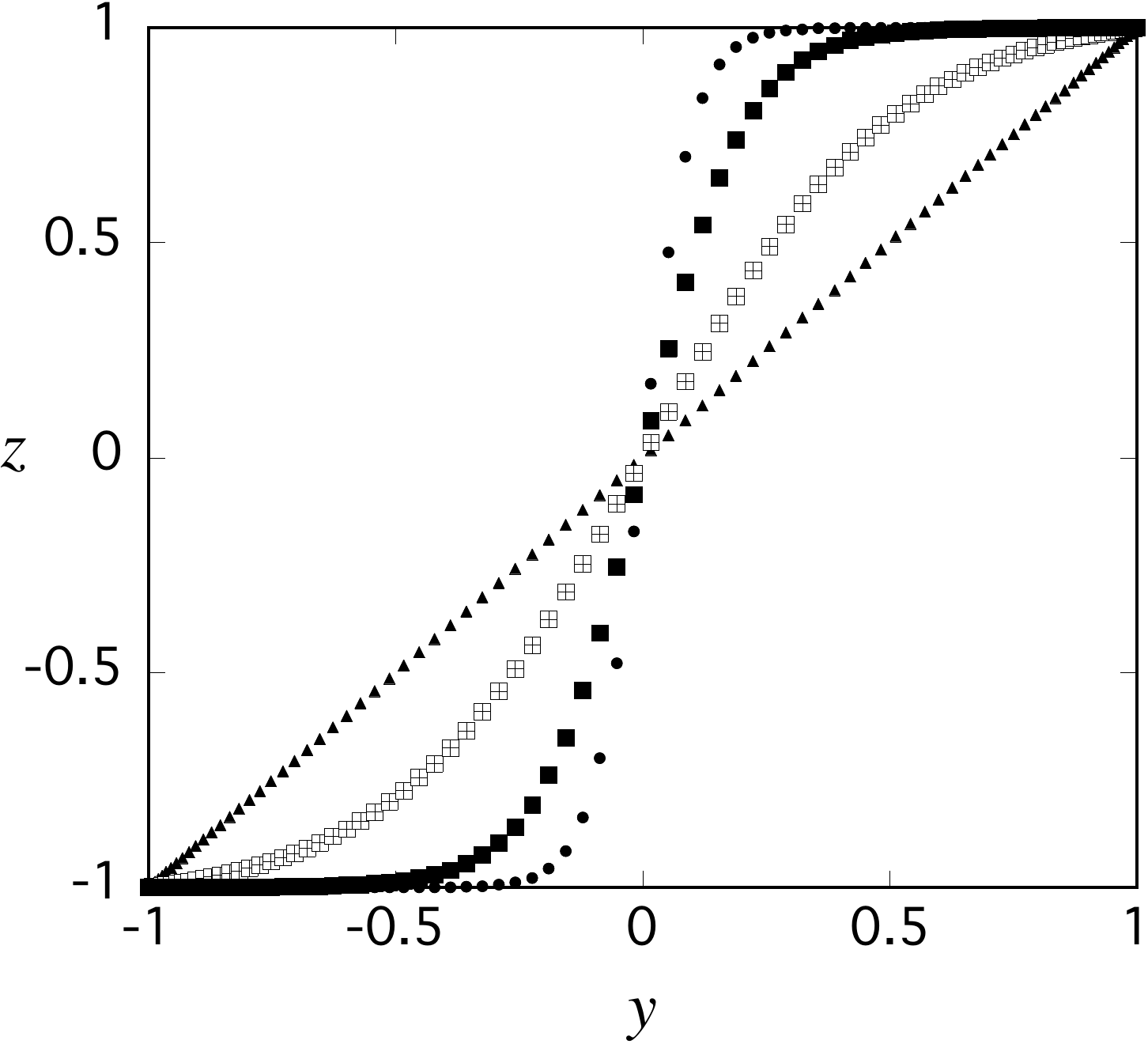}
\caption{\footnotesize\baselineskip=10pt Plot of the mapping function $z(y)$ defined in Eq.\ (\ref{eq1}).  The four curves, each with 90 Gauss points, correspond to $b_0=10$ (small solid triangles), $b_0=1.5$ (open boxes), $b_0=1.2$ (solid boxes), and $b_0=1.1$ (small solid circles). }
\label{fig1}
\end{figure} 

\begin{figure*}
\includegraphics[width=5.5in]{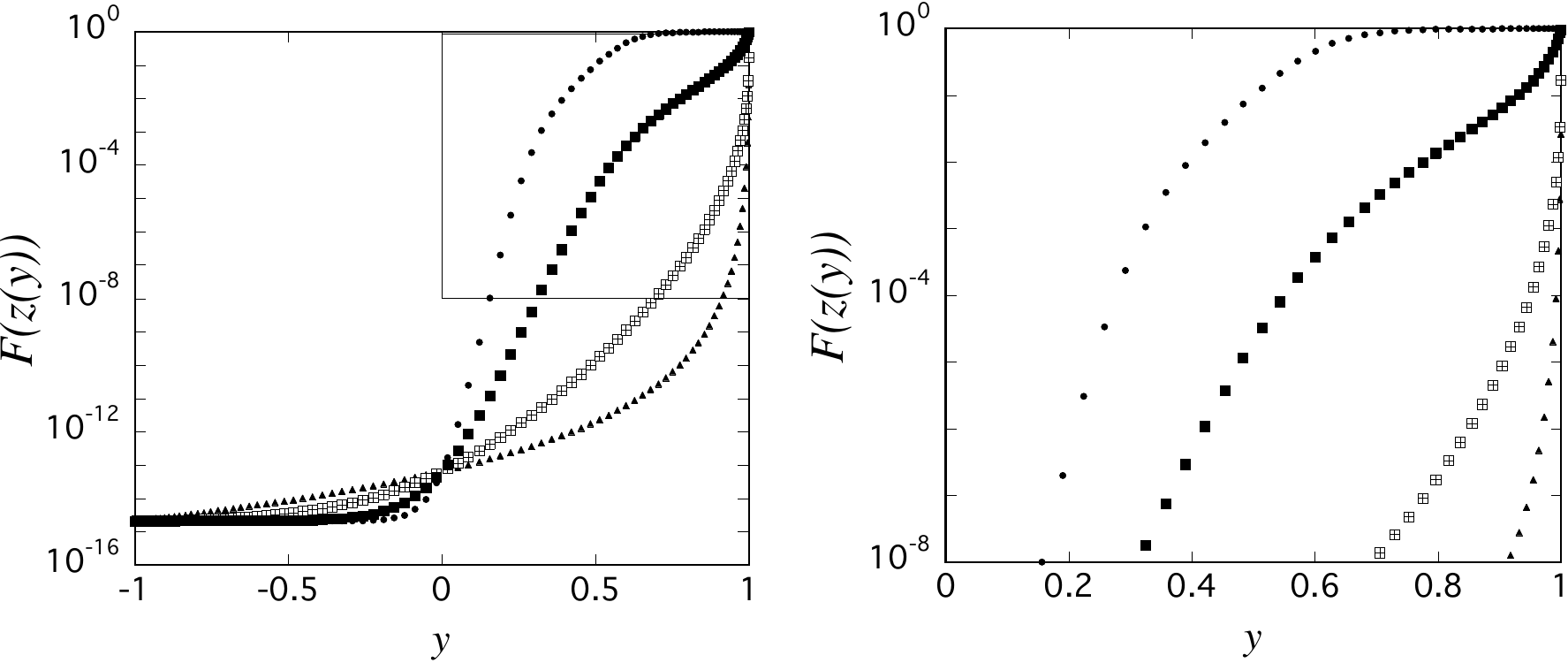}
\caption{\footnotesize\baselineskip=10pt Plot of the meson propagator $F$ (with $m_b=0.14$, $\Lambda_b=4$, $z_0=1$, and $A=1800$) as a function of the mapped variable $y$ defined in Eq.\ (\ref{eq1}).  The four curves, each with 90 Gauss points, correspond to $b_0=10$ (small solid triangles), $b_0=1.5$ (open boxes), $b_0=1.2$ (solid boxes), and $b_0=1.1$ (small solid circles).  The right-hand panel covers the region shown in the upper right corner of the left-hand panel.}
\label{fig2}
\end{figure*} 

\subsection{Mapping for the direct terms}

\begin{table*}
\squeezetable
\caption{\footnotesize\baselineskip=10pt  Convergence of the diagonal element of the {\it direct\/} part of the kernel $v_1^{++}$ at the point $p=p'=p_{\rm max}$.  Columns labeled $y$ use no mapping; those labeled $z(y)$ use the mapping function (\ref{eq1}).  Here the nucleon form factor has been set to 1.  With the correct form factor, all of these matrix elements are reduced by a factor of 7$\times 10^{-15}$.   (All numbers in this table are multiplied by 10$^{-3}$)} 
 \begin{tabular}{ccccccccccccccccccccc}
  &  \multicolumn{2}{c}{ $J=0$} & $\quad$ &  \multicolumn{2}{c}{ $J=1$}  &$\;$ &  \multicolumn{2}{c}{ $J=2$}  &$\;$ &  \multicolumn{2}{c}{ $J=3$} & $\;$ &  \multicolumn{2}{c}{ $J=4$} & $\;$ &  \multicolumn{2}{c}{ $J=5$} & $\;$ &  \multicolumn{2}{c}{ $J=6$}  \\
\hline
$n$ & $y$ & $z(y)$ & & $y$ & $z(y)$ & & $y$ & $z(y)$ & & $y$ & $z(y)$ & & $y$ & $z(y)$ & & $y$ & $z(y)$ & & $y$ & $z(y)$ \\
20 & 0.14321 & -3.3764   & & -0.38372 & -24.288   & & 0.14001 & -3.3825  & &-0.36991 & -24.233  & & 0.13277 & -3.3967  & & -0.34585& -24.134  & & 0.12190 & -3.4184   \\
 30 & 0.62029 & -3.3765    & & -2.0149 & -24.288    & & 0.61385 & -3.3827 & & -1.9816 & -24.232  & & 0.59907 & -3.3968  & & -1.9225 & -24.133  & & 0.57642 &  -3.4186 \\
40 &  1.4365 & -3.3765   & & -5.2403 & -24.288    & & 1.4277 & -3.3827  & & -5.1898 & -24.232   & & 1.4074 & -3.3968   & & -5.0999 & -24.133   & & 1.3762 & -3.4186  \\
60  & 2.5656   & -3.3765   & & -12.496  & -24.288   & & 2.5569  & -3.3827 & & -12.435 & -24.232  & & 2.5368 & -3.3968  & & -12.326 & -24.133  & & 2.5059 & -3.4186 \\
 90  & 1.4055  & -3.3765   & &  -18.312 & -24.288   & & 1.3992  & -3.3827 & & -18.254 & -24.232 & & 1.3847 & -3.3968  & & -18.152 & -24.133  & & 1.3624 & -3.4186 \\
\hline
 \end{tabular} 
\label{tab1aa}
\end{table*} 


\begin{table*}
\squeezetable
\caption{\footnotesize\baselineskip=10pt  Convergence of the diagonal element of the {\it exchange\/} part of the kernel $v_1^{++}$ at the point $p=p'=p_{\rm max}$.   Columns labeled $y$ use no mapping; those labeled $z(y)$ use the mapping function (\ref{eq6}).  Here the nucleon form factor has been set to 1.  With the correct form factor, all of these matrix elements are reduced by a factor of 7$\times 10^{-15}$. } 
 \begin{tabular}{ccccccccccccccccccccc}
  &  \multicolumn{2}{c}{ $J=0$} & $\quad$ &  \multicolumn{2}{c}{ $J=1$}  &$\;$ &  \multicolumn{2}{c}{ $J=2$}  &$\;$ &  \multicolumn{2}{c}{ $J=3$} & $\;$ &  \multicolumn{2}{c}{ $J=4$} & $\;$ &  \multicolumn{2}{c}{ $J=5$} & $\;$ &  \multicolumn{2}{c}{ $J=6$}  \\
\hline
$n$ & $y$ & $z(y)$ & & $y$ & $z(y)$ & & $y$ & $z(y)$ & & $y$ & $z(y)$ & & $y$ & $z(y)$ & & $y$ & $z(y)$ & & $y$ & $z(y)$ \\
20 &  -25.184 & -74.786   & & -5.1601 & -14.525   & & -16.551 & -49.079  & &-2.1999 & -6.2065  & & -2.1638 & -6.2802  & & 0.98501& 2.7617  & & 8.5080 & 25.368   \\
 30 & -37.148 & -81.841    & & -6.0514 & -11.903    & & -24.393 & -53.713 & & -2.5703 & -5.0887  & & -3.1487 & -6.8793  & & 1.1690 & 2.2584  & & 12.580 &  27.757 \\
40 &  -43.422 & -80.907  & & -8.1636 & -13.401   & & -28.504 & -53.097 & & -3.4750 & -5.7208  & & -3.6625 & -6.7959  & & 1.5681 & 2.5541  & & 14.716 & 27.443  \\
60  &  -54.117 & -80.899   & & -9.9713 & -13.594   & & -35.514 & -53.091 & & -4.2542 & -5.8025  & & -4.5432 & -6.7939  & & 1.9029 & 2.5918  & & 18.356 & 27.441 \\
 90  & -67.458  & -80.910   & &  -9.1111 & -13.607   & & -44.267  & -53.098 & & -3.8893 & -5.8080  & & -5.6594 & -6.7948  & & 1.7357 & 2.5944  & & 22.885 & 27.445 \\
\hline
 \end{tabular} 
\label{tab2aa}
\end{table*} 

\begin{figure}
\includegraphics[width=2.8in]{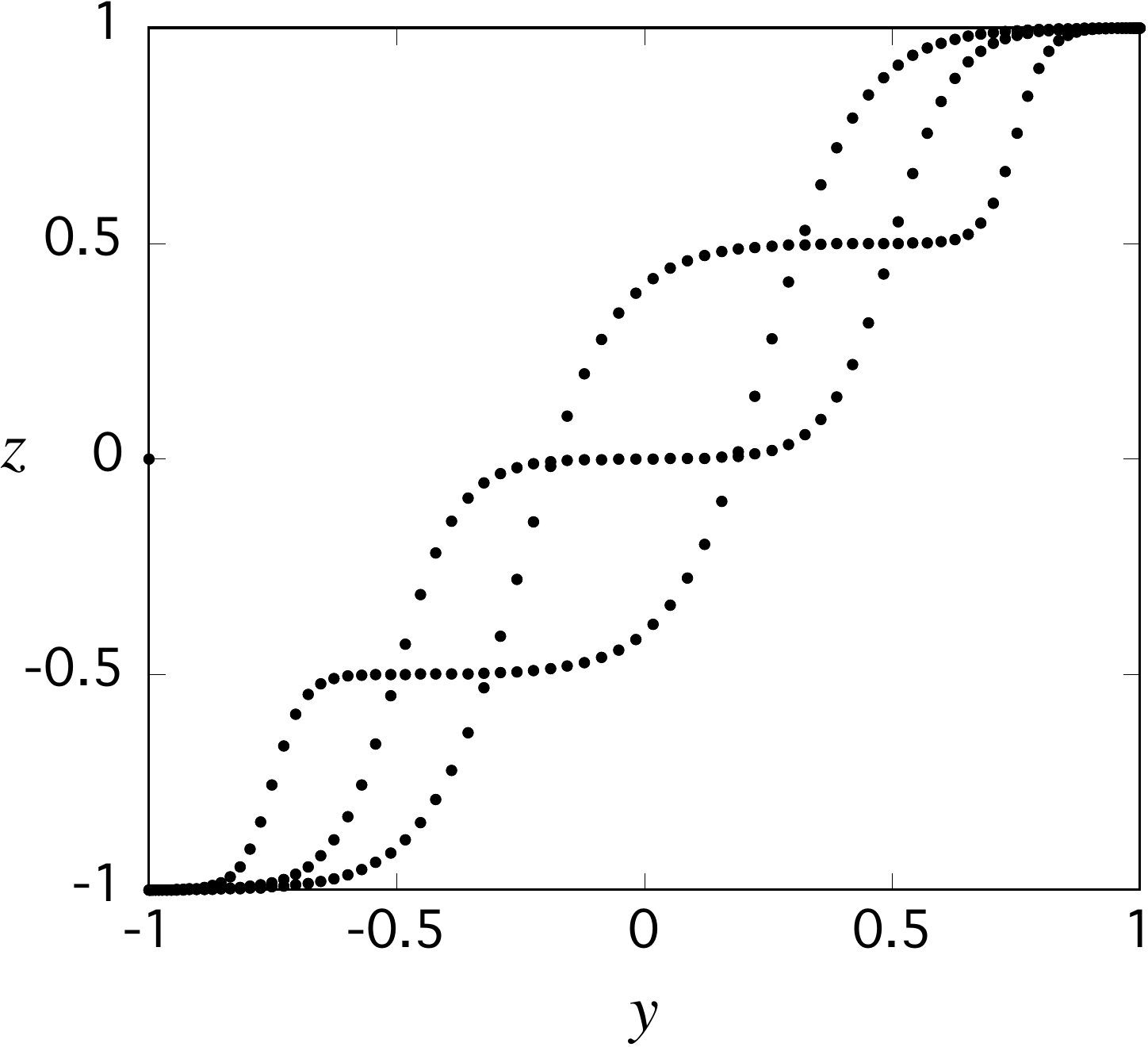}
\caption{\footnotesize\baselineskip=10pt Plot of the mapping function $z(y)$ defined in Eq.\ (\ref{eq6}), with $A=4$.  The three curves, each with 90 Gauss points, correspond to $z_0=-0.5$, $z_0=0$, and $z_0=0.5$.  Each curve flattens out in the region of $z=y=z_0$.}
\label{fig3}
\end{figure} 

\begin{figure}
\includegraphics[width=3.in]{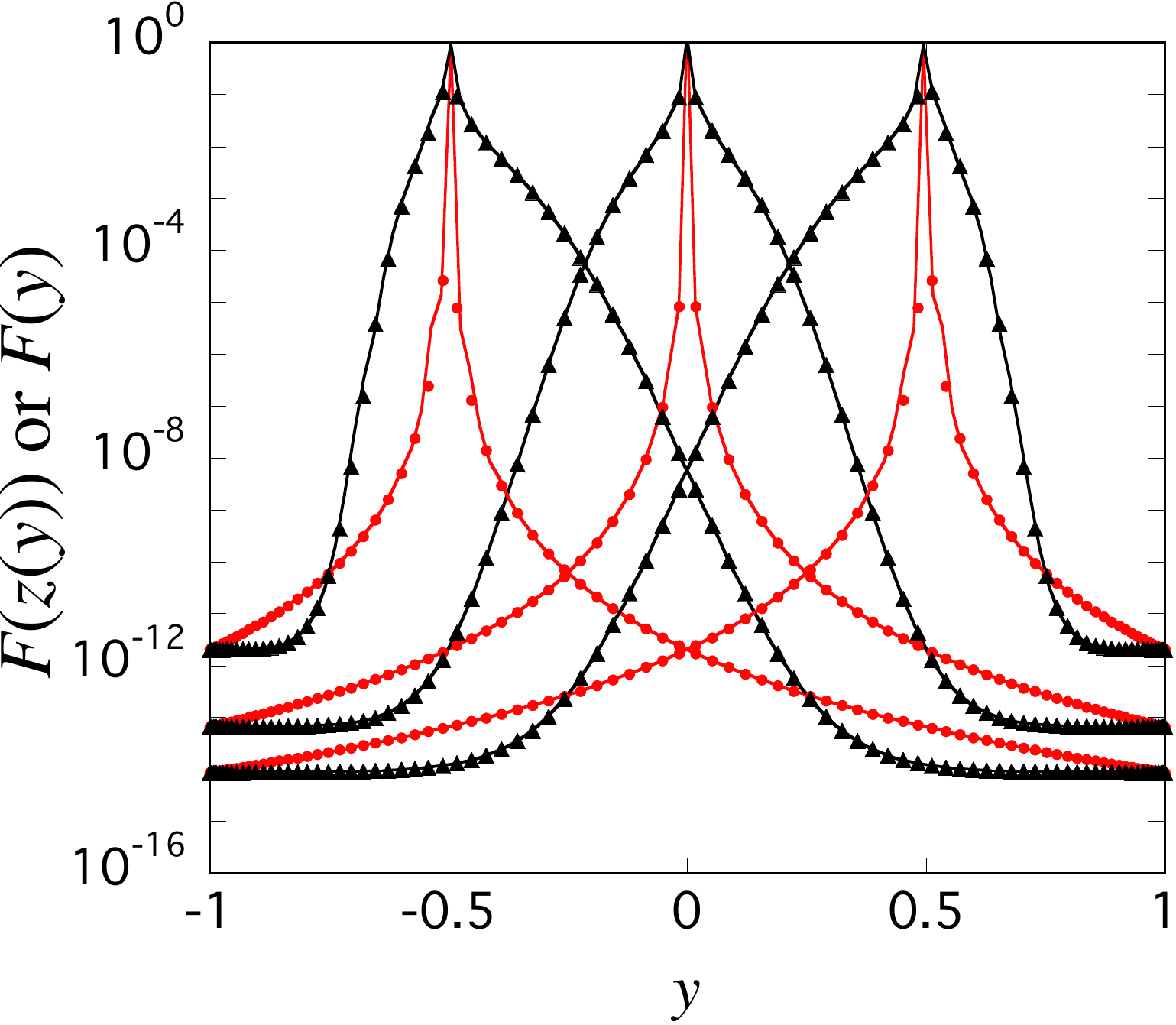}
\caption{\footnotesize\baselineskip=10pt (Color online) Plot of the meson propagator $F$ (with $m_b=0.14$, $\Lambda_b=4$, and $A=1800$) as a function of both the unmapped variable $y$ and the mapped variable $z(y)$, defined in Eq.\ (\ref{eq6}), for {\it each\/} of the choices  $z_0=-0.5$, $z_0=0$, and $z_0=0.5$.  The 90 Gauss points distributed between $-1<y<1$ are shown on each curve.  All the curves peak at $y=z_0$, with the  unmapped functions $F(y)$ having narrower peaks than the mapped function $F(z(y))$. }
\label{fig4}
\end{figure} 

For the direct term the meson propagator is  strongly peaked in the forward direction ($z=1$).   For the recent $NN$ fits, the meson propagators have the form 
\bea
{\cal D}_b(z)\equiv F(z)=\frac{m_b^2}{m_b^2+A|z_0-z|}\left[\frac{\Lambda_b^2}{\Lambda_b^2+A|z_0-z|}\right]^4\qquad \label{eq1a}
\eea
where $m_b$ and $\Lambda_b$ are the meson mass and form factor mass of meson $b$, respectively, and
\bea
A&=&2pp'
\nonumber\\
z_0&=&\frac{p^2+p'^2-(E_p-E_{p'})^2}{2pp'}\equiv 1+\xi \, ,
\eea
with $\xi=0$ if the incoming and outgoing momenta ($p$ and $p'$) are equal in magnitude, and $\xi>0$ when $p\ne p'$.  [Since $z_0\ge1$, the absolute value in (\ref{eq1a}) can be ignored.]  
At the largest momenta used in the numerical solutions, $A\simeq 1800$ GeV$^2$, compared to (for example) the pion mass squared $m_\pi^2\simeq 2\times 10^{-4}$ GeV$^2$.  Under these conditions, numerical evaluation of the integral, with its narrow and large forward peak is difficult.

To carry out the integration numerically, we apply a Gauss-Legendre quadrature rule to discretize the integration variable $z$ into a grid of Gauss points distributed between -1 and +1. However, it is convenient to map the integration variable $z\to y$, where $y$ will vary slowly in the region where $F(z)$ is large.  One possible function that will accomplish this, and still remain bounded between $-1$ and 1 is
\bea
z(y)=\frac{\tanh \left[y/(b_0-1)\right]}{\tanh \left[1/(b_0-1)\right]}\label{eq1}\, .
\eea
When $b_0$ is close to 1, this function spreads out the intervals near $z=\pm1$, so that more Gauss points may be placed in the region of the forward peak.  The mapping function, for four different values of $b_0$, is shown in Fig. \ref{fig1}.   Using these mappings, the propagator becomes a function of the integration variable $y$ (instead of the physical variable $z$), and the integrand, as a function of the mapped variable $y$, is shown  in Fig.\ \ref{fig2} (for the same four choices of $b$).  The angular integral is transformed to
\bea
\int_{-1}^1 dz F(z)=\int_{-1}^1 dy\frac{dz}{dy} F\left[z(y)\right]\, .
\eea

Clearly, without any mapping ($b_0>10$) the forward peak is very poorly sampled.  If we take $b_0=1+\epsilon$, with $\epsilon$ small, but larger than zero (to avoid the ``singularity'' at $b_0=1$), we are able to evaluate the integrals accurately.  In practice, we found that $b_0=1.2$ (one of the cases shown in the figures) gave a good sampling of the forward peak and the integration converged rapidly.

\subsection{Mapping for the exchange terms}

The exchange propagator has  the same form as Eq.~(\ref{eq1a}), except now
\bea
z_0&=&\frac{p^2+p'^2-(E_p+E_{p'}-W)^2}{2pp'}\nonumber\\
&\to&-1+\frac{2W}{p}
\eea
where $W$ is the total energy of the $NN$ system in its center of mass, and the last expression holds if $p=p'\to \infty$.  We see that it is now possible for $|z_0|$ to be less than unity, so that the peak can lie in the region of integration.  

To handle this case we need a mapping where the number of Gauss points is strongly distributed near $z=z_0$, where $F(z)$ is sharply peaked.  It is convenient to use a mapping which maps the points $y=z_0, 1, -1$ into $z=z_0, 1, -1$ respectively.  Such a mapping is 
\bea
z(y)=\begin{cases} z_+(y,b) & z_0<y<1 \cr
z_-(y,b)& -1<y<z_0\end{cases}\label{eq6}
\eea
where
\bea
z_+(y,z_0)&=&\sfrac12(1-z_0)\frac{\tanh[{\zeta}(2y-1-z_0))/(1-z_0)] }{\tanh[{\zeta}]}\nonumber\\
&&+\sfrac12(z_0+1) \nonumber\\
z_-(y,z_0)&=&\sfrac12(1+z_0)\frac{\tanh[{\zeta}(2y+1-z_0))/(1+z_0)] }{\tanh[{\zeta}]}\nonumber\\
&&+\sfrac12(z_0-1) \label{eq:2}
\eea
and ${\zeta}$ is a parameter.   This mapping (with ${\zeta}=4$) is shown in Fig.\ \ref{fig3}.  Note that each curve distributes the 90 Gauss points so that they are densely spaced in the region of $z=z_0$.

Both the mapped and unmapped meson propagators are shown in Fig.\ \ref{fig4}, for each of the cases $z_0=-0.5$, 0, and 0.5.  Note that the unmapped propagator has a very narrow peak at $y=z_0$, and that this peak is described by only a few Gauss points, even when the total number of points is quite large (90 in this case).  The mapped functions, however, have their peaks spread out with many Gauss points distributed around the peak.  From this picture, it is easy to see why the mapping allows a more accurate evaluation of the integrals.  (In practice, a better result was obtained by dividing the interval into $\ell_1=[-1,z_0]$ and $\ell_2=[z_0,1]$, and doing separate Gaussian integrations over each interval. 
The points assigned to each interval, $n_i$, depended on the value of $z_0$.  If $z_0<- 0.5$, we chose $n_1=n/4$ and $n_2=n$, for $-0.5\le z_0<0$, $n_1=n/2$ and $n_2=n-n/4$, for $0\le z_0<0.5$, $n_1=n-n/4$ and $n_2=n/2$, and for $0.5\le z_0$, $n_1=n$ and $n_2=n/4$, so that in all cases we actually used $5n/4$ Gauss points.)

Finally, the angular convergence is illustrated in Tables \ref{tab1aa} and \ref{tab2aa} for a typical matrix element [the isoscalar kernel $v^{++}_1$ defined in Eq.~(\ref{eq:110})] at one of the most difficult points: the diagonal elements with $p=p'=p_{\rm max}$ where $p_{\rm max}\simeq $ 30 GeV, the largest momenta used in the numerical solutions of the equations.  Table \ref{tab1aa} shows the convergence of the direct terms (those with $x_0=1$) and Table \ref{tab2aa} the exchange terms (those with $x_0=-1$).   Note that the direct terms calculated with unmapped integrations (the columns labeled with $y$ in the tables) do not converge, and that the convergence of the exchange terms is very marginal.   With the mappings (columns labeled $z$), the direct terms converge even at $n=20$, but excellent results for the exchange terms requires $n\ge 40$.   Because the high-momentum matrix elements only make a small contribution to the solutions of the equation, smaller values of $n$ give very good results for the phase shifts, and these improved angular integrations only affect the accuracy with which the asymptotic deuteron wave functions can be determined.

\vspace{0.2in}

\section{DEUTERON WAVE FUNCTIONS: SOME DETAILS}\label{app:C}


In this appendix we start by showing in detail how the matrix equation (\ref{eq:CST}) is derived, and then use the equation to study the asymptotic behavior of the wave functions, obtaining the results (\ref{eq:asymp}).  We conclude with a demonstration of how the general normalization condition (\ref{eq:relnorm}) reduces to (\ref{eq:norm1}). 

\begin{widetext}

\subsection{Equations for the partial wave deuteron wave functions} \label{app:C1}

The equation for the deuteron wave functions, in partial wave form, can be extracted from the general result (E36) of Ref.~I.  This equation uses a kernel antisymmetrized under particle exchange (with the superscript $J=1$ suppressed)
\bea
\overline{V}^{+\,\rho_2,+\,\rho'_2}_{{\rm d},\,\lambda_1\lambda_2,\lambda'_1\lambda'_2}(p,k)=
\sfrac{1}{2}\Big\{V^{+\,\rho_2,+\,\rho'_2}_{{\rm dir}\;\lambda_1\lambda_2,\lambda'_1\lambda'_2}({\rm p},p_0;k) +\delta_{_{p_0}}V^{\rho_2\,+,+\,\rho'_2}_{{\rm dir}\;\lambda_1\lambda_2,\lambda'_1\lambda'_2}({\rm p},-p_0;k)\Big\}.
\label{eq:C1}
\eea
We will show how the full kernel with well defined parity, Eq.~(\ref{A31a}), will emerge automatically, but the phase of the amplitude under particle interchange cannot emerge automatically in the CST (as discussed in Refs.~I and II); that  phase  must be imposed by hand from the start by using the symmetrized kernel (\ref{eq:C1}).  The phase \change{$\delta_{p0}$}{$\delta_{p_0}=\pm1$} is related but not equal to the  phase 
under particle exchange [because the helicities $\lambda_1, \lambda_2$ are not exchanged in (\ref{eq:C1})].  

Using this kernel, the deuteron bound state equation is
\bea
\frac{m}{E_p}{\bf \Gamma}^{+\rho_2}_{\lambda_1\lambda_2,\lambda_d}(p)&=&
-\int_0^\infty \frac{k^2dk}{(2\pi)^3} \sum_{\mu_1\mu_2\rho} \overline{V}^{\,+\rho_2,+\rho}_{{\rm d},\lambda_1\lambda_2,\mu_1\mu_2}(p,k)\,G^\rho(k)
 \frac{m}{E_k}{\bf \Gamma}^{+\rho}_{\mu_1\mu_2,\lambda_d}(k)\, . \label{E25}
\eea
The dependence of the amplitudes on the total momentum $P$ has been dropped, and the energy factors of $m/E$, written explicitly, will be absorbed later when we go to the matrix notation of Eq.~(\ref{eq:CST}).  At this stage the sum is over all helicities $\mu_1, \mu_2$ and the intermediate rho-spin $\rho$.  

The symmetries of the vertex function are a direct consequence of the symmetries of the kernel.  Under parity, the deuteron kernel transforms as
\bea
{\cal P} \overline{V}^{\,+\rho_2,+\rho}_{{\rm d},\,\lambda_1\lambda_2,\mu_1\mu_2}(p,k) {\cal P}^{-1}=\rho\rho_2 \overline{V}^{\,+\rho_2,+\rho}_{{\rm d},\, -\!\lambda_1 -\!\lambda_2, -\!\mu_1 -\!\mu_2}(p,k).
\eea
Hence the equation satisfied by the transformed vertex function,
\bea
\frac{m}{E_p}\,\rho_2\,{\bf \Gamma}^{+\rho_2}_{-\!\lambda_1\,-\!\lambda_2,\lambda_d}(p)&=&
-\rho_2\int_0^\infty \frac{k^2dk}{(2\pi)^3} \sum_{\mu_1\mu_2\rho} \overline{V}^{\,+\rho_2,+\rho}_{{\rm d},-\!\lambda_1 \,-\!\lambda_2, \mu_1\mu_2}(p,k)\,G^\rho(k)
 \frac{m}{E_k}{\bf \Gamma}^{+\rho}_{\mu_1\mu_2,\lambda_d}(k)
 \nonumber\\
&=&
-\int_0^\infty \frac{k^2dk}{(2\pi)^3} \sum_{\mu_1\mu_2\rho} \overline{V}^{\,+\rho_2,+\rho}_{{\rm d},\lambda_1 \lambda_2, \mu_1\mu_2}(p,k)\,G^\rho(k)
 \frac{m}{E_k}\,\rho \,{\bf \Gamma}^{+\rho}_{-\!\mu_1 \,-\!\mu_2,\lambda_d}(k) \, ,
\eea
is identical to the original equation, showing that the parity relation (\ref{eq:321}) is satisfied.  
Using this result we may reduce the sum over $\mu_1, \mu_2$ from four to two terms.  Dropping the redundant $\rho_1=\rho_1' = +$ indices, we can write the summation over the two values of $\mu_1$ explicitly and use the parity relations to reduce the equation to a form in which only $\mu_1=+\frac12$ appears:
\bea
\frac{m}{E_p}{\bf \Gamma}^{\rho_2}_{\lambda_1\lambda_2,\lambda_d}(p)&=&
-\int_0^\infty \frac{k^2dk}{(2\pi)^3}
 \frac{m}{E_k} \sum_{\mu_2\rho} G^\rho(k) \bigg\{\overline{V}^{\rho_2,\rho}_{{\rm d},\lambda_1\lambda_2,+\frac12\mu_2}(p,k)\,{\bf \Gamma}^{\rho}_{+\frac12\mu_2,\lambda_d}(k)
 +\overline{V}^{\rho_2,\rho}_{{\rm d},\lambda_1\lambda_2,-\frac12\mu_2}(p,k)\,{\bf \Gamma}^{\rho}_{-\frac12\mu_2,\lambda_d}(k)\bigg\}
 \nonumber\\
 &=&-\int_0^\infty \frac{k^2dk}{(2\pi)^3}
 \frac{m}{E_k} \sum_{\mu_2\rho}  \bigg\{\overline{V}^{\rho_2,\rho}_{{\rm d},\lambda_1\lambda_2,+\frac12\mu_2}(p,k)\,
 +\rho\,\overline{V}^{\rho_2,\rho}_{{\rm d},\lambda_1\lambda_2,-\frac12\,,-\!\mu_2}(p,k)\bigg\}G^\rho(k){\bf \Gamma}^{\rho}_{+\frac12\mu_2,\lambda_d}(k)\bigg\}
 \nonumber\\
  &=&-\int_0^\infty \frac{k^2dk}{(2\pi)^3}
 \frac{m}{E_k} \sum_{\mu_2\rho}  {\bf V}^{\rho_2,\rho}_{{\rm d},\lambda_1\lambda_2,+\frac12\mu_2}(\delta_{p_0},\rho)\,
G^\rho(k){\bf \Gamma}^{\rho}_{+\frac12\mu_2,\lambda_d}(k)\bigg\}\, , \label{E25a}
\eea
where \change{}{in the last line} we have recovered the fully symmetrized kernel of Eq.~(\ref{A31a}).  The deuteron channel must have the phases $\delta_{p0}=\rho_2^\lambda$, appropriate to an isospin 0 state, and $\delta_S=\rho'_2=\rho$.  Using the notation ${\bf V}={\bf V}^{\rho_2,\rho}_{\lambda_2\mu_2,\delta_{_{p_0}}}$  (and, for the helicities, $\pm\frac12\to\pm$) leads to the identifications
\bea
{\bf V}_{++,+}^{\rho_2\,+}(p,k)&=&v_9^{\rho_2\,+}
\qquad{\bf V}_{++,+}^{\rho_2\,-}(p,k)=v_1^{\rho_2\,-}
\qquad
{\bf V}_{+-,+}^{\rho_2\,+}(p,k)=v_{11}^{\rho_2\,+}
\qquad{\bf V}_{+-,+}^{\rho_2\,-}(p,k)=v_3^{\rho_2\,-}
\nonumber\\
{\bf V}_{-+,+}^{+\,+}(p,k)&=&v_{12}^{+\,+}
\qquad{\bf V}_{-+,+}^{+\,-}(p,k)=v_8^{+\,-}
\qquad
{\bf V}_{--,+}^{+\,+}(p,k)=v_{10}^{+\,+}
\qquad{\bf V}_{--,+}^{+\,-}(p,k)=v_6^{+\,-}
\nonumber\\
{\bf V}_{-+,-}^{-\,+}(p,k)&=&v_{16}^{-\,+}
\qquad{\bf V}_{-+,-}^{-\,-}(p,k)=v_4^{-\,-}
\qquad
{\bf V}_{--,-}^{-\,+}(p,k)=v_{14}^{-\,+}
\qquad{\bf V}_{--,-}^{-\,-}(p,k)=v_2^{-\,-}
\label{eq:C10}
\eea
%
where the kernels are written in the order for coupling to $\ket{{\bf \Gamma}^+_{++}}, \ket{{\bf  \Gamma}^-_{++}}, \ket{{\bf  \Gamma}^+_{+-}},$ $\ket{{\bf \Gamma}^-_{+-}}$, respectively.   Note that they reproduce the matrix (\ref{coupled}), finishing this discussion.

\subsection{Asymptotic behavior of the wave functions}

Starting from Eq.~(\ref{E25a}),  the wave function (\ref{eq:d1a}) has the following general behavior at large p
\bea
&&\psi^\rho_{\lambda_1\lambda_2,\lambda_d}({\rm p})=N_d\, G^\rho(p) \frac{m}{E_p} {\bf \Gamma}^\rho_{\lambda_1\lambda_2,\lambda_d}(p)
\simeq  G^\rho(p) \int_0^\infty d{\rm k}\, {\bf V}_{\rm d}(p,k) C(k) ,\qquad \label{eq:C11}
\eea
\end{widetext}
where, in this section, we return to the notation ${\rm k}\equiv |{\bf k}|$ for the magnitude of the three-momentum, and ${\bf V}_{\rm d}$ is the appropriate partial wave projection of the Feynman OBE amplitudes [shown in detail in Eq.~(\ref{eq:C10})].    All we need know about the function $C(k)$ is that it provides the convergence for the $k$ integral.  The kernels include a factor of $m/E_p$, which contributes to the large p behavior.


The kernels ${\bf V}_{\rm d}$ are multiplied by nucleon form factors [recall Eq.~(\ref{eq:nff})], one of which depends on the intermediate momentum ($k$) and the other on the final momentum ($p$).   The final state form factor, $h(p)$, plays an important role in the asymptotic behavior of the wave function.  It depends only on $p^2$, which is the square of the mass of the  off-shell  nucleon
\bea
p^2&=&(M_d-E_p)^2-{\rm p}^2
\nonumber\\
p^2 &\rightarrow & -2M_d{\rm p} \quad \mbox{for p}\gg m \, .
\eea
Hence, 
\bea
h(p^2) \rightarrow \frac{(\Lambda_N^2-m^2)^4}{16M^4_d\,{\rm p}^4} \quad \mbox{for p}\gg m,  \label{eq:313}
\eea
so the form factor behaves like p$^{-4}$.  This contributes a k$^{-4}$ behavior to the function $C(k)$, which, together with the internal  nucleon form factor $h(k)$ that is part of ${\bf V}_d$, assures that the integral over k will converge.  Hence we may conclude that the wave functions go like
\bea
&&\psi^\rho_{\lambda_1\lambda_2,\lambda_d}({\rm p})\to  G^\rho(p) {\bf V}_{\rm d}(p,\bar k) C(\bar k) \, , \label{eq:C10a}
\eea
where the mean value theorem has been used to write the integral over k, with $\bar k$ a four-vector  whose three-vector part has some fixed (unknown) value $\bar {\rm k}$ in the interval [0,$\infty)$.

To finish this discussion we examine the asymptotic behavior of the partial wave amplitudes for one of the OBE terms in the kernel.  If the form factors $h(p)$ and $h(k)$ are removed, this integral has the generic form
\bea
D_b({\rm p},\bar{\rm k})=\int_{-1}^1 dz\, g_b({\rm p})\,\frac{m}{E_p} \frac{f(\Lambda_b,q)}{m_b^2+|q^2|} d(z)\qquad \label{eq:336}
\quad 
\eea
where $q^2$ is the momentum transferred by the boson with mass $m_b$, form factor $f(\Lambda_b,q)$, and momentum dependent coupling $g_b($p)$\to {\rm p}^\ell$ (as p$\to \infty$, where $\ell$ is yet to be determined). We have isolated the $m/E_p$ term incorporated into the definition of ${\bf V}_{\rm d}$, and $d(z)$ includes the $z$ dependence that arises from the generic rotation matrices of the type tabulated in Table \ref{tab:3}.  In Ref.~I the meson form factors have the form
\bea
f(\Lambda_b,q)=\left[\frac{\Lambda_b^2}{\Lambda_b^2+|q^2|}\right]^{n_b} \, ,
\eea
where $n_b=4$ for all mesons.

The asymptotic behavior of the integral (\ref{eq:336}) depends on whether or not it is a {\it direct\/} or an {\it exchange\/} term.  
The momentum transferred by the mesons in the direct terms  
\bea
q^2&&=(E_p-E_k)^2-({\bf p}-{\bf k})^2
\nonumber\\
&&\to2m^2-2{\rm p}(E_k-{\rm k}z) \quad ({\rm as}\;{\rm p}\to\infty,\;{\rm k\; fixed})\qquad
\eea
and is never positive, so the integrand is  a smooth function of $z$.  The asymptotic behavior, for a fixed k=k$_0$, is therefore
\bea
D_b^{\rm dir}({\rm p}, \bar{\rm k})\to\frac{{\rm p}^\ell}{{\rm p}^{n_b+2}}
\eea

As shown below, these terms are vanishingly small compared to the exchange terms.

The exchange terms have a very different structure.  Here the momentum transferred by the meson is 
\bea
q^2&=&(M_d-E_p-E_k)^2-({\bf p}+{\bf k})^2
\nonumber\\
&\to&m^2+(M_d-E_k)^2-{\rm k}^2
-2{\rm p}\,(M_d-E_k+{\rm k}z) 
\nonumber\\
&\to& -2{\rm p}(M_d-E_k+{\rm k}z)\quad\;{\rm as\;p} \to\infty\,.
\eea
When p is very large, this is zero along a line in the ${\rm k},z$ plane given approximately by $M_d-E_k+{\rm k}z=0$. Along this line the integrand has a cusp which sits at
\bea
z_c=-\frac{M_d-E_k}{{\rm k}}
\eea
in the region of integration ($-1\le z\le1$) for values of ${\rm k}$ in the interval 
\bea
\frac{M_d^2-m^2}{2M_d}\equiv {\rm k}_{\rm min}<{\rm k}<\infty\, .
\eea
Contributions from the cusp clearly dominate the integral, because in that region the integrand is constant.  Hence, letting $m_b=\Lambda_b$ in order to simplify the determination of the asymptotic p-dependence, the exchange integral is approximately
%
\begin{align}
\frac{D_b^{\rm ex}({\rm p, \bar k})}{{\rm p}^{\ell-1}}&\simeq\int_{-1}^1 dz \left[\frac{\Lambda_b^2}{\Lambda_b^2+|q^2|}\right]^{n_b+1} d(z)
\nonumber\\
&\simeq\int_{-1}^{z_c} dz \left[\frac{\Lambda_b^2}{\Lambda_b^2+2{\rm p\bar k}(z_c-z)}\right]^{n_b+1}d(z)
\nonumber\\&\qquad 
+\int^{1}_{z_c} dz \left[\frac{\Lambda_b^2}{\Lambda_b^2+2{\rm p\bar k}(z-z_c)}\right]^{
n_b+1}d(z)
\nonumber\\
&\simeq\frac{2\Lambda_b^2}{{n_b}{\rm p\bar k}}d(z_c) ,
\end{align}
%
where we have integrated by parts, keeping only the leading term (the corrections go like higher powers of 1/p).  

Collecting these results together we see that the reduced partial wave amplitude, $D_b$, behaves like
\bea
D({\rm p,k}_0)\to \begin{cases} {\rm p}^{\ell-(2+n_b)} & {\rm direct}\cr
{\rm p}^{\ell-2} & {\rm exchange}  \end{cases}\qquad \label{eq:falloff}
\eea
where $n_b$ is the power of the form factor at the $NNb$ vertex and $\ell$ the power of the p dependence of the $NNb$ vertex function.  

It remains to determine the power $\ell$.  First, note that the single particle $\rho$-spin matrix elements given in Table \ref{tab:Dirac} each contribute a factor of $\sqrt{{\rm p}/(2m)}$ [from the high momentum limits (\ref{eq:A5})]; their product gives one power of p implicitly contained in Eqs.~(\ref{eq:A44})-(\ref{eq:A47}).

\begin{figure*}
\centerline{
\includegraphics[width=5.5in]{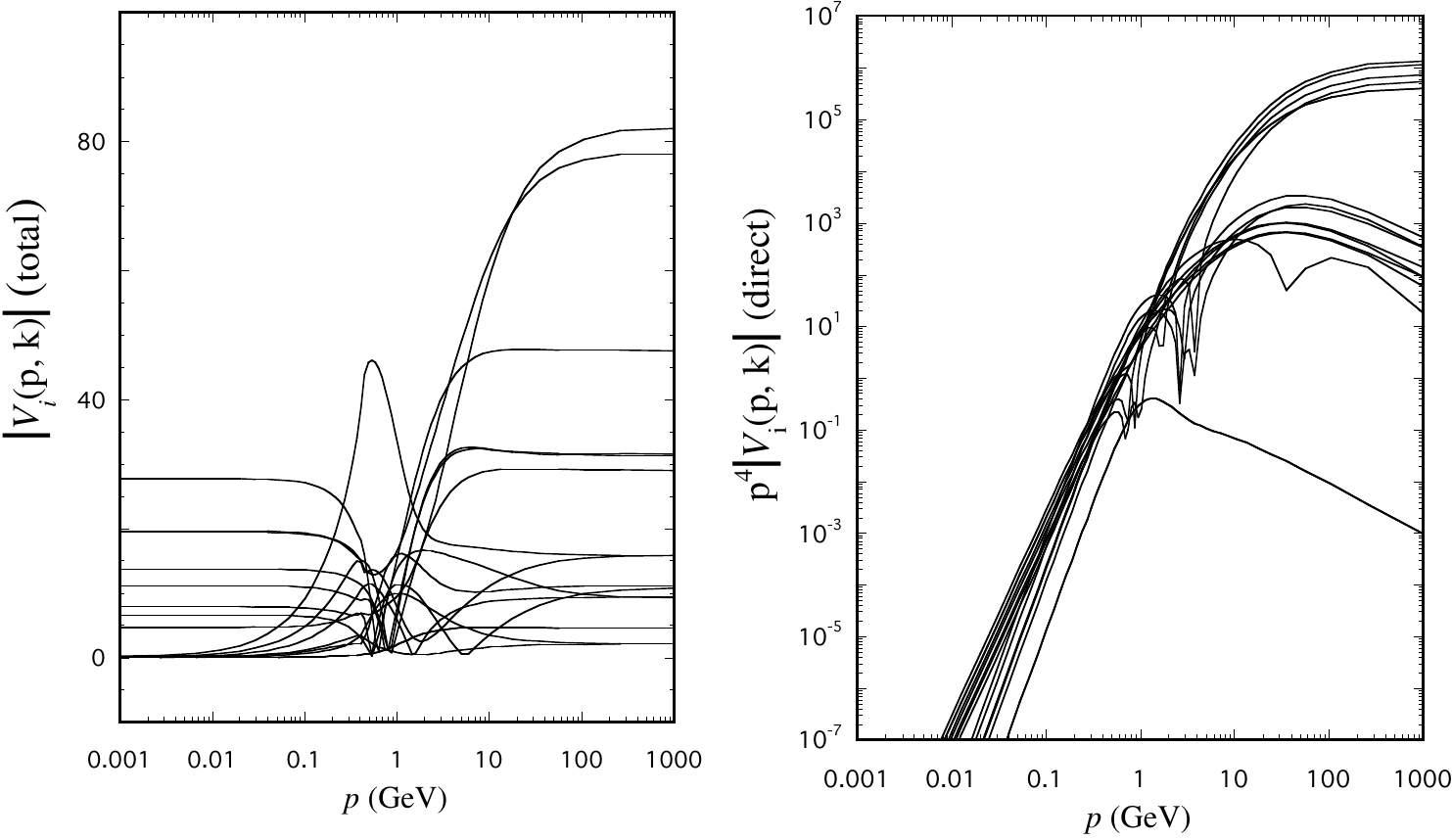}
}
\caption{\footnotesize\baselineskip=10pt The p dependence (for k fixed at 816 MeV) of the 16 model WJC-1 deuteron kernels $V_i($p, k) defined in Eq.~(\ref{eq:C10}) (but with the nucleon form factors set to unity).  The left panel shows clearly that  the {\it total\/} kernel (direct plus exchange) approaches a constant at large p, as predicted by (\ref{eq:falloff}) with $\ell=2$.  The left panel shows that the asymptotic p dependence of 5 of the {\it direct\/} kernels behaves like p$^{-4}$, again predicted by (\ref{eq:falloff}) with $n_b=4$ and $\ell=2$.  The other direct kernels decrease more rapidly, presumably because they have an $\ell<2$.  }
\label{fig:asy}
\end{figure*} 


Next, examine any additional p dependence coming from the two particle $\rho$-spin matrix elements given in Eqs.~(\ref{eq:A44})-(\ref{eq:A47}).  There are terms of order p coming from the off-shell energy factors $\Delta E_i$ defined in (\ref{eq:A24}), but the condition (\ref{eq:A25}) shows that these factors cannot be multiplied together to make a term of order p$^2$.  Special terms of order p$^2$ seem to be present in the vector meson exchange terms, but we will show below that these terms cancel.  Hence the final result is the product of two terms of order p giving $\ell=2$.

To show the cancellation of the special p$^2$ terms contributing to vector-meson exchange, consider the three terms in Eq.~(\ref{eq:A46}) which give rise to them: the product of $R_{v1}$'s (first term), the p$^2$ multipying the $D_4$ terms (fourth term), and    the last term.  It is sufficient to look at the exchange-term matrix elements where particle 1 is off-shell in the final state, so that $\Delta E_2=0$.  At large p use the fact that $\Delta E_1\to(\rho_1+1)$p to obtain
\bea
R_{v1}R_{v1}&\to& {\rm p}^2 \frac{g_v^2 \kappa_v}{(2m)^2} \Big[\rho_1\kappa_v-(\rho_1+1)\nu_v\Big] {\bf 1}_1{\bf 1}_2
\nonumber\\
D_4 D_4&\to& {\rm p}^2\left( \frac{g_v\kappa_v}{2m} \right)^2 {\bf 1}_1{\bf 1}_2
\nonumber\\
R_{v3}D_4&\to&{\rm p}^2 \,\frac{g_v^2\kappa_v}{(2m)^2} (\kappa_v-\nu_v)(\rho_1+1)\,2\lambda_1{\bm \tau}_1^1{\bf 1}_2\, .\qquad
\eea
Adding these together, and replacing $\rho_1$ by $\rho_2$ [in accordance with the notation for the exchange terms in Eq.~(\ref{eq:110})] gives the following compact result
\bea
g_v({\rm p})&\simeq& {\rm p}^2 \,\frac{g_v^2\kappa_v}{(2m)^2} (\kappa_v-\nu_v)(\rho_2+1)
\nonumber\\
&&\qquad\times\Big[{\bf 1}_1+2\lambda_1{\bm \tau}_1^1\Big]{\bf 1}_2\, .
\eea
Note that this is zero if $\rho_2=-1$, insuring immediately that 8 of the deuteron kernels (\ref{eq:C10}) have no special terms of order p$^2$.  To see that this is also true of the other 8 cases set  $\rho_2=1$ (and recall that $\rho'_1=1$) and substitute for the matrix elements ${\bf 1}$ and ${\bm \tau}^1$ using the results from Table \ref{tab:Dirac} (with $\rho_2\leftrightarrow \rho_1$) and the high momentum limits (\ref{eq:A5})
\bea
g_v({\rm p})&\simeq&2{\rm p}^2\sqrt{\frac{ {\rm p}}{2m}} \,\frac{g_v^2\kappa_v}{(2m)^2} (\kappa_v-\nu_v)
\nonumber\\
&&\qquad\times(1-2\lambda_1)(c'-s'){\bf 1}_2\, .
\eea
This is zero if $\lambda_1=\frac12$, which is true of all of the amplitudes (and, in particular, the remaining 8).  We have proved that the special p$^2$ terms coming from vector meson exchange cancel, and that $\ell=2$.  The conclusion is that the wave functions should go like 
\bea
z^\rho({\rm p})\to G^\rho(p) h(p^2) D^{\rm ex}({\rm p, k}_0)
\eea
which gives the result (\ref{eq:asymp}) as expected.

To confirm the predictions of Eq.~(\ref{eq:C10}), Fig.~\ref{fig:asy} shows how the 16 deuteron kernels $V_i($p, k) vary with p for a fixed k .


\subsection{Reduction of Eq.~(\ref{eq:relnorm}) to Eq.~(\ref{eq:norm1})}\label{app:norm}



Substituting the expansion (\ref{eq:34a}) into Eq.~(\ref{eq:relnorm}), and using the orthogonality of the Dirac spinors, compactly written as
\bea
\bar u^{\rho}(-{\bf p},\lambda)\gamma^0 u^{\rho'}(-{\bf p},\lambda')&&=\frac{E_p}{m}\delta_{\lambda'\lambda}\delta_{\rho'\rho}
\eea
gives
%
\bea
\delta_{\lambda_d\lambda'_d}&=&\int d^3 p\; \psi^{\rho\,\dagger}_{\lambda_1\lambda_2,\lambda'_d}({\bf p})\,\psi^\rho_{\lambda_1\lambda_2,\lambda_d}({\bf p})
\nonumber\\&&
-\int\!\! \int \frac{d^3 p d^3 p'}{(2\pi)^3}  {\psi}^{\rho \dagger}_{\lambda_1\lambda_2,\lambda_d}({\bf p})
\,\Delta \overline {V}^{\rho\rho'}_{\lambda_1\lambda_2,\lambda'_1\lambda'_2}(p,p';P)
\nonumber\\&& \quad\qquad\qquad\times
{\psi}^{\rho'}_{\lambda'_1\lambda'_2,\lambda'_d}({\bf p}')\, ,\qquad
\label{eq:relnorm2}
\eea
%
where the $\overline{V}^{\rho\rho'}\equiv \overline{V}^{+\rho,+\rho'}$ includes the factors of $m^2/E_pE_{p'}$ [cf. Eq.~(E13) of Ref.~I].

Next we express the helicity amplitudes in terms of the wave functions $z_\ell({\rm p})$ using (\ref{B6}).  To this end recall Eq.~(\ref{A3}) for the two-component helicity spinors and use the convenient relation $i\sigma_2\chi_{_\lambda}=-2\lambda \chi_{_{-\lambda}}$ and  the matrix elements
\bea
\chi^\dagger_{_{-\lambda_2}}\bm\sigma\cdot\bm\xi_{_{\lambda_d}} \,i\sigma_2\chi_{_{\lambda_1}}&&=\sqrt{2}^{\,|\lambda|}d^1_{\lambda_d,\lambda}(\theta)
\nonumber\\
\chi^\dagger_{_{-\lambda_2}}\bm\sigma\cdot{\bf \hat p}\, i\sigma_2\chi_{_{\lambda_1}}&&=\chi^\dagger_{_{-\lambda_2}}\chi_{_{-\lambda_1}}=\delta_{\lambda_1,\lambda_2}
\nonumber\\
{\bf \hat p}\cdot \bm\xi_{_{\lambda_d}}&&=d^1_{\lambda_d,0}(\theta),
\eea
where $\lambda=\lambda_1-\lambda_2$ and the rotation matrices are
\bea
&&d^1_{00}(\theta) = \cos\theta
\nonumber\\
&&d^1_{10}(\theta)=
d^1_{0,-1}(\theta)=-d^1_{0,1}(\theta)=-d^1_{-1,0}(\theta)=-\sfrac1{\sqrt{2}}\sin\theta
\nonumber\\
&&d^1_{11}(\theta)=d^1_{-\!1,\,-\!1}(\theta)=\sfrac12(1+\cos\theta)
\nonumber\\
&&d^1_{-\!1,1}(\theta)=d^1_{1,\,-\!1}(\theta)=\sfrac12(1-\cos\theta).
\eea
The result is
\bea
&&\psi_{\lambda_1\lambda_2,\lambda_d}^+({\bf p})=\frac{1}{\sqrt{8\pi}} \Bigg\{\left(u({\rm p})-\frac{1}{\sqrt{2}}\,w({\rm p})\right)\,
\sqrt{2}^{\,|\lambda|}d^1_{\lambda_d,\lambda}(\theta)
\nonumber\\
&&\qquad\qquad\qquad+\delta_{\lambda_1,\lambda_2} \frac{3w({\rm p})}{\sqrt{2}}\,d^1_{\lambda_d,0}(\theta)\Bigg\}
\nonumber\\
&&\psi_{\lambda_1\lambda_2,\lambda_d}^-({\bf p})=-2\lambda_1\sqrt{\frac{3}{8\pi}} \Bigg\{\delta_{\lambda_1,\lambda_2} \left(v_s({\rm p})-\frac{v_t({\rm p})}{\sqrt{2}}\right)\,d^1_{\lambda_d,0}(\theta)
\nonumber\\
&&\qquad\qquad\qquad+\frac{v_t({\rm p})}{\sqrt{2}}\,\sqrt{2}^{\,|\lambda|}d^1_{\lambda_d,\lambda}(\theta)\Bigg\}.\qquad
\eea
Explicitly, for each individual case, using $\lambda=\pm\frac12$
\bea
\psi_{\lambda\lambda,\lambda_d}^+({\bf p}) &&=\frac{1}{\sqrt{8\pi}}\left(u({\rm p})+\sqrt{2}\,w({\rm p})\right)\,d^1_{\lambda_d,0}(\theta)
\nonumber\\
\psi_{\lambda,-\!\lambda,\lambda_d}^+({\bf p}) &&=\frac{1}{\sqrt{4\pi}}\left(u({\rm p})-\frac1{\sqrt{2}}\,w({\rm p})\right)\,d^1_{\lambda_d,\,2\lambda}(\theta)
\nonumber\\
\psi_{\lambda\lambda,\lambda_d}^-({\bf p}) &&=-2\lambda\sqrt{\frac{3}{8\pi}} \,v_s({\rm p})\,d^1_{\lambda_d,0}(\theta)
\nonumber\\
\psi_{\lambda,\,-\!\lambda,\lambda_d}^-({\bf p}) &&=-2\lambda\sqrt{\frac{3}{8\pi}} \,v_t({\rm p})\,d^1_{\lambda_d,2\lambda}(\theta)\, .
\eea

We now can reduce the first term in the normalization condition (\ref{eq:relnorm2}).  Taking $\lambda_d=\lambda'_d$ and using the normalization condition (\ref{eq:djnorm}) for the $d^1$ functions gives
\bea
1=&&2\pi\int_0^\infty p^2 dp\int_0^\pi \sin\thetaÊd\theta\Big\{2| \psi_{\lambda\lambda,\lambda_d}^+({\bf p})|^2 +2|\psi_{\lambda,\,-\!\lambda,\lambda_d}^+({\bf p}) |^2 
\nonumber\\
&&\qquad\qquad+2|\psi_{\lambda\lambda,\lambda_d}^-({\bf p})|^2 +2|\psi_{\lambda\lambda,\lambda_d}^+({\bf p})|^2\Big\}
\nonumber\\
=&&\frac{1}{3}\int_0^\infty p^2 dp\Big\{\left(u({\rm p})+\sqrt{2}\,w({\rm p})\right)^2 
\nonumber\\
&&\qquad+2\left(u({\rm p})-\frac1{\sqrt{2}}\,w({\rm p})\right)^2 +3 v_s^2({\rm p})+3v_t^2({\rm p})\Big\}
\nonumber\\
=&&\int_0^\infty p^2 dp\Big\{u^2({\rm p})+w^2({\rm p})+ v_t^2({\rm p})+v_s^2({\rm p})\Big\}
\eea
in agreement with Eq.~(\ref{eq:norm1}).  The derivative terms can be similarly reduced and expressed in terms of the kernels given in Eq.~(\ref{eq:C10}).

Before leaving this section we note that the partial-wave deuteron wave functions follow from the normalization condition (\ref{eq:djnorm}) and the definitions (\ref{eq:320a})
\begin{align}
\psi^+_{++,\lambda_d}({{\rm p}}) &=\sqrt{\sfrac38}\int_{-1}^1 dz \,|d^1_{\lambda_d,0}(\theta)|^2\,
\left(u+\sqrt{2}w\right)
\nonumber\\
&=\sfrac1{\sqrt{6}}\left(u+\sqrt{2}w\right)
\nonumber\\
\psi^+_{+-,\lambda_d}({{\rm p}}) &=\sqrt{\sfrac38}\int_{-1}^1 dz \,|d^1_{\lambda_d,1}(\theta)|^2\,
\left(\sqrt{2}u-w\right)
\nonumber\\
&=\sfrac1{\sqrt{6}}\left(\sqrt{2}u-w\right)
\nonumber\\
\psi^-_{++,\lambda_d}({{\rm p}}) &=-
\sfrac3{\sqrt{8}} \int_{-1}^1 dz \,|d^1_{\lambda_d,0}(\theta)|^2\,v_s
=-\sfrac1{\sqrt{2}} \; v_s
\nonumber\\
\psi^-_{+-,\lambda_d}({{\rm p}}) &=-
\sfrac3{\sqrt{8}} \int_{-1}^1 dz \,|d^1_{\lambda_d,1}(\theta)|^2\, v_t
=-\sfrac1{\sqrt{2}}  \; v_t\, .
\label{eq:120}
\end{align}
%


\subsection{Coordinate-space expansion functions} \label{app:Rspace}

The coordinate-space wave functions discussed in Sec.~\ref{sec:Rspace} are constructed from the spherical Bessel transforms (\ref{eq:besseltrans}), which can be written in terms of the $j_0$ Bessel transform 
\bea
G^i_\ell({r})&=& {r}\sqrt{\sfrac2\pi}\int_0^\infty {p}^2d{p}\,j_\ell({p}{r})\,G^i_\ell({p})
\nonumber\\
&=&(-1)^\ell \sqrt{\sfrac2{\pi}}r^{\ell+1}\left(\frac{1}{r}\frac{d}{dr}\right)^\ell
\int_0^\infty {p}^2d{p}\,j_0({p}{r})\,\frac{G^i_\ell({p})}{p^\ell}
\nonumber\\
&=&(-1)^\ell\frac{r^{\ell+1}}{M_i^\ell}\left(\frac{1}{r}\frac{d}{dr}\right)^\ell \Big[\frac{G_0^i(r)}{r}\Big]\, .\qquad
\label{eq:besseltrans2}
\eea
Hence, for S and P states it is sufficient to calculate the following Fourier sine transforms:
\bea
G^i_0(r)&=&\sqrt{\sfrac2\pi}\int_0^\infty p\, dp \sin(pr)G^i_0(p)=
\nonumber\\&=&
A_i\Bigg\{e^{-z_i}
-e^{-Z_i}\left[1+\sfrac12 Z_i\left(1-R_i^2\right)\right]\Bigg\}
\nonumber\\
G^n_0(r)&=&\sqrt{\sfrac2\pi}\int_0^\infty p\, dp \sin(pr)G^n_0(p)
\nonumber\\
&=&\frac2{3\pi} M_n^2\,Z_n^2 K_1(Z_n)
\eea
where $z_i, Z_i, R_i$, and $A_i$ were given in Eq.~(\ref{eq:343a})
and the modified Bessel functions of the second kind are
\bea
K_n(z)=\frac{ z^n}{(2n-1)!!}\int_1^\infty dt\,e^{-zt} (t^2-1)^{n-\frac12} .\qquad\quad
\eea

The P-state wave functions are obtained by differentiation, as outlined in Eq.~(\ref{eq:besseltrans2}).  They are
\bea
G^i_1(r)&=&-\frac{r}{M_i} \frac{d}{dr}\Big[\frac{G_0^i(r)}{r}\Big]=
A_i\Bigg\{R_ie^{-z_i}\Big[1+\frac1{z_i}\Big]
\nonumber\\
&&\qquad-e^{-Z_i}\left[1+\frac1{Z_i}+\sfrac12 Z_i\left(1-R_i^2\right)\right]\Bigg\},\qquad
\nonumber\\
G^n_1(r)&=&-\frac{r}{M_n} \frac{d}{dr}\Big[\frac{G_0^n(r)}{r}\Big]
=\frac2{3\pi} M_n^2 Z_n^2 K_0(Z_n).
\eea

Finally, the D-state wave functions are computed by first transforming the functions 
\bea
G^i_{0d}(p)&=&\sqrt{\frac2\pi}\frac{m_i^2\,M_{i}^{6}}{(m_i^2+p^2)(M_{i}^2+p^2)^{3}}
\nonumber\\
G^n_{0d}(p)&=&\sqrt{\frac2\pi}\frac{M_n^{7}}{(M_n^2+p^2)^{7/2}} 
\eea
which gives
\begin{widetext}
\bea
G^i_{0d}(r)&=&\sqrt{\sfrac2\pi}\int_0^\infty p\, dp \sin(pr)G^i_{0d}(p)
=B_i\Bigg\{e^{-z_i}-e^{-Z_i} -\sfrac18 Z_i e^{-Z_i} 
\bigg[\left(5-6 R_i^2 +R_i^4\right)+ Z_i\left(1-R_i^2\right)^2\bigg]\Bigg\}
\nonumber\\
G^n_{0d}(r)&=&\sqrt{\sfrac2\pi}\int_0^\infty p\, dp \sin(pr)G^n_{0d}(p)
=\frac2{15\pi} M_n^2\,Z_n^3 K_2(Z_n)
\eea
where $B_i$ was given in Eq.~(\ref{eq:343a}).  Differentiating these following (\ref{eq:besseltrans2}) gives
\bea
G^i_2(r)&=&\frac{r^2}{M_n^2} \left(\frac{d}{dr}\frac1r \right)^2 G_{0d}^i(r)
=B_i\Bigg\{R_i^2e^{-z_i}\Big[1+\frac3{z_i}+\frac3{z_i^2}\Big]
\nonumber\\&&\qquad
-\,e^{-Z_i}\Big[1+\frac3{Z_i}+\frac3{Z_1^2}
+\sfrac12(1-R_i^2)(1+Z_i)+\sfrac18 Z_i^2\left(1-R_i^2\right)^2\Big]\Bigg\},
\nonumber\\
G^n_2(r)&=&\frac{r^2}{M_n^2}\left[\frac{d}{dr}\frac1r \right]^2\!\! G_{0d}^n(r)
=\frac2{15\pi} M_n^2 Z_n^3 K_0(Z_n).\quad
\eea
\end{widetext}

At large $r$ the functions $G_\ell^i$ go like
\bea
\lim_{r\to\infty}G_0^i(r)&=&A_i\,e^{-z_i}
\nonumber\\
\lim_{r\to\infty}G_1^i(r)&=&A_iR_i\,e^{-z_i} \Big[1+\frac1{z_i}\Big]
\nonumber\\
\lim_{r\to\infty}G^i_2(r)&=&
B_i \,R_i^2\,e^{-z_i}\Big[1+\frac3{z_i}+\frac3{z_i^2}\Big].
\eea
This is the asymptotic behavior expected for solutions of the Schr\"odinger equation with orbital angular momentum $\ell$.   The tail functions fall-off like exponentials multiplied by a fractional power of $r$
\bea
\lim_{r\to\infty}G_0^n(r)&=&\frac13\sqrt{\frac2\pi} M_n^2 Z_n^{3/2}\,e^{-Z_n}
\nonumber\\
\lim_{r\to\infty}G_1^n(r)&=&\frac13\sqrt{\frac2\pi} M_n^2 Z_n^{3/2}\,e^{-Z_n}
\nonumber\\
\lim_{r\to\infty}G_2^n(r)&=&\frac1{15}\sqrt{\frac2\pi} M^2_n Z_n^{5/2}\,e^{-Z_n}\, .
\eea

At small $r$ the functions $G_\ell^i$ have the expected $r^{\ell+1}\sim Z_i^{\ell+1}$ behavior
\bea
\lim_{r\to0}G_0^i(r)&=&\sfrac12A_i Z_i\Big(1-R_i\Big)^2
\nonumber\\
\lim_{r\to0}G_1^i(r)&=&\sfrac16 A_iZ_i^2 (1-R_i)^2(1+2R_i)
\nonumber\\
\lim_{r\to0}G_2^i(r)&=&\sfrac1{40} B_i Z_i^3 (1-R_i)^3(1+3R_i+\sfrac83 R_i^2).\qquad
\eea
However, the tail functions contribute some nonanalytic behavior at small $r$:
\bea
\lim_{r\to0}G_0^n(r)&=&\frac2{3\pi} M_n^2 Z_n \, .
\nonumber\\
\lim_{r\to0}G_1^n(r)&=&-\frac2{3\pi} M_n^2 Z_n^2\Big(\log Z_n - \log2 +\gamma\Big)
\nonumber\\
\lim_{r\to0}G_2^n(r)&=&-\frac2{15\pi} M_n^2 Z_n^3\Big(\log Z_n - \log2 +\gamma\Big),\qquad\quad
\eea
where $\gamma=0.5772$ is Euler's constant.



\end{document}